\documentclass[12pt]{report}

\usepackage{titlesec}  
\usepackage{fancyhdr}  
\usepackage{graphicx}  
\usepackage{amsmath}   
\usepackage{amssymb}   
\usepackage{booktabs}
\usepackage{multirow}
\usepackage{amsmath}
\usepackage{bm}

\usepackage{graphicx}
\usepackage{array}
\usepackage{colortbl}
\usepackage{makecell}

\usepackage{xcolor}
\definecolor{mygray}{gray}{0.9}

\definecolor{Gray}{gray}{0.9}

\usepackage{caption}

\usepackage{hyperref}  
\hypersetup{
    colorlinks=true,    
    linkcolor=black,     
    citecolor=black,     
    filecolor=black,  
    urlcolor=black,      
    linktocpage=true    
}

\usepackage{framed}

\usepackage{lipsum}    

\titleformat{\chapter}[hang]{\Large\bfseries}{Chapter \thechapter:}{1em}{}
\titleformat{\section}[hang]{\Large\bfseries}{\thesection}{1em}{}
\titleformat{\subsection}[hang]{\large\bfseries}{\thesubsection}{1em}{}
\titleformat{\subsubsection}[hang]{\normalsize\bfseries}{\thesubsubsection}{1em}{}

\begin{document}

\begin{titlepage}
    \centering
    \vspace*{2cm}
    {\Huge\bfseries Automated Retinal Image Analysis and Medical Report Generation through Deep Learning \\[1.5cm]}
    {\Large Ph.D. Candidate: \\ Jia-Hong Huang \\[1cm]}
    {\large Ph.D. Supervisor: \\ Prof. Dr. Marcel Worring \\[2cm]}
    {\large Doctor of Philosophy Thesis \\[0.5cm]}
    {\large Department of Computer Science \\[0.5cm]}
    {\large University of Amsterdam \\[2cm]}
\end{titlepage}

\pagenumbering{roman}
\tableofcontents
\newpage
\listoffigures
\newpage
\listoftables
\newpage

\pagenumbering{arabic}

\chapter{Introduction}\label{ch:introduction}

Medical imaging, particularly retinal imaging, plays a critical role in diagnosing and managing retinal diseases, capturing detailed visual information, providing context for patient conditions, representing temporal changes in disease progression, offering clinical realism, facilitating precise analysis, and enabling effective communication among healthcare providers \cite{yang2018novel,huck2019auto,jing2018automatic,huang2021deepopht}. The increasing prevalence of retinal diseases and the advancements in imaging technologies have resulted in substantial growth of retinal image data generated daily in clinical settings \cite{yang2018novel,liu2019synthesizing,huck2019auto,huang2021deepopht}. While this abundance of medical imagery serves as a valuable source of information for diagnosing and treating retinal diseases, it also presents challenges for clinicians who need to efficiently and accurately interpret these images and generate medical reports to make informed treatment decisions.
In response to this challenge, automated medical report generation techniques have emerged, aiming to extract the most significant and relevant information from retinal images and present it in a concise, interpretable format \cite{huang2021deepopht,huang2021contextualized,wu2023expert}.

The quality of automated medical reports is crucial for their utility in clinical practice and research. Inaccurate or incomplete reports can lead to misdiagnosis and suboptimal treatment, negatively impacting patient outcomes. In research, poorly generated reports can hinder study interpretations and slow medical advancements. Therefore, developing techniques that generate accurate, comprehensive, and relevant reports from retinal images, while minimizing irrelevant or misleading information, is essential for ensuring their reliability and effectiveness in both clinical and research contexts.

\begin{figure}[t!]
\begin{center}
\includegraphics[width=1.0\linewidth]{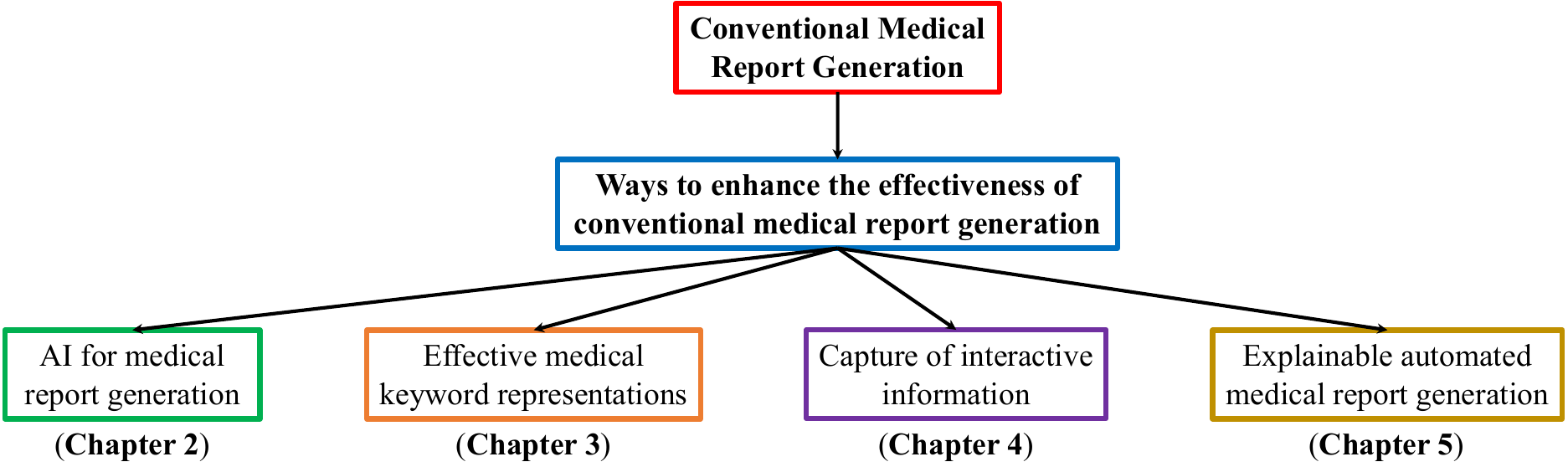}
\end{center}
   \caption{An overview of the thesis.}
\label{overview}
\end{figure}

In this thesis, illustrated in Figure \ref{overview} for an overview, we therefore pose the central research question (CRQ): 
\colorlet{shadecolor}{lightgray!50!} 
\begin{shaded}
\begin{changemargin}{0.25cm}{0.25cm} 
\textbf{CRQ:} \textit{How can we improve automatic medical report generation to optimize the efficiency and effectiveness of traditional treatment processes for retinal diseases?} 
\end{changemargin}
\end{shaded}

\begin{figure}[t!]
\centering
  \includegraphics[width=1.0\linewidth]{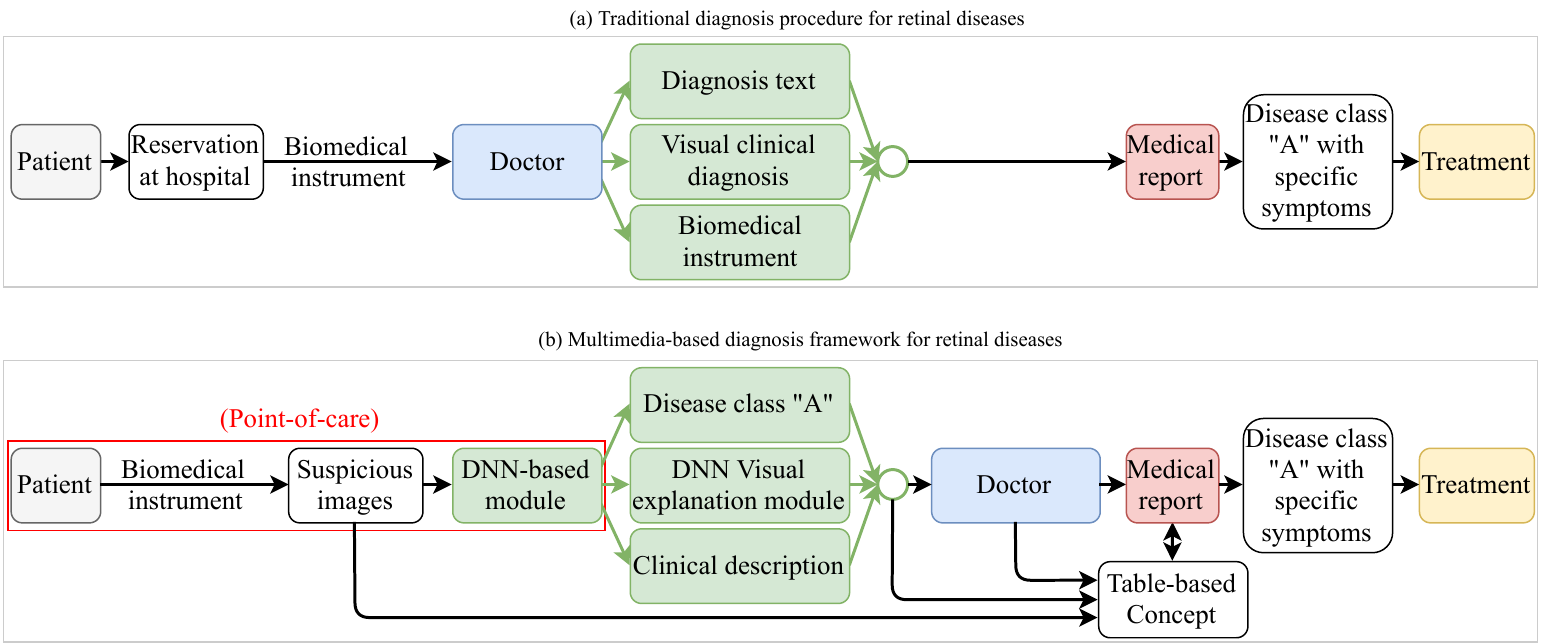}
  \caption{
    Figure (a) illustrates an existing traditional medical treatment procedure for retinal diseases \cite{tukey2014impact}, where doctors are primarily responsible for most tasks. In contrast, Figure (b) integrates our AI-based medical diagnosis method into the traditional treatment process, aiming to enhance efficiency in line with the point-of-care (POC) concept \cite{pai2012point}. The proposed method consists of DNN-based and DNN visual explanation modules. The outputs from the DNN-based module include the ``Disease Class A'' and ``Clinical Description.'' Meanwhile, the DNN visual explanation module provides a visualization of the information generated by the DNN-based module for classification tasks. For a more detailed explanation, please refer to Section \hyperref[me:method_2.4]{2.4}. Note that DNN stands for deep neural networks in this figure.
}
  \label{fig:figure_2_2}
\end{figure}

In the traditional treatment process for retinal diseases, as illustrated in Figure \ref{fig:figure_2_2}, methods for generating medical reports from retinal images rely on manual interpretation by clinicians, which is time-consuming and prone to human error. Reviewing numerous retinal images and patient records to create a comprehensive report is labor-intensive and subject to variability in expertise. Additionally, manual methods may struggle to consistently capture subtle nuances and complex patterns in retinal images, which are critical for accurate diagnosis \cite{jing2018automatic,huck2019auto}. To address these limitations, we propose leveraging Artificial Intelligence (AI) to automate parts of the traditional process. AI can quickly and accurately analyze vast amounts of image data, identifying patterns and anomalies that humans may miss. Integrating AI into the traditional diagnostic process aims to improve accuracy, reduce clinicians' workload, and enhance patient care.
This leads to the research question: 
\colorlet{shadecolor}{lightgray!50!} 
\begin{shaded}
\begin{changemargin}{0.25cm}{0.25cm} 
\textbf{RQ1:} \textit{How can we leverage AI to improve the existing retinal disease treatment procedure?} 
\end{changemargin}
\end{shaded}
The manual interpretation of retinal images by ophthalmologists is a cornerstone of traditional retinal disease diagnosis, yet it is often inefficient and subject to human error \cite{jing2018automatic,huck2019auto}. In Chapter \ref{ch:ch2} of this thesis, we investigate this issue and propose an AI-based method to improve the conventional retinal disease treatment procedure, building upon our previous work \cite{huang2021deepopht}.
Our approach consists of a DNN-based module, which includes a retinal disease identifier (RDI) and a clinical description generator (CDG), along with a DNN visual explanation module. To train and validate the effectiveness of our DNN-based module, we have developed a large-scale retinal disease image dataset. Additionally, we provide a manually labeled retinal image dataset, annotated by ophthalmologists, to qualitatively demonstrate the effectiveness of the proposed AI-based method. Our method is capable of creating meaningful retinal image descriptions and visual explanations that are clinically relevant. The primary challenge lies in encoding and utilizing the interactive information between text-based keywords and retinal images to generate accurate and comprehensive medical reports. 
To address this, we present a multi-modal deep learning method capable of handling both image and text information, optimizing the generation of medical reports. We fuse these modalities using an average-based method to ensure the quality of the generated reports. Evaluating the model's performance is crucial, and we employ metrics such as BLEU, CIDEr, and ROUGE to assess the quality of the generated reports.  
Our experimental results show that the proposed method is effective both quantitatively and qualitatively, offering significant improvements in the diagnosis and treatment of retinal diseases.

In the proposed AI-based method, the text-based input serves as critical guidance that significantly impacts the quality of the resulting reports. Hence, developing effective methods for encoding the provided text-based input is imperative. This prompts the next research question: 
\colorlet{shadecolor}{lightgray!50!} 
\begin{shaded}
\begin{changemargin}{0.25cm}{0.25cm} 
\textbf{RQ2:} \textit{How to improve the effectiveness of medical keyword representations to better capture the nuances of medical terminology?} 
\end{changemargin}
\end{shaded}
Multi-modal medical image captioning, which combines visual and textual information, enhances the utility and effectiveness of automatically generated medical descriptions. Integrating expert-defined keywords with retinal images generates comprehensive descriptions that capture both content and contextual nuances. The success of this approach relies on proficiently encoding both textual inputs and medical images.
In Chapter \ref{ch:ch4}, we address the enhancement of medical keyword representations by introducing a novel end-to-end deep multi-modal medical image captioning model. This model employs contextualized representations, textual feature reinforcement, and masked self-attention to more effectively encode medical keywords alongside images. This chapter builds upon our earlier work cited in \cite{huang2021contextualized}. This model includes a contextualized medical description generator, multi-modal attention mechanisms, and a specially designed attention network. Unlike traditional single-image input methods, multi-modal inputs effectively capture semantic meaning and context, enhancing the overall quality of medical image captioning. Experimental results show that our model generates more accurate and meaningful descriptions for retinal images compared to baseline methods.

In multi-modal medical image captioning, both text-based keywords and retinal images are integral to generating medical reports, with the interaction between these elements significantly influencing the quality of the reports. Therefore, it is crucial to develop effective approaches for encoding this interactive information. This leads to the next research question: 
\colorlet{shadecolor}{lightgray!50!} 
\begin{shaded}
\begin{changemargin}{0.25cm}{0.25cm} 
\textbf{RQ3:} \textit{How can we enhance the capture of interactive information between keywords and images in multi-modal medical report generation?} 
\end{changemargin}
\end{shaded}
Automatically generating medical reports from retinal images has become a transformative solution to reduce the workload of ophthalmologists and improve clinical efficiency \cite{huck2019auto,huang2021deepopht,jing2018automatic}. In Chapter \ref{ch:ch5}, we present a novel context-driven network, primarily based on our earlier work \cite{huang2022non,huang2021deep,huang2021longer}, specifically designed for automating the generation of these reports. Central to our approach is TransFuser, a novel non-local attention-based multi-modal feature fusion technique. TransFuser effectively captures the intricate interactions between textual keywords and retinal images, generating precise and informative medical descriptions. By addressing the nuanced complexities inherent in medical image interpretation, our approach provides a robust framework that enhances diagnostic accuracy and supports more informed clinical decision-making in ophthalmology.

Existing automated medical report generation methods typically rely on deep learning. However, the lack of explainability in deep models can reduce user trust, particularly in the medical domain. Consequently, the final research question emerges: 
\colorlet{shadecolor}{lightgray!50!} 
\begin{shaded}
\begin{changemargin}{0.25cm}{0.25cm} 
\textbf{RQ4:} \textit{How can we improve the explainability of automated medical report generation for retinal images?} 
\end{changemargin}
\end{shaded}
The interpretability of machine learning (ML)-based medical report generation systems for retinal images remains a significant challenge, hindering their widespread acceptance \cite{ghassemi2021false}. Chapter \ref{ch:ch6} addresses this crucial issue, emphasizing the need for trust in ML-based systems. Building on our previous work \cite{wu2023expert}, this chapter explores the complexities of defining interpretability in a way that makes ML-based medical report generation systems human-comprehensible.
Common post-hoc explanation methods, like heat maps and saliency maps, highlight important image areas but fail to convey the specific features the model finds useful. This discrepancy can lead to biases due to varying human interpretations of these highlighted regions.
To tackle these issues, we propose using expert-defined keywords and a specialized attention-based strategy to enhance interpretability in medical report generation systems for retinal images. These keywords are effective carriers of domain knowledge and are inherently understandable, improving the system's interpretability. By integrating these keywords with an attention-based mechanism, our approach enhances both interpretability and performance.
Our method achieves state-of-the-art performance on text evaluation metrics such as BLEU, ROUGE, CIDEr, and METEOR. This chapter demonstrates that incorporating expert-defined keywords and a robust attention-based strategy can create a more trustworthy and interpretable ML-based medical report generation system, overcoming a critical barrier to clinical acceptance.

\vspace{+6pt}\noindent In conclusion, this thesis aims to enhance automated medical reports by refining the representation of medical terminology and improving the interaction between keywords and images. Additionally, we focus on increasing the explainability of automated reports. These advancements collectively enhance the quality and utility of automated medical reports for retinal disease diagnosis and treatment, thereby optimizing the efficiency and effectiveness of traditional treatment processes.

\chapter{Report Generation with Deep Models and Visual Explanations}\label{ch:ch2}

\textit{
In Chapter \ref{ch:ch2}, we propose an AI-based method designed to enhance the traditional retinal disease treatment process, aiming to improve diagnostic efficiency and accuracy for ophthalmologists. Our approach consists of a DNN-based module, which includes a RDI, a CDG, and a DNN visual explanation module. To train and validate the effectiveness of this module, we have developed a large-scale retinal disease image dataset. Additionally, we provide a manually labeled retinal image dataset, annotated by ophthalmologists, to serve as ground truth and demonstrate the method's effectiveness qualitatively. Experimental results show that our proposed method is both quantitatively and qualitatively effective, generating meaningful retinal image descriptions and clinically relevant visual explanations. 
}

\colorlet{shadecolor}{lightgray!50!} 
\begin{shaded}
\begin{changemargin}{0cm}{0.25cm} 
\small
This Chapter is based on:
\begin{itemize}
  \item \textit{``DeepOpht: Medical Report Generation for Retinal Images via Deep Models and Visual Explanation''}, published in \textit{IEEE/CVF Winter Conference on Applications of Computer Vision}, 2021~\cite{huang2021deepopht}, by \textbf{Jia-Hong Huang}, Chao-Han Huck Yang, Fangyu Liu, Meng Tian, Yi-Chieh Liu, Ting-Wei Wu, I-Hung Lin, Kang Wang, Hiromasa Morikawa, Hernghua Chang, Jesper Tegner, and Marcel Worring.
\end{itemize}

\end{changemargin}
\end{shaded}

\section{Introduction}
The World Health Organization (WHO) estimates that common retinal diseases like Age-related Macular Degeneration (AMD) and Diabetic Retinopathy (DR) will soon affect over 500 million people globally \cite{pizzarello2004vision}. The traditional process of diagnosing retinal diseases and generating medical reports is time-consuming, which implies that ophthalmologists will face an increasing workload soon.

The current state of the art in AI focuses on deep learning, which holds great promise for assisting ophthalmologists and enhancing traditional retinal disease treatment procedures. Deep learning models, such as convolutional neural networks (CNNs) for computer vision and recurrent neural networks (RNNs) for natural language processing, have achieved—and sometimes surpassed—human-level performance. Therefore, it is an opportune time to introduce AI-based medical diagnosis methods to support ophthalmologists.

\begin{figure}[t!]
\begin{center}
\includegraphics[width=1.0\linewidth]{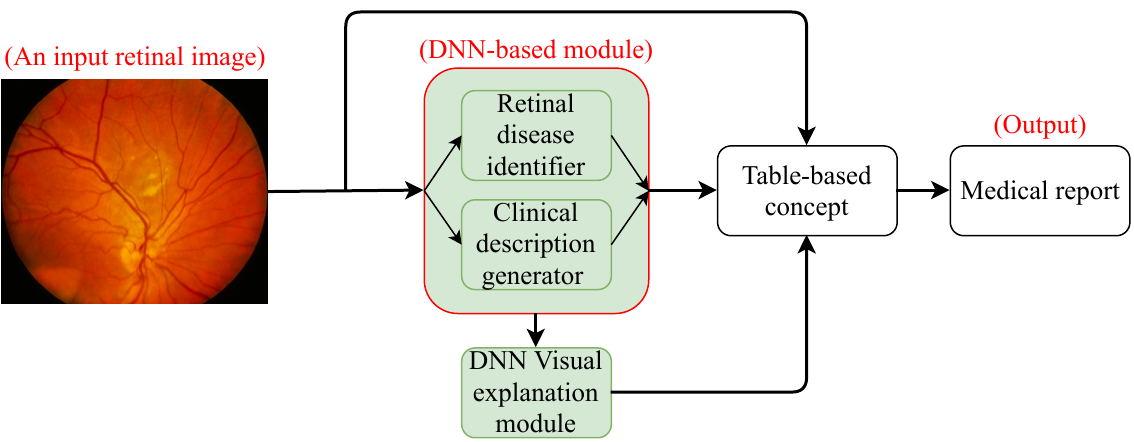}
\end{center}
   \caption{
   Illustration of the proposed AI-based medical diagnosis method for the ophthalmology domain. It comprises DNN-based and DNN visual explanation modules. The DNN-based module includes two sub-modules: a RDI and a CDG, reinforced by our proposed keyword-driven method, as detailed in Section \hyperref[me:method_2.4]{2.4}. The input to our method is a retinal image, and the output is a table-based medical report \cite{zahalka2014towards}. Figure \ref{fig:figure_2_2} demonstrates how this AI-based method can enhance the traditional retinal disease treatment process.
}
\label{fig:figure_2_1}
\end{figure}
In Chapter \ref{ch:ch2}, we propose an AI-based method for automatic medical report generation from retinal images, as illustrated in Figure \ref{fig:figure_2_1}. This approach aims to enhance the traditional retinal disease diagnosis process, shown in Figure \ref{fig:figure_2_2}, by increasing efficiency and accuracy for ophthalmologists. Our method leverages deep learning models, including a RDI and a CDG, to automate parts of the traditional diagnostic procedure, improving overall efficiency.

To train our deep learning models and validate the effectiveness of our RDI and CDG, we introduce a new large-scale retinal disease image dataset called DeepEyeNet (DEN). Additionally, we provide a retinal image dataset manually labeled by ophthalmologists as ground truth, demonstrating the effectiveness of our proposed AI-based model qualitatively. This dataset allows us to show that the activation maps of our deep models align with clinically recognized image features associated with identified diseases. Our experimental results indicate that the proposed AI-based method effectively enhances the traditional retinal disease treatment procedure. Our main contributions are summarized as follows:

\vspace{+3pt}
\noindent\textbf{Contributions.}
\begin{itemize}
    \item We propose an AI-based method for generating medical reports from retinal images to enhance the traditional retinal disease treatment procedure and assist ophthalmologists in increasing diagnostic efficiency and accuracy. This approach leverages deep learning models, including a RDI and a CDG, to automate portions of the conventional treatment process.
    
    \item We introduce the DEN dataset, a large-scale retinal disease image collection comprising 15,709 images, to train our deep learning models and quantitatively validate the effectiveness of the proposed RDI and CDG.
    
    \item We offer an additional dataset containing 300 retinal images labeled by ophthalmologists to demonstrate the effectiveness of our method qualitatively. This dataset allows for visual confirmation that the activation maps of our models align with image features that are clinically recognized by ophthalmologists.
    
\end{itemize}

\section{Related Work}
In this section, we categorize the related works into four main areas: retinal disease classification, image captioning, visual explanations for neural networks, and comparisons of retinal datasets.

\subsection{Retinal Disease Classification}

Optical Coherence Tomography (OCT), Fluorescein Angiography (FA), and Color Fundus Photography (CFP) are the three most widely utilized imaging techniques for diagnosing retinal diseases \cite{yanagihara2020methodological}. OCT is a cutting-edge biomedical imaging technology that offers high-resolution, non-invasive, real-time images of the retina, revealing its intricate structure in detail \cite{de2018clinically, lang2013retinal, gerckens2003optical}. The algorithm proposed by \cite{bagci2008thickness} facilitates the segmentation and detection of six distinct retinal layers in OCT images: the Nerve Fiber Layer (NFL), Ganglion Cell Layer (GCL) + Inner Plexiform Layer (IPL), Inner Nuclear Layer (INL), Outer Plexiform Layer (OPL), Outer Nuclear Layer (ONL) + Photoreceptor Inner Segments (PIS), and Photoreceptor Outer Segments (POS).

FA is commonly used to elucidate the pathophysiological progression of Retinopathy of Prematurity (ROP) after intravitreal anti-Vascular Endothelial Growth Factor (Anti-VEGF) treatment \cite{lepore2018follow}. Meanwhile, CFP remains a simple and cost-effective imaging technique frequently employed by trained medical professionals. Image preprocessing plays a vital role in the automated analysis of CFP images, and \cite{youssif2006comparative} proposed a method to reduce the vignetting effect caused by non-uniform illumination in retinal images.

In Chapter \ref{ch:ch2}, we primarily employ DNN methods \cite{he2016deep, simonyan2014very, yang2020net} to further advance retinal disease classification, focusing on a multi-label task that integrates language information \cite{yang2018novel}.

\subsection{Image Captioning}

Recently, computer vision researchers have introduced the task of image captioning, with early works by \cite{karpathy2015deep}, \cite{vinyals2015show}, and \cite{fang2015captions}. In \cite{karpathy2015deep}, the authors present a model that embeds both visual and language information into a common multimodal space. Meanwhile, \cite{fang2015captions} utilizes a natural language model to combine relevant words associated with different parts of an image to generate captions. The approach by \cite{vinyals2015show} involves using CNNs to extract image features, which are then fed as input at the first time step of RNNs for caption generation.

Further advancements include the deliberate residual attention network proposed by \cite{gao2019deliberate}, which consists of a first-pass residual attention layer that prepares visual attention and hidden states to generate preliminary captions, followed by a second-pass layer that refines them using global features. This method shows promise for producing improved captions.

In \cite{liu2017improved}, the authors note that traditional image captioning models often rely on maximum likelihood estimation for training, which can lead to a poor correlation between log-likelihood scores and human quality assessments. Standard syntactic evaluation metrics such as METEOR \cite{banerjee2005meteor}, BLEU \cite{papineni2002bleu}, and ROUGE \cite{lin2004rouge} also show weak correlations. They propose using a policy gradient method to optimize a linear combination of CIDEr \cite{vedantam2015cider} and SPICE \cite{anderson2016spice} for better alignment with human evaluation.

Additionally, \cite{hendricks2016generating} introduces a method that emphasizes the discriminative properties of visible objects, predicting class labels while explaining the rationale behind them. Their model leverages a loss function based on reinforcement learning and sampling to enhance caption generation. Despite these advancements, as noted by \cite{vinyals2015show}, \cite{karpathy2015deep}, and \cite{gao2019deliberate}, existing models primarily generate rough descriptions of images. In Chapter \ref{ch:ch2}, we leverage keywords to enhance the reasoning capabilities of our proposed CDG.

\begin{table*}
\begin{center}
    \caption{
    Summary of available retinal datasets. Our proposed DEN dataset is significantly larger than other retinal image datasets. DEN includes three types of labels: disease names, keywords, and clinical descriptions. In contrast, most existing retinal datasets primarily consist of image data and tend to be smaller in size. It is important to note that ``Text*'' refers to both clinical descriptions and keywords, as detailed in Section \hyperref[dia:data_2.3]{2.3}, while ``Text'' signifies clinical descriptions alone. Thus, our DEN dataset is unique in its comprehensive labeling approach.
}
\rotatebox{270}{
\scalebox{0.7}{
    \begin{tabular}{c|c|c|c|c}
    \toprule
    \textbf{Name of Dataset} & \textbf{Field of View} & \textbf{Resolution} & \textbf{Data Type} & \textbf{Number of Images}\\ \midrule
    VICAVR \cite{vazquez2013improving} & $45^{\circ}$ & $768 \times 584$ & Image & 58\\ \midrule
    VARIA \cite{ortega2009retinal} & $20^{\circ}$ & $768 \times 584$ & Image & 233\\ \midrule
    STARE \cite{hoover2003locating} & $\approx 30^{\circ}-45^{\circ}$ & $700 \times 605$ & Image + Text & 397\\ \midrule
    CHASE-DB1 \cite{fraz2012ensemble} & $\approx 25^{\circ}$ & $999 \times 960$ & Image & 14\\ \midrule
    RODREP \cite{adal2015accuracy} & $45^{\circ}$ & $2000 \times 1312$ & Image & 1,120\\ \midrule
    HRF \cite{odstrvcilik2009improvement} & $45^{\circ}$ & 3504*2336 & Image & 45\\ \midrule
    e-ophtha  \cite{decenciere2013teleophta} & $\approx 45^{\circ}$ & $2544 \times 1696$ & Image & 463\\ \midrule    
    ROC \cite{niemeijer2010retinopathy} & $\approx 30^{\circ}-45^{\circ}$ & $768 \times 576 \sim 1386 \times 1391$ & Image & 100\\ \midrule
    REVIEW \cite{al2008reference} & $\approx 45^{\circ}$ & $1360 \times 1024 \sim 3584 \times 2438$ & Image & 14\\ \midrule
    ONHSD \cite{computing2012understanding} & $45^{\circ}$ & $640 \times 480$ & Image & 99\\ \midrule   
    INSPIRE-AVR \cite{niemeijer2011inspire} & $30^{\circ}$ & $2392 \times 2048$ & Image & 40\\ \midrule    
    DIARETDB1 \cite{kauppi2007diaretdb1} & $50^{\circ}$ & $1500 \times 1152$ & Image + Text & 89\\ \midrule
    DIARETDB0 \cite{kauppidiaretdb0} & $50^{\circ}$ & $1500 \times 1152$ & Image & 130\\ \midrule  
    MESSIDOR \cite{decenciere2014feedback} & $45^{\circ}$ & $1440 \times 960 \sim 2304 \times 1536$ & Image + Text & 1,200\\ \midrule
    Drishti-GS \cite{sivaswamy2014drishti} & $\approx 25^{\circ}$ & $2045 \times 1752$ & Image & 101\\  \midrule
    FIRE \cite{hernandez2017fire} & $45^{\circ}$ & $2912 \times 2912$ & Image & 129\\  \midrule  
    DRIONS-DB \cite{carmona2008identification} & $\approx 30^{\circ}$ & $600 \times 400$ & Image & 110\\  \midrule 
    IDRiD \cite{porwal2018indian} & $50^{\circ}$ & $4288 \times 2848$ & Image & 516\\  \midrule 
    DRIVE \cite{staal2004ridge} & $45^{\circ}$ & $565 \times 584$ & Image & 40\\ \midrule
    \textbf{DEN} & $\approx \textbf{30}^{\circ}-\textbf{60}^{\circ}$ & \textbf{various} & \textbf{Image + Text*} & \textbf{15,709}\\ 
    \bottomrule 
    
    \end{tabular}}
    }

    \label{table:table_2_1}
\end{center}
\end{table*}

\subsection{Neural Networks Visual Explanation}
Several popular tools have been developed for visualizing CNNs \cite{zhou2015cnnlocalization,selvaraju2017grad,yang2019causal}. One such tool is Class Activation Mapping (CAM), introduced in \cite{zhou2015cnnlocalization}, which enables CNNs trained for classification to also perform object localization without requiring bounding boxes. In our previous works \cite{liu2019synthesizing, yang2018auto}, we utilize CAM to visualize predicted class scores on retinal images, highlighting the discriminative regions detected by the CNN.

Another influential tool is Gradient-weighted Class Activation Mapping (Grad-CAM), proposed in \cite{selvaraju2017grad}. Grad-CAM provides visual explanations for CNN-based models by making the network's decision process more transparent. This technique allows users to see which parts of an image are most influential in a model's prediction. Expanding on this, \cite{chattopadhay2018grad} introduces Grad-CAM++, a method that offers improved visual explanations and better object localization, especially in scenarios involving multiple objects in a single image \cite{yang2020wavelet}.

Additionally, \cite{zeiler2014visualizing} presents a CNN visualization technique that reveals insights into the function of intermediate layers and the overall operation of the classifier. This approach aids in the understanding and refinement of CNN architectures.

In contrast to these methods, \cite{li2018tell} proposes a novel approach that directly supervises the visual explanations produced by the network. Their end-to-end model guides the network to focus on specific, expected regions, demonstrating that such supervision enhances the quality of visual explanations.

While the aforementioned techniques focus primarily on visualizing image data, there are also methods for visualizing multimedia data, including text and images. For example, \cite{zahalka2014towards,rooij2012efficient,huang2020query,gan2015devnet} have proposed various multimedia visualization techniques. \cite{zahalka2014towards} introduces five key concepts for multimedia visualization: basic grid, similarity space, similarity-based, spreadsheet, and thread-based concepts.

In our work, we leverage CAM to demonstrate that the activation maps of our deep learning models align with clinically recognized features associated with specific diseases in retinal images. Furthermore, we utilize a table-based visualization approach, akin to the static spreadsheet concept, to display our medical reports effectively.

\subsection{Comparison of Retinal Datasets}

Retinal disease research has a long history, with numerous datasets established for analysis, including \cite{staal2004ridge,porwal2018indian,carmona2008identification,hernandez2017fire,sivaswamy2014drishti,decenciere2014feedback,kauppidiaretdb0,kauppi2007diaretdb1,niemeijer2011inspire,computing2012understanding,al2008reference,niemeijer2010retinopathy,decenciere2013teleophta,odstrvcilik2009improvement,adal2015accuracy,fraz2012ensemble,hoover2003locating,ortega2009retinal,vazquez2013improving}. For instance, the DRIVE dataset \cite{staal2004ridge} comprises 40 retinal images obtained from a DR screening program in the Netherlands, divided into training and test sets. The training set includes a single manual segmentation of the vasculature, while the test set features two manual segmentations.

The IDRiD dataset \cite{porwal2018indian} contains 516 retinal fundus images, with ground truth annotations for signs of Diabetic Macular Edema (DME) and DR, including (i) pixel-level labels for typical DR lesions and the optic disc, (ii) image-level disease severity grading for DR and DME, and (iii) coordinates for the optic disc and fovea center. The DRIONS-DB dataset \cite{carmona2008identification} consists of 110 color digital retinal images, showcasing various visual characteristics such as cataracts, light artifacts, and rim blurring. The FIRE dataset \cite{hernandez2017fire} includes 129 retinal images, forming 134 image pairs categorized based on their characteristics. The Drishti-GS dataset \cite{sivaswamy2014drishti} contains 101 images, split into 50 training and 51 testing images. The MESSIDOR dataset \cite{decenciere2014feedback} offers 1,200 color fundus images with medical diagnoses, though it lacks manual annotations for lesions or their positions.

The DIARETDB0 dataset \cite{kauppidiaretdb0} consists of 130 color fundus images, with 20 normal images and 110 displaying signs of DR. The DIARETDB1 dataset \cite{kauppi2007diaretdb1} contains 89 color fundus images, of which 84 exhibit at least mild non-proliferative signs of DR, while five are normal. The INSPIRE-AVR dataset \cite{niemeijer2011inspire} includes 40 images of the optic disc and vessels with arterio-venous ratio references. The ONHSD dataset \cite{computing2012understanding} features 99 retinal images primarily for segmentation tasks, and the REVIEW dataset \cite{al2008reference} consists of 14 images also focused on segmentation. The ROC dataset \cite{niemeijer2010retinopathy} aims to enhance computer-aided detection and diagnosis of DR in diabetic patients. The e-ophtha dataset \cite{decenciere2013teleophta} is specifically designed for scientific research in DR. The HRF dataset \cite{odstrvcilik2009improvement} contains 15 images each of healthy patients, DR patients, and those with glaucoma, accompanied by binary gold standard vessel segmentation images. The RODREP dataset \cite{adal2015accuracy} consists of 1,120 repeated color fundus photos from 70 patients in a DR screening program. The CHASE-DB1 dataset \cite{fraz2012ensemble} contains 14 images primarily for retinal vessel analysis. The STARE dataset \cite{hoover2003locating} comprises 397 images aimed at developing an automated diagnostic system for eye diseases. The VARIA dataset \cite{ortega2009retinal} includes 233 images for authentication purposes, while the VICAVR dataset \cite{vazquez2013improving} contains 58 images used for calculating the artery/vein (A/V) ratio.

In Chapter \ref{ch:ch2}, we introduce a large-scale retinal image dataset, DEN, to train our deep learning models and validate our RDI and CDG. For convenience, a summary of the aforementioned retinal datasets is provided in Table \ref{table:table_2_1}.

\section{Dataset Introduction and Analysis}\label{dia:data_2.3}
In this section, we describe our proposed DEN dataset, focusing on the types of retinal images, labeling, and key statistics. Several members of our team are experienced ophthalmologists who contributed their expertise to curate the DEN dataset, which is organized around 265 unique retinal symptoms based on clinical definitions and professional knowledge. The dataset comprises two types of retinal images: grayscale FA and colorful CFP. 

In total, the dataset includes 15,709 images, with 1,811 FA images and 13,898 CFP images. Following the conventions of large-scale datasets for deep learning research, we have created standard splits, allocating 60\% of the data for training, 20\% for validation, and 20\% for testing, resulting in 9,425, 3,142, and 3,142 images, respectively.

Each retinal image in the dataset is associated with three labels: the name of the disease, keywords, and a clinical description. The dataset encompasses 265 different retinal diseases or symptoms, including both common and rare conditions. For the keyword and clinical description labels, there are 15,709 entries for each. The keyword labels provide critical information relevant to the diagnosis process, while the clinical description labels correspond to the captions for each retinal image. Importantly, all labels have been defined by retina specialists or ophthalmologists.

\begin{figure}[t!]
\begin{center}
\includegraphics[width=1.0\linewidth]{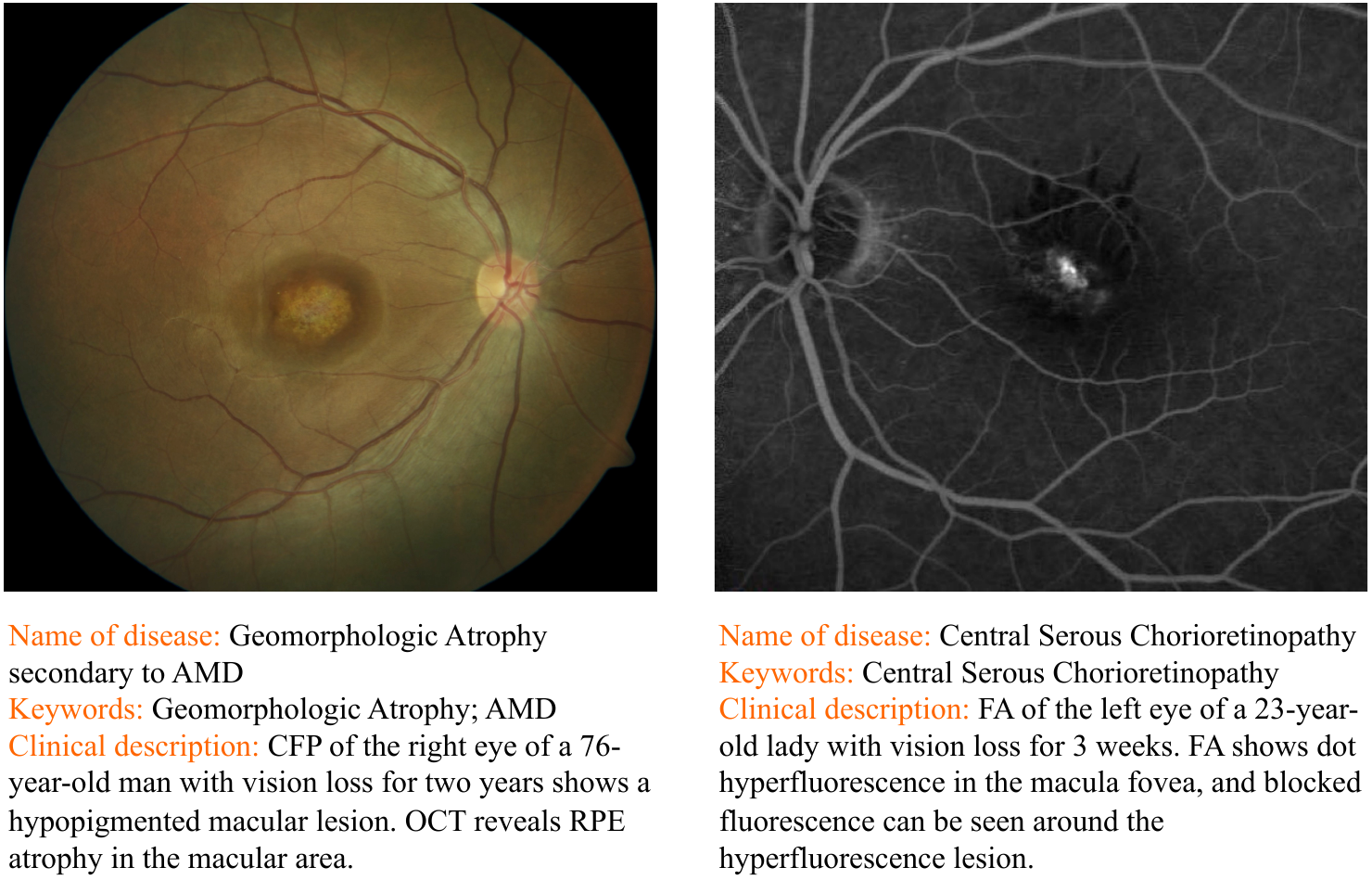}
\end{center}
   \caption{
   Examples from our DEN dataset. Each image is accompanied by three labels: the name of the disease, keywords, and a clinical description. All labels have been defined by ophthalmologists to ensure accuracy and relevance.
}
\label{fig:figure_2_3}
\end{figure}

\begin{figure}[t!]
\begin{center}
\includegraphics[width=1.0\linewidth]{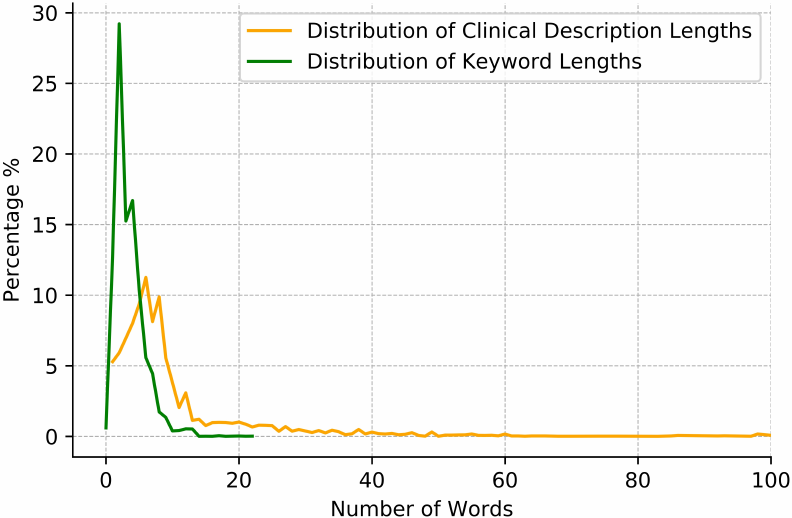}
\end{center}
   \caption{
   Illustration of the word length distribution for the keyword and clinical description labels. The majority of word lengths in our DEN dataset range between 5 and 10 words.
}
\label{fig:figure_2_4}
\end{figure}

To illustrate our dataset, we present several examples from the DEN dataset in Figure \ref{fig:figure_2_3}. Additionally, Figure \ref{fig:figure_2_4} displays the word length distribution of the keyword and clinical description labels. Notably, the longest keywords in our dataset exceed 15 words, while the clinical descriptions can reach over 50 words. In contrast, existing datasets for natural image captioning or Visual Question Answering (VQA) \cite{antol2015vqa,chen2015microsoft,lin2014microsoft,huang2019assessing,huang2017robustness} typically feature captions averaging around 10 words, highlighting the challenges presented by our dataset.

\begin{figure}[t!]
\begin{center}
\includegraphics[width=1.0\linewidth]{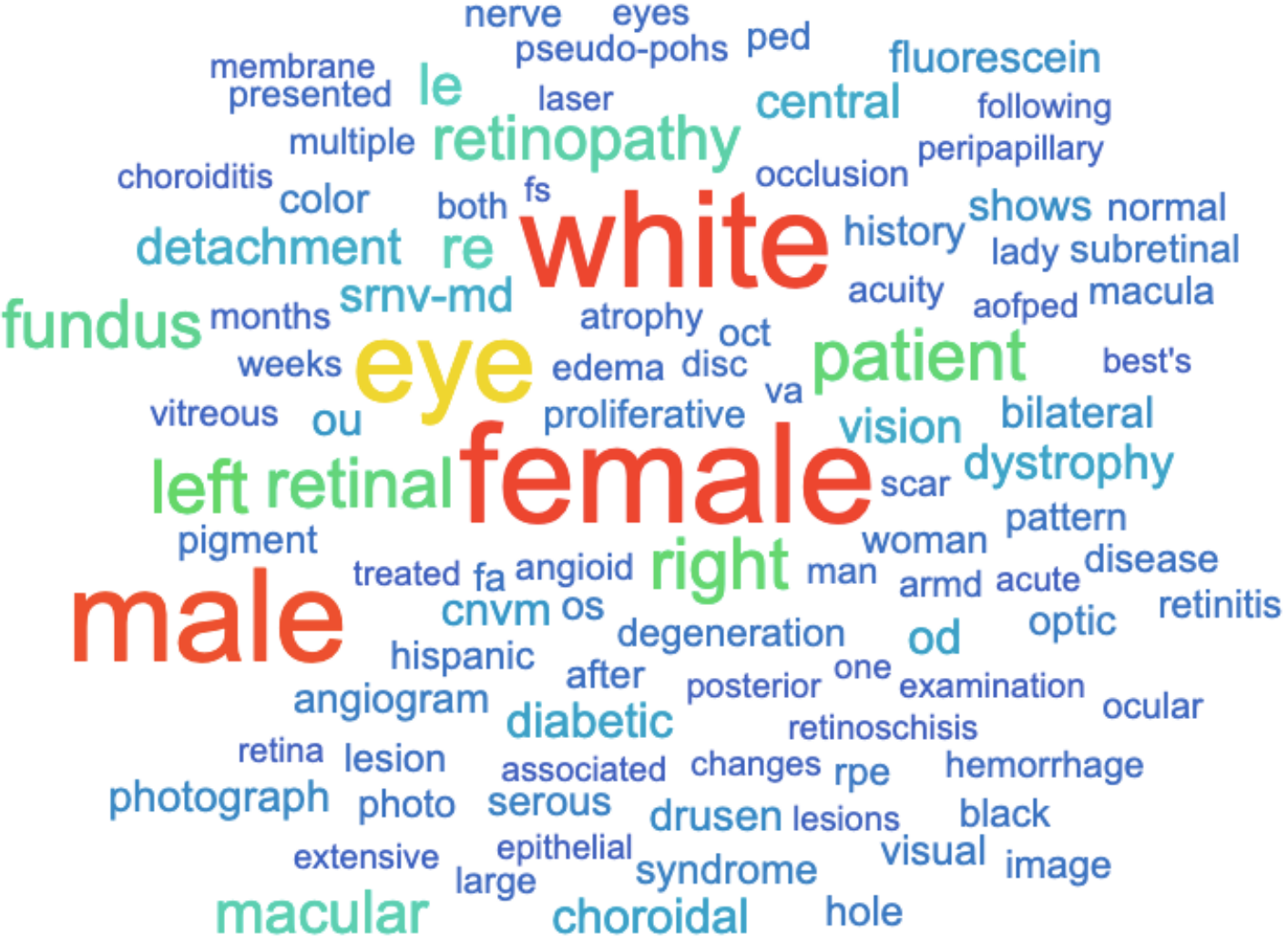}
\end{center}
  \caption{
  The Venn-style word cloud for the clinical description labels. The size of each word reflects its normalized frequency. This visualization reveals the presence of specific abstract concepts, which pose a challenge for image captioning algorithms to produce high-quality descriptions.
}
\label{fig:figure_2_5}
\end{figure}

Furthermore, we provide a Venn-style word cloud visualization of the clinical description labels in Figure \ref{fig:figure_2_5}, revealing the presence of specific abstract concepts that further enhance the complexity of the dataset.

\begin{figure}[ht]
\begin{center}
\includegraphics[width=1.0\linewidth]{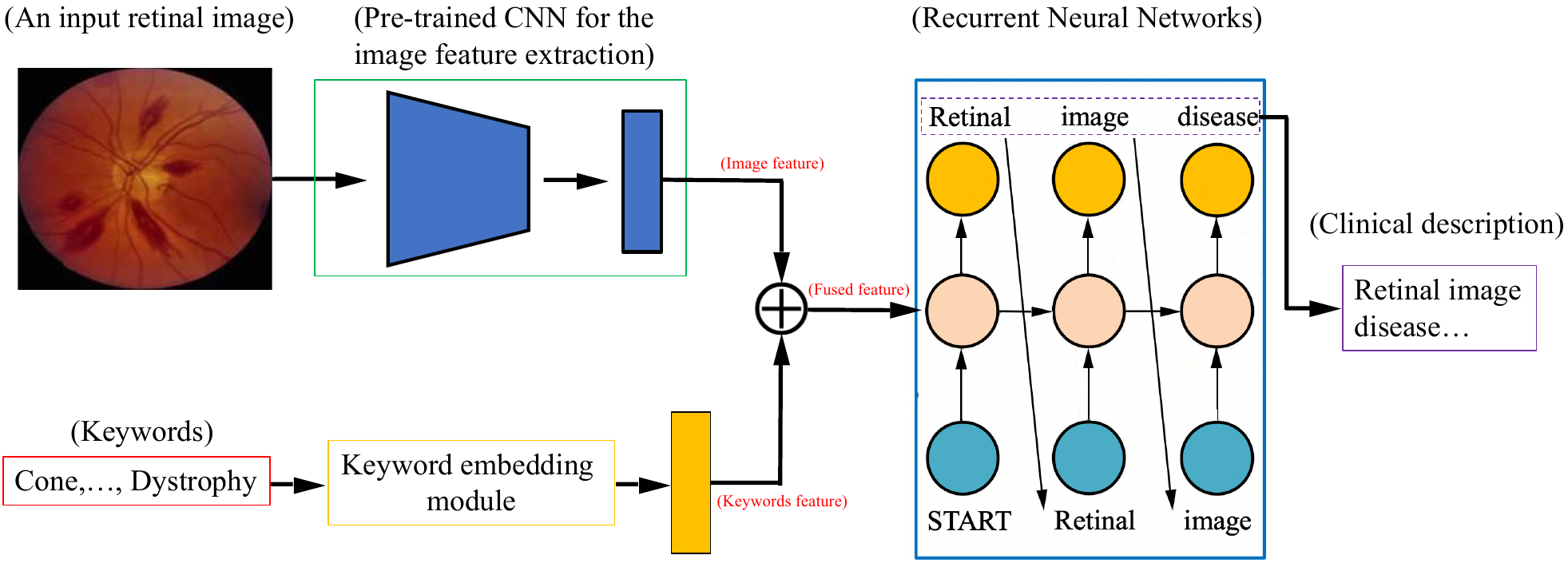}
\end{center}
   \caption{
   Conceptual illustration of the CDG, using our proposed keyword-driven method. In the generator, we utilize a pre-trained CNN model to extract features from the retinal images, serving as the image encoder. Subsequently, we employ an long short-term memory (LSTM) model, a type of RNN, as the decoder to generate one word at each time step. Ultimately, the collected words come together to form a complete clinical description.
 }
\label{fig:figure_2_6}
\end{figure}

\section{Methodology}\label{me:method_2.4}
In this section, we describe our proposed AI-based method for automatic medical report generation. The method primarily consists of a DNN-based module and a DNN visual explanation module.

\subsection{DNN-based Module}\label{med:method_4.1}

The DNN-based module comprises two components: a RDI and a CDG, which we will introduce in the following subsections. We hypothesize that an effective RDI and CDG will enhance the conventional retinal disease treatment process, improving both diagnostic efficiency and accuracy for ophthalmologists.

\noindent\textbf{RDI.}
In our RDI sub-module, we utilize two types of deep learning models based on \cite{he2016deep, simonyan2014very}, both pre-trained on ImageNet and subsequently trained on the proposed DEN dataset. While most medical images, such as chest radiology scans, are primarily grayscale \cite{laserson2018textray}, retinal images in our dataset are predominantly colorful. Leveraging ImageNet pre-training aids in extracting superior lower-level features. Thus, we anticipate that this pre-training will enhance model performance.

\noindent\textbf{CDG.}
To generate clinical descriptions for input retinal images, we utilize a pre-trained CNN model—such as MobileNetV2, VGG16, VGG19, or InceptionV3—as our image feature encoder, along with an LSTM network as the decoder to produce text, as illustrated in Figure \ref{fig:figure_2_6}. When generating descriptions with the LSTM unit, we implement a beam search mechanism to optimize the final output. In ophthalmology, commonly used keywords—often unordered—assist practitioners in creating medical reports. Inspired by this, we incorporate keywords to enhance our CDG sub-module. As depicted in Figure \ref{fig:figure_2_6}, we employ a keyword embedding module, such as a bag-of-words approach, to encode this keyword information. With the integration of keywords to bolster the CDG, we effectively utilize two types of input features: image features and text features. In our approach, we apply an averaging method to fuse these two feature types, as shown in Figure \ref{fig:figure_2_6}.

\subsection{DNN Visual Explanation Modules}\label{med:method_4.2}

Several existing DNN visual explanation methods have been proposed, including those by \cite{zhou2015cnnlocalization,selvaraju2017grad,hu2019silco}. One notable technique is CAM introduced by \cite{zhou2015cnnlocalization}, which enables classification-trained CNNs to perform object localization without the need for bounding boxes. This method uses class activation maps to visualize predicted class scores on images, highlighting the discriminative parts identified by CNN. To enhance traditional retinal disease treatment procedures, we integrate the DNN visual explanation module into our AI-based method. Additionally, we leverage this module to validate the effectiveness of our approach, as detailed in Section \hyperref[exp:experiment_2.5]{2.5}.

\begin{table}[t!]
\caption{
Quantitative results of various RDI models evaluated on our DEN dataset. The RDI model based on \cite{simonyan2014very} with ImageNet pre-training demonstrates the best performance. Here, ``Pre-trained'' indicates that the model is initialized using weights from ImageNet, while ``Random init'' means the model's weights are initialized randomly. Prec@k measures the frequency with which the ground truth label appears among the top $k$ ranked labels after the \textup{softmax} layer. We focus on Prec@1 and Prec@5, as they are crucial for shortlisting disease candidates in real-world applications. It's important to note that, given our dataset includes $265$ retinal disease candidates and limited training data, achieving high Prec@1 performance is challenging. This issue of limited data is common in the medical field.
}
\centering
\scalebox{0.7}{
\begin{tabular}{c|cccc}
\toprule
\multirow{3}{*}{Model} & \multicolumn{4}{c}{Precision}                                                          \\ \cline{2-5} 
                      & \multicolumn{2}{c|}{Pre-trained}                       & \multicolumn{2}{c}{Random init} \\ \cline{2-5} 
                      & Prec@1         & \multicolumn{1}{c|}{Prec@5}         & Prec@1         & Prec@5         \\ \midrule
Jing, et al. \cite{jing2018automatic}                 & 32.72          & \multicolumn{1}{c|}{63.75}          & 29.11          & 60.68          \\
He, et al. \cite{he2016deep}              & 37.09 & \multicolumn{1}{c|}{63.36} & \textbf{36.60}          & 62.87          \\
Simonyan, et al. \cite{simonyan2014very}                 & \textbf{54.23}          & \multicolumn{1}{c|}{\textbf{80.75}}          & 35.93          & \textbf{73.73}          \\
\bottomrule
\end{tabular}}

\label{table:table_2_2}
\end{table}

\subsection{Creating Medical Reports}

\begin{figure*}[t!]
\begin{center}
\includegraphics[width=1.0\linewidth]{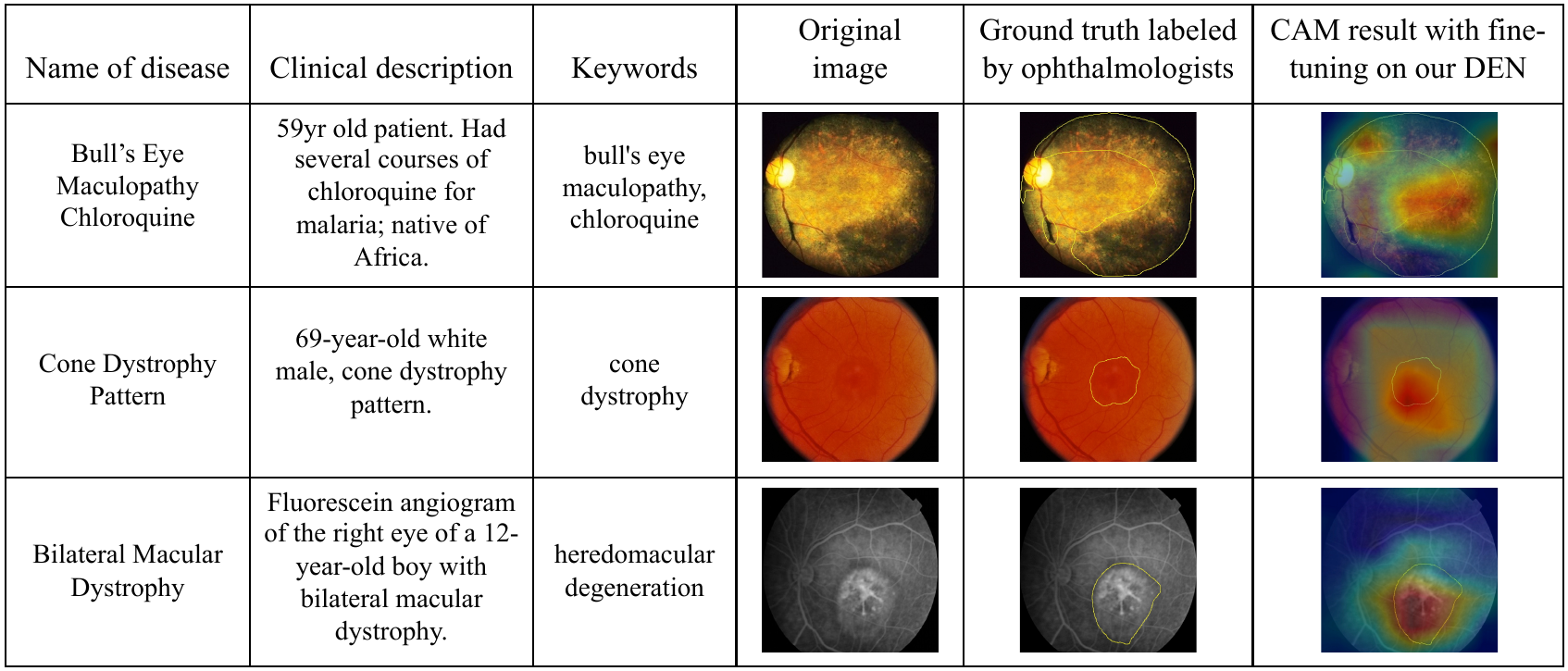}
\end{center}
   \caption{
   Illustration of a medical report structured using a table-based concept \cite{zahalka2014towards}. Given that retinal diseases may share implicit common properties or relationships, we have grouped diseases with similar characteristics in the table. This organized presentation aims to provide ophthalmologists with enhanced insights into the data.
}
\label{fig:figure_2_7}
\end{figure*}

According to \cite{zahalka2014towards,rooij2012efficient}, effective multimedia data visualization enhances the ability to extract insights from data efficiently. In essence, multimedia visualization involves visually organizing various data types, often leading to a deeper understanding and additional information derived from the visualized content. In this study, we incorporate five types of multimedia data: the disease name, keywords, clinical descriptions, retinal images, and CAM result images. To effectively present this medical report, we employ a table-based concept similar to a static spreadsheet \cite{zahalka2014towards}, as illustrated in Figure \ref{fig:figure_2_7}. This visualization aims to enable ophthalmologists to efficiently extract insights from the image and text data, ultimately enhancing diagnostic accuracy.

\begin{figure*}[t!]
\begin{center}
\includegraphics[width=1.0\linewidth]{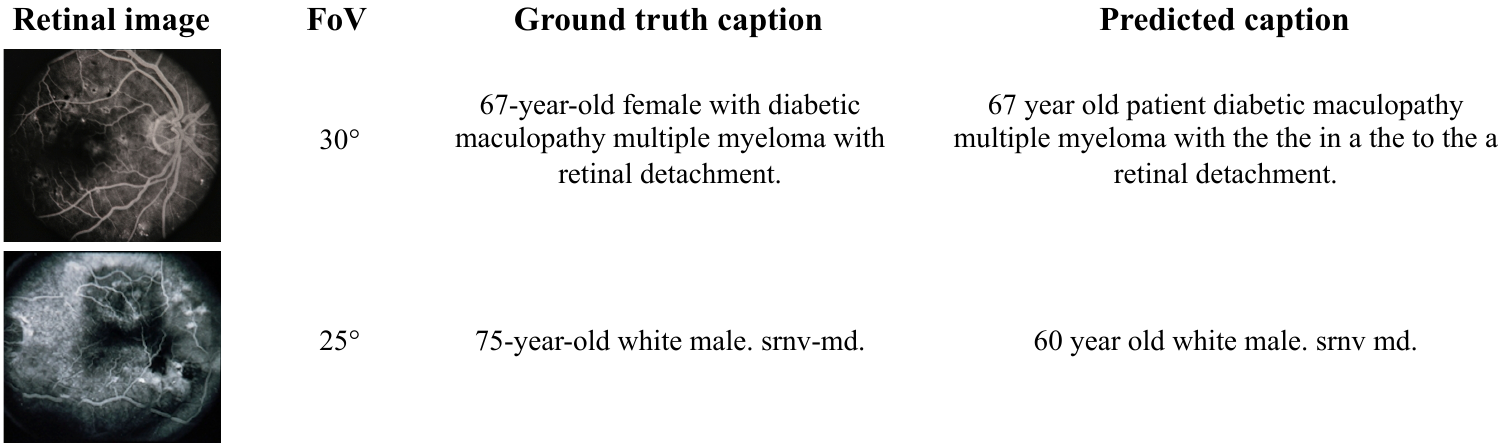}
\end{center}
  \caption{
  Results generated by our CDG. Our models can produce meaningful clinical descriptions that are valuable for ophthalmologists. However, it is important to note that accurately generating information such as ``age`` and ``gender'' remains challenging for automated algorithms. The first row, which contains a correct prediction for ``age,'' is an exception rather than the norm.
  }
\label{fig:figure_2_8}
\end{figure*}

\section{Experiments}\label{exp:experiment_2.5}
In this section, we evaluate the effectiveness of our proposed method by comparing it to baseline models.


\subsection{RDI Verification}
In our experiment, we aim to demonstrate that the RDI model with ImageNet pre-training outperforms the baseline RDI model without it. We fine-tune both ImageNet-pre-trained and non-ImageNet-pre-trained DNN models with different architectures on the DEN dataset. Two training strategies are employed: for the RDI model based on \cite{he2016deep}, we start with a learning rate of $0.1$, reducing it fivefold every 50 epochs. For the RDI model based on \cite{simonyan2014very}, we begin with a learning rate of $0.001$, applying the same decay schedule.

As shown in the evaluation results in Table \ref{table:table_2_2}, the RDI model based on \cite{simonyan2014very} with ImageNet pre-training demonstrates superior performance compared to the others. We hypothesize that the RDI models based on \cite{he2016deep} and \cite{jing2018automatic} may be too complex for the DEN dataset. Although DEN is a large-scale dataset from the retinal field perspective, it still contains only 8,512 training images across 265 classes of both common and rare retinal diseases and symptoms. This limited number of training images makes it challenging to achieve high Prec@1 accuracy for both human doctors and AI models. Therefore, we consider both Prec@1 and Prec@5 in our analysis, with the latter being more relevant for real-world applications.

\begin{table*}[t!]
\caption{
Evaluation results for our keyword-driven and non-keyword-driven CDGs. We highlight the best scores for both types of generators in each column. The term ``w/o'' refers to the non-keyword-driven baseline generators, while ``w/'' denotes our proposed keyword-driven generators. ``BLEU-avg'' indicates the average score of BLEU-1, BLEU-2, BLEU-3, and BLEU-4. Notably, the model based on ``Jing et al. \cite{jing2018automatic}'' achieves the highest performance among all non-keyword-driven models, and the keyword-driven version of the same model also outperforms all others. Overall, all keyword-driven models utilizing the average feature fusion method demonstrate superior performance compared to the non-keyword-driven models, indicating that the use of keywords to enhance the CDGs is effective.
}
\centering
\rotatebox{270}{
\scalebox{0.7}{
\begin{tabular}{c|c|c|c|c|c|c|c|c}
\toprule
\multicolumn{2}{c|}{Model}                  & BLEU-1 & BLEU-2 & BLEU-3 & BLEU-4 & BLEU-avg & CIDEr & ROUGE \\ \midrule
\multirow{2}{*}{Karpathy, et al. \cite{karpathy2015deep}}           & w/o & 0.067  & 0.029  & 0.005  & 0.002  & 0.026    & 0.031 & 0.085 \\ \cline{2-9}
                                    & w/ &  \textbf{0.169}     & \textbf{0.103}   & \textbf{0.060}      & \textbf{0.017}      & \textbf{0.087}      & \textbf{0.120}            & \textbf{0.202} \\ \midrule
\multirow{2}{*}{Vinyals, et al. \cite{vinyals2015show}}              & w/o & 0.054  & 0.018  & 0.002  & 0.001  & 0.019    & 0.056 & 0.083 \\ \cline{2-9} 
                                    & w/ &  \textbf{0.144}     & \textbf{0.092}   & \textbf{0.052}      & \textbf{0.021}      & \textbf{0.077}      & \textbf{0.296}            & \textbf{0.197} \\ \midrule
\multirow{2}{*}{Jing, et al. \cite{jing2018automatic}}              & w/o & 0.130  & 0.083  & 0.044  & 0.012  & 0.067    & 0.167 & 0.149 \\ \cline{2-9} 
                                    & w/ &  \textbf{0.184}     & \textbf{0.114}      & \textbf{0.068}       & \textbf{0.032}    & \textbf{0.100}     & \textbf{0.361}            & \textbf{0.232} \\ \midrule
\multirow{2}{*}{Li, et al. \cite{li2019knowledge}}        & w/o & 0.111  & 0.060  & 0.026  & 0.006  & 0.051    & 0.066 & 0.129 \\ \cline{2-9} 
                                    & w/ & \textbf{0.181}    & \textbf{0.107}  & \textbf{0.062}   & \textbf{0.032}   & \textbf{0.096}   & \textbf{0.453}   & \textbf{0.230} \\ \bottomrule
\end{tabular}}
}
\label{table:table_2_3}
\end{table*}

\subsection{CDG Verification}

In \cite{huang2017vqabq,huang2019novel,huang2017robustness}, the authors note that evaluating image description generators is highly subjective and that no single metric can definitively assess text-to-text similarity. Since different metrics exhibit varying characteristics, we employ six commonly used metrics—BLEU-1, BLEU-2, BLEU-3, BLEU-4 \cite{papineni2002bleu}, ROUGE \cite{lin2004rouge}, and CIDEr \cite{vedantam2015cider}—to evaluate the outputs generated by our CDG. Table \ref{table:table_2_3} presents the evaluation results of our CDGs based on these six metrics. All CDG modules utilizing the keyword-driven approach outperform the non-keyword-driven baselines, demonstrating the effectiveness of incorporating keywords.

Moreover, as indicated in Table \ref{table:table_2_3} and supported by findings from \cite{jing2018automatic,wang2018tienet}, the evaluation scores for medical image captioning are significantly lower than those for natural image captioning using the same metrics. This discrepancy can be attributed to several factors: typically, medical image captions are longer and contain more abstract terms or concepts than their natural counterparts, which complicates the generation of accurate descriptions. Additionally, the inherent characteristics of the commonly used text-to-text similarity metrics \cite{huang2017vqabq,huang2019novel,huang2017robustness} may contribute to these lower scores.

Furthermore, Figure \ref{fig:figure_2_8} illustrates some of the generated clinical descriptions. While our CDG module does not consistently produce correct values for ``age'' or ``gender,'' it effectively generates accurate descriptions of significant features within retinal images.
Based on the assumptions outlined in Subsections 
\hyperref[med:method_4.1]{2.4.1} and \hyperref[med:method_4.2]{2.4.2}, we demonstrate that our proposed AI-based method is quantitatively effective.

\subsection{Evaluation by DNN Visual Explanation Module}\label{visual:DVE_2.5.3}

\begin{figure}[t!]
\begin{center}
   \includegraphics[width=1.0\linewidth]{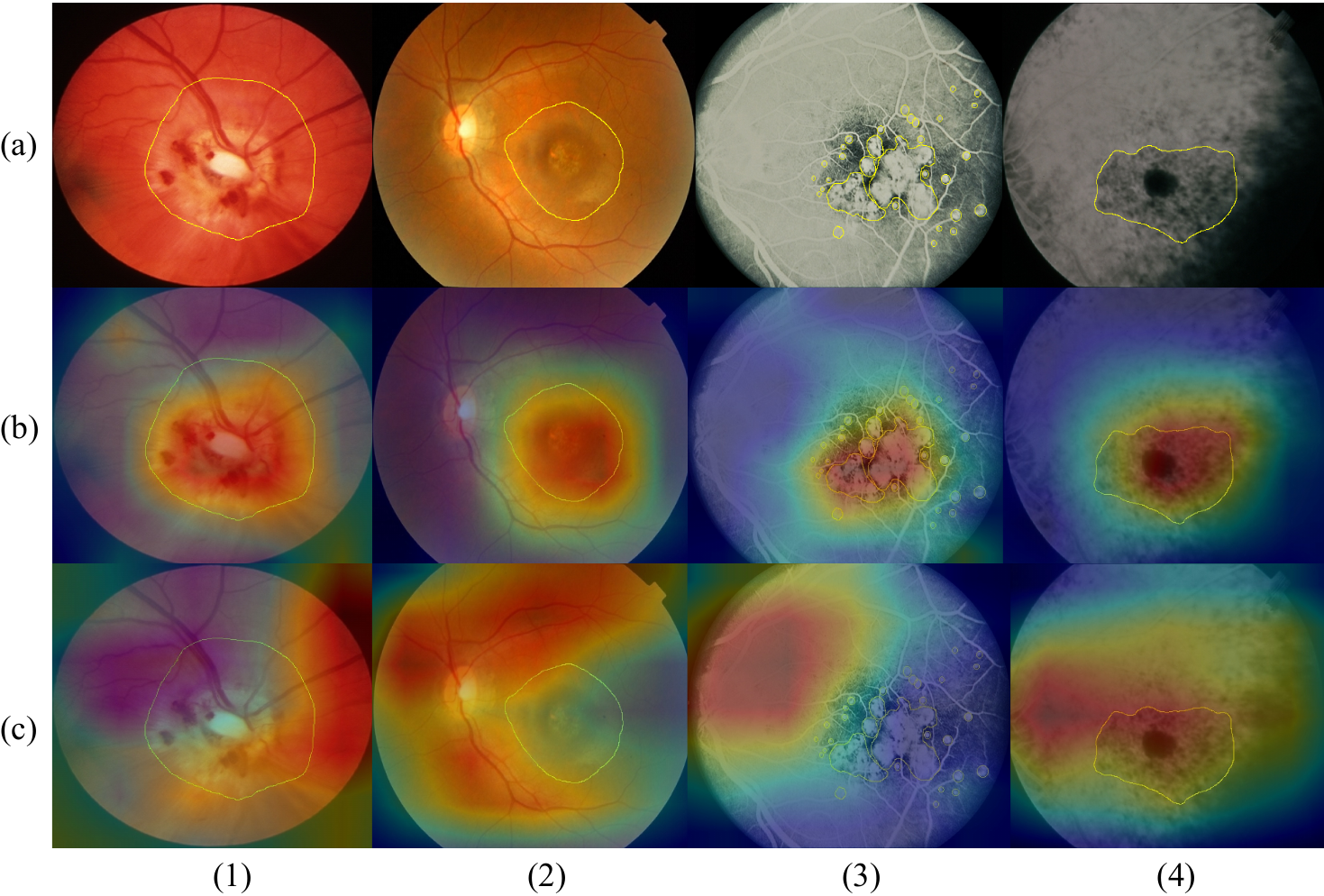}
\end{center}
    \caption{
    Qualitative results generated by CAM. For a detailed explanation, please refer to Subsection  \hyperref[visual:DVE_2.5.3]{2.5.3}.
    }
\label{fig:figure_2_9}
\end{figure}

The DNN visual explanation module evaluation centers on the premise that if the activation maps generated by CAM \cite{zhou2015cnnlocalization} are accepted by ophthalmologists, it indicates the qualitative effectiveness of our proposed method. To validate this, we created an additional retinal image dataset comprising 300 images, each labeled by ophthalmologists, and utilized the CAM visualization tool to analyze the learned features in comparison to the ground truth retinal images. The qualitative results are presented in Figure \ref{fig:figure_2_9}.

In row (a) of the figure, we display four different raw images of retinal diseases, each accompanied by a yellow outline drawn by an ophthalmologist to highlight the lesion areas. The images correspond to four distinct diseases: Optic Neuritis, Macular Dystrophy, Albinotic Spots in the Macula, and Stargardt Cone-Rod Dystrophy, denoted by numbers (1) to (4). Row (b) illustrates the visualization results generated by our DNN-based model using CAM. Row (c) is produced using the same method, but with a key difference: while both rows utilize the same pre-trained ImageNet weights, row (b) is fine-tuned on our DEN dataset, whereas row (c) is not.

The comparison between rows (b) and (c) demonstrates that the DNN-based model effectively learns robust features from retinal images when trained on our DEN dataset. Furthermore, the results in row (b) indicate that the features learned by the DNN align well with the clinical knowledge of ophthalmologists. This suggests that the activation maps generated by our deep models correspond to image features that ophthalmologists recognize as relevant to the identified diseases. Overall, these experimental results provide strong evidence that our proposed method is qualitatively effective.

\section{Conclusion}
In summary, we propose an AI-based method for automatically generating medical reports for retinal images, aimed at enhancing traditional retinal disease treatment procedures. This method comprises a DNN-based module, which includes the RDI and CDG sub-modules, as well as a DNN visual explanation module. To train our deep models and validate the effectiveness of the RDI and CDG, we introduce a large-scale retinal disease image dataset, DEN. Additionally, we provide another retinal image dataset that has been manually labeled by ophthalmologists to qualitatively assess the proposed method. The experimental results demonstrate that our method effectively improves the conventional treatment procedures for retinal diseases.

\chapter{Contextualized Features for Multi-modal Retinal Image Captioning}\label{ch:ch4}

\textit{
Medical image captioning involves automatically generating descriptions to convey the content of a given medical image. Traditional models typically rely on a single medical image input, which makes it challenging to produce abstract medical concepts or descriptions. This limitation hinders the effectiveness of medical image captioning. To address this issue, multi-modal medical image captioning incorporates textual inputs, such as expert-defined keywords, as key drivers in generating medical descriptions. Consequently, effectively encoding both the textual input and the medical image is crucial for successful multi-modal captioning. In Chapter \ref{ch:ch4}, we propose a new end-to-end deep multi-modal medical image captioning model that utilizes contextualized keyword representations, textual feature reinforcement, and masked self-attention. Evaluated against the existing commonly used multi-modal medical image captioning dataset, our experimental results demonstrate that the proposed model is effective, achieving improvements of +53.2\% in BLEU-avg and +18.6\% in CIDEr compared to the state-of-the-art methods.
}

\colorlet{shadecolor}{lightgray!50!} 
\begin{shaded}
\begin{changemargin}{0cm}{0.25cm} 
\small
This Chapter is based on:

\begin{itemize}
  \item \textit{``Contextualized Keyword Representations for Multi-modal Retinal Image Captioning''}, published in \textit{ACM International Conference on Multimedia Retrieval}, 2021~\cite{huang2021contextualized}, by \textbf{Jia-Hong Huang}, Ting-Wei Wu, and Marcel Worring.
\end{itemize}

\end{changemargin}
\end{shaded}

\begin{figure}[ht]
\begin{center}
\includegraphics[width=1.0\linewidth]{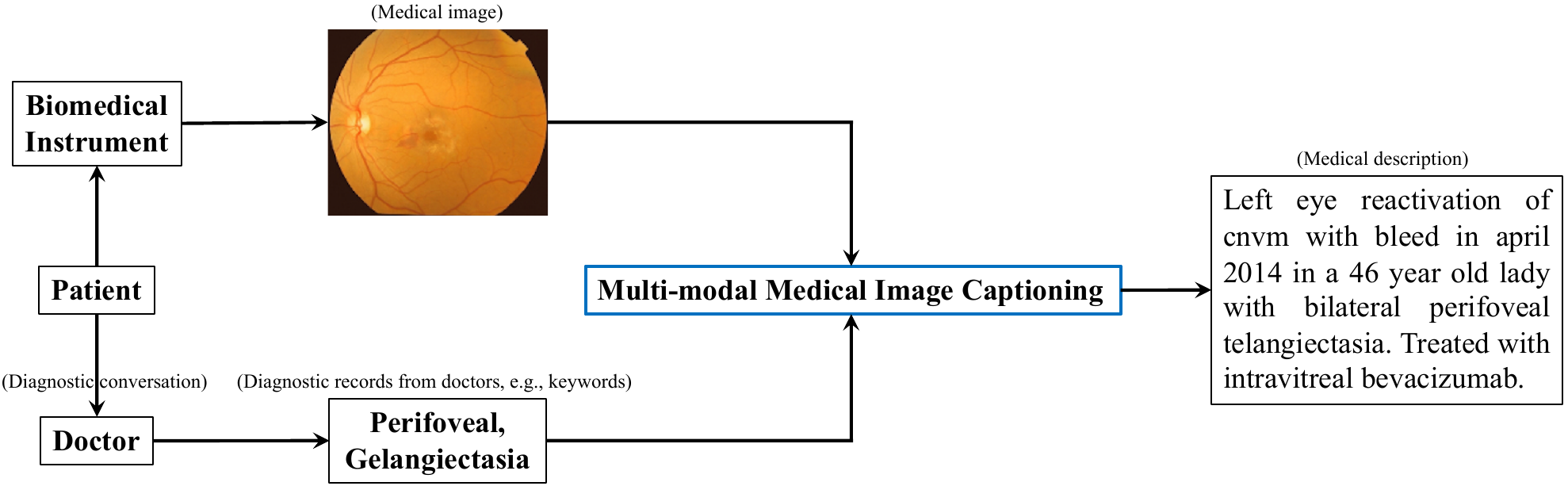}
\end{center}
   \caption{
   Illustration of multi-modal medical image captioning.  A multi-modal medical image captioning algorithm utilizes a medical image, such as a retinal image, alongside text-based diagnostic records, like a set of keywords, to generate a comprehensive medical description. The inclusion of these additional input keywords enhances the effectiveness of traditional medical image captioning models.
   }
\label{fig:figure_4_1}
\end{figure}

\section{Introduction} 
Medical image captioning involves the automatic generation of a report or description that conveys the content of a given medical image \cite{laserson2018textray,jing2018automatic,li2018hybrid,huang2021deepopht}. However, traditional medical image captioning methods \cite{laserson2018textray,li2018hybrid,jing2018automatic} rely solely on the image information to create these descriptions. This approach makes it challenging to derive abstract medical concepts \cite{laserson2018textray,jing2018automatic}, ultimately limiting the effectiveness of medical image captioning.

Multi-modal medical image captioning has emerged as a method to enhance conventional medical image captioning \cite{huang2021deepopht}. The primary objective of this approach is to generate descriptions for a given medical image by incorporating additional text-based information provided by physicians, such as keywords, to assist in the medical description generation, as illustrated in Figure \ref{fig:figure_4_1}. Keywords often appear in doctors' textual diagnosis records during the early diagnostic process \cite{huang2021deepopht}. Traditional medical image captioning typically utilizes a single input modality—the image—while an effective multi-modal medical image captioning system includes both the medical image and a set of keywords \cite{huang2021deepopht}. Since keywords play a crucial role as one of the main inputs in this context, their effective embedding is essential. In \cite{huang2021deepopht}, the Bag of Words (BoW) approach is employed to encode the keyword inputs for multi-modal medical image captioning. Although BoW has demonstrated success in natural language processing tasks, such as language modeling, it has notable limitations, as highlighted by the authors of \cite{scott1998text,soumya2014text}. First, from a time and space complexity perspective, the sparse representations produced by BoW can be challenging to model. Second, BoW may fail to capture the semantic meanings of the keywords effectively.

In Chapter \ref{ch:ch4}, we propose a novel approach to enhance the performance of multi-modal medical image captioning models. According to \cite{ethayarajh2019contextual}, static word embedding methods, such as skip-gram with negative sampling \cite{mikolov2013distributed}, offer a superior alternative to the BoW approach for encoding textual inputs. However, a significant limitation of static word embeddings is that all meanings of a polysemous word are represented by a single vector, as skip-gram generates only one representation per word \cite{ethayarajh2019contextual}. In contrast, contextualized word representations, such as those generated by Generative Pretrained Transformer-2 (GPT-2), are more effective for capturing nuanced meanings \cite{ethayarajh2019contextual}. To encode textual input, our proposed method leverages GPT-2 to better represent the input keywords. For visual feature extraction, we utilize a pre-trained CNN, such as VGG16 or VGG19, trained on ImageNet \cite{simonyan2014very,russakovsky2015imagenet}, to effectively encode the input medical images. Our experiments are based on the existing multi-modal retinal image captioning dataset introduced by \cite{huang2021deepopht}, which primarily consists of colorful retinal images. Thus, employing a CNN pre-trained on ImageNet proves beneficial for capturing low-level features, such as color \cite{huang2021deepopht}. The experimental results indicate that our multi-modal medical image captioning method generates more accurate and meaningful descriptions for retinal images compared to baseline models.

\vspace{+3pt}
\noindent\textbf{Contributions.}
\begin{itemize}

    \item We propose a new end-to-end deep model for multi-modal medical image captioning that incorporates contextualized keyword representations, a textual feature reinforcement module, and a masked self-attention mechanism.

    \item The proposed method is rigorously validated through experiments on an established multi-modal retinal image captioning dataset. The results demonstrate that our model outperforms those based on static word embeddings or the BoW approach, with significant improvements in performance as indicated by higher BLEU and CIDEr scores.

\end{itemize}

\section{Related Work}
This section reviews relevant literature on image captioning, multi-modal medical image captioning, word embedding techniques, and retinal image datasets.

\subsection{Image Description Generation}

The objective of conventional image captioning is to automatically generate textual descriptions for natural images \cite{vinyals2015show,fang2015captions,karpathy2015deep}. In \cite{vinyals2015show}, an encoder-decoder architecture is introduced, utilizing CNNs as the encoder to extract image features, while RNNs served as the decoder to produce descriptions based on these features. The work by \cite{fang2015captions} integrates a language model to combine potential words related to various parts of an image, facilitating the generation of image descriptions. Additionally, \cite{karpathy2015deep} proposes an approach that embeds both language and visual information into a shared space. In \cite{hendricks2016generating}, the authors focus on distinguishing properties of visible objects by jointly predicting class labels and providing explanations for the appropriateness of these labels concerning input images. Their model utilizes reinforcement learning and a loss function to refine the generated captions.

According to \cite{liu2017improved}, existing image captioning models are trained using maximum likelihood estimation. However, this method has limitations, as the log-likelihood scores of certain descriptions may not correlate well with human quality assessments. In \cite{gao2019deliberate}, a deliberate residual attention image captioning model is proposed, which employs first-pass residual-based attention to generate hidden states and visual attention, followed by refining preliminary descriptions through second-pass attention based on global features. This method has shown the potential to generate more accurate image captions. Furthermore, \cite{pan2020x} introduces a unified attention block that leverages bilinear pooling to enhance reasoning and selectively utilize visual information.

Existing medical image captioning methods, such as those proposed by \cite{laserson2018textray,li2018hybrid,jing2018automatic}, largely rely on traditional natural image captioning approaches. However, the methods often fail to generalize effectively to medical datasets. An image captioning model that incorporates contextual information, such as keywords, represents a promising avenue for improving medical image descriptions \cite{huang2021deepopht}. In Chapter \ref{ch:ch4}, we introduce a new context-driven model aimed at enhancing the performance of medical image captioning.

\subsection{Multi-modal Medical Image Captioning}
Recently, the multi-modal task of VQA has gained attention \cite{malinowski2015ask,malinowski2014multi,malinowski2017ask,huang2019assessing,huang2018robustness,huang2017vqabq,huang2017robustness,huang2019novel}. The primary goal of VQA is to provide a text-based answer to a question posed in text form, based on the content of a given image. In VQA, textual and visual inputs interact, with one modality aiding the other \cite{agrawal2017vqa}. This concept of leveraging multiple input modalities can also be applied to develop multi-modal models for related tasks \cite{jing2018automatic} or multi-modal medical image captioning \cite{huang2021deepopht}.

For instance, \cite{jing2018automatic} introduces a multi-modal approach to generate medical descriptions for lung X-ray images. However, their model relies solely on image input to produce intermediate outputs, such as text-based tags, to support the subsequent generation of medical descriptions. This reliance on model-generated intermediate products can lead to inaccuracies or biased information, potentially confusing the model during training. To address this issue, \cite{huang2021deepopht} proposes a multi-modal model that incorporates both a set of expert-defined keywords and retinal images to enhance the quality of medical descriptions. The inclusion of expert-defined keywords ensures that the input's correctness and quality are maintained \cite{huang2021deepopht}.

While integrating expert-defined keywords as a multi-modal input can enhance model performance, effective encoding of textual inputs presents another challenge. In Chapter \ref{ch:ch4}, we introduce a new model that employs an effective method for textual input embedding to tackle this challenge.

\subsection{Word Embeddings}
In Chapter \ref{ch:ch4}, we categorize word embedding methods into three distinct types: BoW, static word embeddings, and contextualized word representations.

\noindent\textbf{BoW.} 
The BoW model \cite{harris1954distributional} is a simplified representation widely used in information retrieval, natural language processing, and computer vision. In the BoW approach, a sentence or document is represented as a collection of its words, preserving their frequency. Each word's occurrence is used as a feature for model training. However, as noted by \cite{scott1998text,soumya2014text,huang2021deepopht}, the BoW method often fails to capture semantic meaning effectively, making it inadequate for multi-modal medical image captioning tasks.

\noindent\textbf{Static Word Embeddings.} 
According to \cite{ethayarajh2019contextual}, Skip-gram with negative sampling \cite{mikolov2013distributed} and GloVe \cite{pennington2014glove} are two of the most prominent models for generating static word embeddings. These models learn word embeddings iteratively; however, they both effectively factorize a word-context matrix that captures co-occurrence statistics \cite{levy2014neural,levy2014linguistic}. A significant limitation of static word embeddings is that all meanings of a polysemous word are represented by a single vector, resulting in a loss of semantic distinction for words with multiple senses.

\noindent\textbf{Contextualized Word Representations.} 
To address the limitations of static word embeddings, the authors of \cite{peters2018deep,devlin2018bert,radford2019language} have introduced various deep neural language models that generate context-sensitive word representations. These models can be fine-tuned for a wide range of downstream natural language processing tasks, with the internal representations of words derived from the entire textual input. Consequently, these representations are referred to as contextualized word representations. An approach presented in \cite{liu2019linguistic} suggests that these representations capture task-agnostic and highly transferable language properties. In \cite{peters2018deep}, a method is introduced to generate contextualized representations for each token by concatenating the internal states of a 2-layer biLSTM. Conversely, \cite{radford2019language} describes a uni-directional transformer-based language model \cite{vaswani2017attention}, while \cite{devlin2018bert} presents a bi-directional transformer-based model. According to \cite{ethayarajh2019contextual}, contextualized word representations are more effective than static embeddings. Therefore, in Chapter \ref{ch:ch4}, we propose to leverage contextualized word representations to develop a new model for multi-modal medical image captioning.

\subsection{Retinal Image Datasets}
Recently, numerous medical image datasets have been developed for research purposes, particularly in the field of retinal images \cite{hoover2003locating,staal2004ridge,kauppi2007diaretdb1,al2008reference,carmona2008identification,ortega2009retinal,niemeijer2011inspire,computing2012understanding,huang2021deep,fraz2012ensemble,huang2020query,vazquez2013improving,huang2021gpt2mvs,sivaswamy2014drishti,hu2019silco,yang2018novel,yang2018auto,liu2019synthesizing,decenciere2014feedback,huang2021longer,adal2015accuracy,hernandez2017fire,porwal2018indian,huang2021deepopht}. This subsection reviews the existing retinal image datasets mentioned above.

In \cite{hoover2003locating}, the STARE dataset is introduced, comprising 397 images used to develop an automatic system for diagnosing eye diseases. The DRIVE dataset, presented by \cite{staal2004ridge}, includes 40 retina images, evenly divided into training and testing sets. The training set features a single manual segmentation of the vasculature, while the testing set provides two manual segmentations.

The DIARETDB1 dataset \cite{kauppi2007diaretdb1} consists of 89 color fundus images, with 84 showing at least mild non-proliferative signs of DR and the remaining five being normal. The REVIEW dataset \cite{al2008reference} includes 14 images specifically designed for segmentation tasks. DRIONS-DB \cite{carmona2008identification} contains 110 color retinal images with various visual characteristics, such as cataracts, light artifacts, and concentric peripapillary atrophy.

The VARIA dataset \cite{ortega2009retinal} includes 233 images from 139 individuals for authentication purposes. The INSPIRE-AVR dataset \cite{niemeijer2011inspire} consists of 40 color retinal images focused on vessels and the optic disc, along with a reference standard for the arterio-venous ratio. ONHSD \cite{computing2012understanding} includes 99 retinal images intended for segmentation tasks. The CHASE-DB1 dataset \cite{fraz2012ensemble} contains 14 retinal images used for vessel analysis, while the VICAVR dataset \cite{vazquez2013improving} comprises 58 images designed to calculate the A/V ratio.

The Drishti-GS dataset \cite{sivaswamy2014drishti} includes 101 images, split into 50 for training and 51 for testing. The MESSIDOR dataset \cite{decenciere2014feedback} features 1,200 color eye fundus images without manual annotations, such as lesion contours or positions. The RODREP dataset \cite{adal2015accuracy} consists of 1,120 color fundus images repeated across 70 patients within a DR screening program at the Rotterdam Eye Hospital. The FIRE dataset \cite{hernandez2017fire} contains 129 retinal images organized into 134 image pairs categorized by specific characteristics. Finally, the IDRiD dataset \cite{porwal2018indian} provides 516 retinal fundus images with ground truths related to DR signs, Diabetic Macular Edema (DME), and normal retinal structures. This dataset includes pixel-level labels of DR lesions and the optic disc, coordinates for the optic disc and fovea center, and image-level disease severity grading for DR and DME.

While numerous retinal image datasets are available, not all are tailored for multi-modal retinal image captioning. Chapter \ref{ch:ch4} primarily relies on the DEN dataset introduced by \cite{huang2021deepopht}, which is large-scale and specifically designed for multi-modal deep learning research.

\begin{figure*}[t!]
  \includegraphics[width=\textwidth]{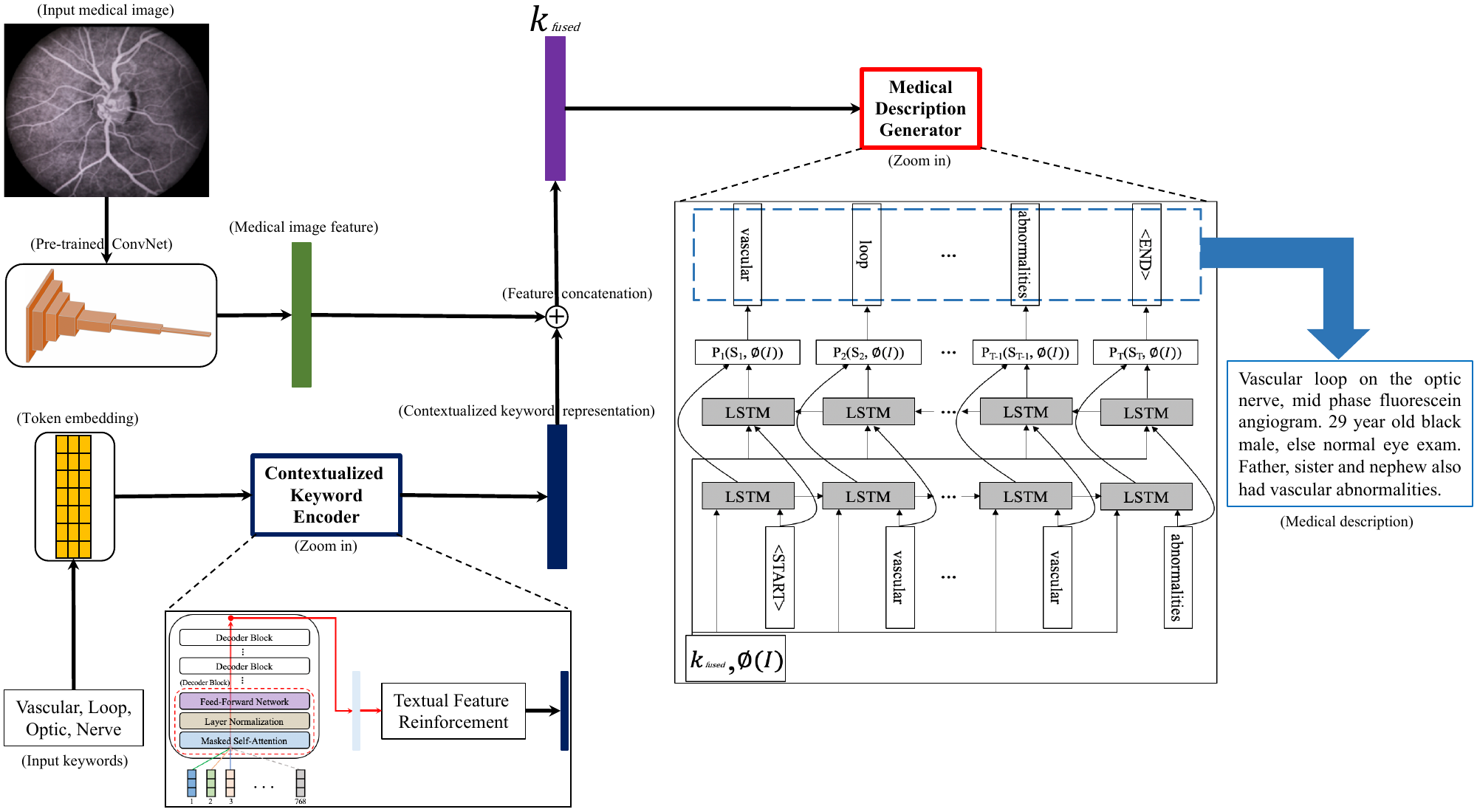}
  \caption{
  Flowchart of our proposed multi-modal medical image captioning model. A pre-trained CNN, such as VGG16 or VGG19 trained on ImageNet, is employed to extract features from the medical image input (represented in dark green). The ``Token Embedding'' layer processes a set of input keywords, producing the input for the ``Contextualized Keyword Encoder,'' which consists of a series of decoder blocks and ``Textual Feature Reinforcement.'' Each decoder block includes masked self-attention, layer normalization, and a feed-forward network (indicated by the red dashed line box). The ``Textual Feature Reinforcement,'' comprising a stack of fully connected layers, generates the contextualized keyword representation (shown in dark blue). It is important to note that the keyword encoder receives 768 color-coded, brick-stacked vectors as input. The symbol $\oplus$ denotes the concatenation of the medical image features and the contextualized keyword representation. In the ``Medical Description Generator,'' which formulates the medical description, $k_{\textup{fused}}$ represents a fused feature vector (illustrated in purple), $\phi(I)$ denotes the image feature vector, and $P_{i}(S_{i}, \phi(I))$ indicates a probability distribution, where $i=1,2,..., T$. Please refer to Section \hyperref[med:method_4.3]{4.3} for further details.
          }
  \label{fig:figure_4_2}
\end{figure*}

\section{Methodology}\label{med:method_4.3}
This section provides a detailed description of the proposed multi-modal medical image captioning model. The method comprises two main components: a contextualized keyword encoder and a medical description generator. The flowchart of the proposed model is illustrated in Figure \ref{fig:figure_4_2}.

\subsection{Contextualized Keyword Encoder}

A transformer comprises two main components: a transformer-encoder and a transformer-decoder \cite{vaswani2017attention}. This architecture has been successfully applied to tasks such as language modeling and machine translation. Both the transformer-encoder and transformer-decoder consist of stacks of multiple basic transformer blocks. Our proposed model draws inspiration from the GPT-2 structure, specifically its transformer-decoder-like architecture, known for its parallelization and masked self-attention capabilities. These characteristics are utilized to develop the proposed contextualized keyword encoder for embedding keywords. The details of the contextualized keyword encoder are as follows:

\begin{equation}
    x_n = W_e*k_n, n \in \{0,...,N-1\}, 
	\label{eq:text_embed1}
\end{equation}
where $x_n$ represents an input token or keyword embedding, $W_e \in \mathbb{R}^{E_s \times V_s}$ denotes the token embedding matrix, $E_s$ is the size of the word embedding, $V_s$ is the vocabulary size, $k_n$ signifies an input token or keyword, and $N$ is the number of input tokens or keywords.

\noindent\textbf{Masked Self-attention Mechanism.}
The mechanism of masked self-attention is outlined as follows:

\begin{equation}
    Q = W_q*x_n+b_q,
	\label{eq:text_embed2}
\end{equation}
where $Q$ represents the current word's representation \cite{vaswani2017attention}. A linear layer, denoted as $W_q \in \mathbb{R}^{H_s \times E_s}$ with a bias term $b_q$ and an output size of $H_s$, is used to generate $Q$.

\begin{equation}
    K = W_k*x_n+b_k,
	\label{eq:text_embed3}
\end{equation}
where $K$ denotes the key vector \cite{vaswani2017attention}. A linear layer, represented as $W_k \in \mathbb{R}^{H_s \times E_s}$ with a bias term $b_k$ and an output size of $H_s$, is employed to generate $K$.

\begin{equation}
    V = W_v*x_n+b_v,
	\label{eq:text_embed4}
\end{equation}
where $V$ represents the value vector \cite{vaswani2017attention}. A linear layer, denoted as $W_v \in \mathbb{R}^{H_s \times E_s}$ with a bias term $b_v$ and an output size of $H_s$, is utilized to generate $V$.

\begin{equation}
    \textup{MaskAtten}(Q,K,V) = \textup{softmax}(m(\frac{QK^T}{\sqrt{d_k}}))V,
	\label{eq:attention}
\end{equation}
where $m(\cdot)$ represents the masked self-attention function, and $d_k$ denotes the scaling factor \cite{vaswani2017attention}.

\noindent\textbf{Layer Normalization.}
Layer normalization is computed as shown in Equation (\ref{eq:layernorm}).

\begin{equation}
    Z_{\textup{Norm}} = \textup{LayerNorm}(\textup{MaskAtten}(Q,K,V)),
	\label{eq:layernorm}
\end{equation}
where $\textup{MaskAtten}(Q,K,V)$ refers to the output from Equation (\ref{eq:attention}), and $\textup{LayerNorm}(\cdot)$ represents the layer normalization function.

\noindent\textbf{Contextualized Keyword Representation.}
By applying the previous Equations (\ref{eq:text_embed1}-\ref{eq:layernorm}), we derive the contextualized keyword representation $F$ as follows:

\begin{equation}
    F = \textup{FFN}(Z_{\textup{Norm}}) = \sigma(W_1Z_{\textup{Norm}}+b_1)W_2+b_2,
	\label{eq:ffn}
\end{equation}
where $\textup{FFN}(\cdot)$ represents a position-wise feed-forward network (FFN), and $\sigma$ denotes an activation function. The parameters $W1$, $W2$, $b1$, and $b2$ are learnable components of the FFN. In practice, a stack of fully connected layers is employed to enhance $F$. Refer to Figure \ref{fig:figure_4_2} for further details.

\subsection{Loss Function}

In Chapter \ref{ch:ch4}, we approach medical description generation as a classification problem, specifically focusing on word-by-word classification based on probability vectors. Consequently, we employ the categorical cross-entropy loss function to develop our proposed method, as indicated in Equation (\ref{eq:loss}).

\begin{equation}
    \textup{Loss} = -\frac{1}{N}\sum_{i=0}^{N-1}\sum_{c=0}^{C-1}\mathbf{1}_{y_{i}\in C_{c}}\textup{log}(P_{\textup{model}}\left [y_{i}\in C_{c} \right ]),
    \label{eq:loss}
\end{equation}
where $N$ represents the number of observations, $C$ denotes the number of categories, $\mathbf{1}_{y_i \in C_c}$ is an indicator function that indicates whether the $i^{\textup{th}}$ observation belongs to the $c^{\textup{th}}$ category, and $P_{\textup{model}}[y_i \in C_c]$ is the probability predicted by the model that the $i^{\textup{th}}$  observation belongs to the $c^{\textup{th}}$ category.

When there are more than two categories, the neural network model produces a probability vector with $C$ dimensions. Each element in this vector represents the probability that the input belongs to the corresponding category. It is important to note that when the number of categories is two, the categorical cross-entropy loss simplifies to binary cross-entropy loss, which is a specific case of categorical cross-entropy. In this scenario, the neural network outputs a single probability $\hat{y}_i$, while the probability for the other category is $1-\hat{y}_i$.

\subsection{Medical Description Generator}

The medical description generator utilizes the CNN medical image encoder $\phi$ from \cite{huang2021deepopht} to extract image features. These extracted features are then fed into a subsequent bidirectional LSTM model at each time step. Here, $p(S_t|I, S_0,..., S_t-1)$ represents the probability of the current word given all preceding words, while $S=(S_0,..., S_T)$ denotes the true sentence that describes the input image $I$.

The medical description generator is structured as follows:

\begin{equation}
    e_t = W_d\times \phi(I), t \in \{0,...,T\},
	\label{eq:4_5}
\end{equation}
where $W_d \in \mathbb{R}^{E \times F}$ represents a fully connected layer, $E$ denotes the size of the word embedding, and $F$ indicates the size of the image feature

\begin{equation}
    x_t = W_eS_t, t \in \{0,...,T\},
	\label{eq:5_5}
\end{equation}
where each word is denoted by its bag-of-words ID $S_t$, while both the sentence $S$ and the image $I$ are mapped into the same high-dimensional space.

\begin{equation}
    P_{t} = \textup{BiLSTM}([e_t, k_{\textup{fused}}, x_t]), t \in \{0,...,T\},
    \label{eq:6_5}
\end{equation}
where in Equation (\ref{eq:6_5}), at each time step, the image contents $e_t$, the fused multi-modal feature $k_{\textup{fused}}$, and the ground truth word vector $x_t$ are input into the network to enhance its memory of the images.

\begin{table*}
\caption{
State-of-the-art comparisons. The proposed method is compared with the state-of-the-art model ``DeepOpht with BoW'' \cite{huang2021deepopht} using evaluation metrics such as BLEU \cite{papineni2002bleu}, CIDEr \cite{vedantam2015cider}, ROUGE \cite{lin2004rouge}, and METEOR \cite{banerjee2005meteor}. The results show that the proposed approach outperforms the model in \cite{huang2021deepopht} by +53.2\% in BLEU-avg and +18.6\% in CIDEr. These findings are based on a VGG16 image feature extractor pre-trained on ImageNet, with a beam search algorithm employed during the testing phase. Note that BLEU-avg refers to the average scores of BLEU-1, BLEU-2, BLEU-3, and BLEU-4. The term ``beam'' indicates the number of beams used in the beam search, while $*$ signifies the unavailability of data. Similar notations are used in Table \ref{table:table_4_2}. Overall, the proposed method exceeds the baseline performance even with beam=1 in both BLEU and CIDEr and is competitive in ROUGE-L.
}
\centering
\rotatebox{270}{
\scalebox{0.68}{
\begin{tabular}{c|c|c|c|c|c|c|c|c}
\toprule
Model                  & BLEU-1 & BLEU-2 & BLEU-3 & BLEU-4 & BLEU-avg & CIDEr & ROUGE-L & METEOR \\ \midrule
DeepOpht with BoW (beam=3) \cite{huang2021deepopht}  & 0.144  & 0.092  & 0.052  & 0.021  & 0.077  & 0.296 & \textbf{0.197} & $*$ \\ \midrule

Proposed Model with GloVe (beam=1)    & 0.173  & 0.111  & 0.072  & 0.048  & 0.101   & 0.243 & 0.164 & 0.153 \\ \midrule
                                    
                                    
Proposed Model with GPT-2 (beam=1)   & \textbf{0.192}  & \textbf{0.130}  & \textbf{0.088}  & \textbf{0.060}  & \textbf{0.118}  & \textbf{0.351} & 0.188 & \textbf{0.167} \\ \bottomrule
                                
                                 
\end{tabular}}
}
\label{table:table_4_1}
\end{table*}

\begin{table*}
\caption{
The proposed method is compared with the state-of-the-art model ``DeepOpht with BoW'' \cite{huang2021deepopht}. The evaluation is conducted using a VGG19 image feature extractor pre-trained on ImageNet, along with a beam search algorithm during the testing phase. The results indicate that the proposed method outperforms the model in \cite{huang2021deepopht} by +30\% in BLEU-avg and +8\% in CIDEr.
}
\centering
\rotatebox{270}{
\scalebox{0.70}{
\begin{tabular}{c|c|c|c|c|c|c|c|c}
\toprule
Model                  & BLEU-1 & BLEU-2 & BLEU-3 & BLEU-4 & BLEU-avg & CIDEr & ROUGE-L & METEOR \\ \midrule
DeepOpht with BoW (beam=3) \cite{huang2021deepopht}     & 0.184  & 0.114  & 0.068  & 0.032  & 0.100  & 0.361 & \textbf{0.232} & $*$ \\ \midrule

Proposed Model with GloVe (beam=1)   & 0.192  & 0.132  & 0.093  & 0.067  & 0.121   & 0.356 & 0.186 & 0.169 \\ \midrule 
                                    
Proposed Model with GloVe (beam=3)   & 0.201  & 0.139  & 0.098  & 0.071  & 0.127  & 0.359 & 0.203 & 0.184 \\ \midrule 
                                    
Proposed Model with GPT-2 (beam=1)     & 0.203  & 0.137  & 0.093  & 0.065  & 0.125  & 0.356 & 0.197 & 0.174 \\ \midrule 
             
Proposed Model with GPT-2 (beam=3)     & \textbf{0.203}  & \textbf{0.142}  & \textbf{0.100}  & \textbf{0.073}  & \textbf{0.130}  & \textbf{0.389} & 0.211 & \textbf{0.188} \\ \bottomrule      
                                 
\end{tabular}}
}
\label{table:table_4_2}
\end{table*}

\section{Experiments and Analysis}
This section introduces the dataset and evaluation metrics utilized in the experiments, followed by a detailed description of the experimental setup. Subsequently, we analyze the effectiveness of the proposed multi-modal medical image captioning model. Finally, we present several randomly selected qualitative results.

\subsection{Dataset and Evaluation Metrics}

\noindent\textbf{Dataset.}
In \cite{huang2021deepopht}, a state-of-the-art model is introduced alongside a large-scale retinal image dataset featuring unique expert-defined keyword annotations for multi-modal medical image captioning. The dataset comprises 1,811 grayscale FA images and 13,898 colorful CFP images. Each image is associated with two corresponding labels: a clinical description and expert-defined keywords. In this dataset, the longest keyword sequence contains over 15 words, while the clinical descriptions can reach up to 50 words. The average word length of the keywords and clinical descriptions is between 5 and 10 words. The dataset encompasses 265 different retinal diseases and symptoms, including both common and rare conditions. The expert-defined keywords are derived from the analysis and diagnosis records of ophthalmologists and retinal specialists, providing insights into potential retinal diseases, symptoms, and patient characteristics. The dataset is divided into 60\% for training, 20\% for validation, and 20\% for testing \cite{huang2021deepopht}.

\noindent\textbf{Evaluation Metrics.}
This study employs the same medical description evaluation metrics as those used in \cite{huang2021deepopht} to assess the performance of the proposed model: BLEU \cite{papineni2002bleu}, CIDEr \cite{vedantam2015cider}, and ROUGE \cite{lin2004rouge}. Additionally, we include METEOR \cite{banerjee2005meteor}, a commonly used text evaluation metric, to further evaluate the proposed method. The definitions of these evaluation metrics are provided below:

\begin{equation}
    \textup{BP}= \left\{\begin{matrix}
                        1 & \textup{if} & c>r \\ 
                        \textup{exp}(1-\frac{r}{c}) & \textup{if} & c\leq r
                        \end{matrix}\right.;
    \textup{BLEU}= \textup{BP}\cdot \textup{exp}\left ( \sum_{n=1}^{N} w_{n}\textup{log}p_{n} \right ),
    \label{eq:bleu}
\end{equation}
where $r$ represents the effective length of the ground truth text, $c$ indicates the length of the predicted text, and $\textup{BP}$ refers to the brevity penalty. The geometric average of the modified $n$-gram precisions, $p_{n}$, is calculated using $n$-grams up to a length of $N$, with positive weights $w_{n}$ that sum to 1.

\begin{equation}
    \begin{split}
    \textup{CIDEr}_{n}(c_{i},S_{i})= \frac{1}{m}\sum_{j}\frac{\boldsymbol{g}^{\boldsymbol{n}}(c_{i})\cdot               
        \boldsymbol{g}^{\boldsymbol{n}}(s_{ij})}{\left \| \boldsymbol{g}^{\boldsymbol{n}}(c_{i}) \right \|\left \| \boldsymbol{g}^{\boldsymbol{n}}(s_{ij}) \right \|};\\
    \textup{CIDEr}(c_{i},S_{i})= \sum_{n=1}^{N}w_{n}\textup{CIDEr}_{n}(c_{i},S_{i}),
    \end{split}
    \label{eq:f1-cider}
\end{equation}
where $c_{i}$ represents the predicted text, while $S_{i}$ = \{$s_{i1}$,..., $s_{im}$\} denotes a set of ground truth descriptions. The $\textup{CIDEr}_{n}(c_{i},S_{i})$ score for $n$-grams of length $n$ is calculated as the average cosine similarity between $c_{i}$ and $S_{i}$, incorporating both precision and recall. The vector $\boldsymbol{g}^{\boldsymbol{n}}(c_{i})$ is formed by the $g_{k}(c_{i})$ values corresponding to all $n$-grams of length $n$, with $\left \| \boldsymbol{g}^{\boldsymbol{n}}(c_{i}) \right \|$ representing the magnitude of this vector. A similar definition applies to $\boldsymbol{g}^{\boldsymbol{n}}(s_{ij})$. Higher-order (longer) $n$-grams are utilized to capture grammatical structures and richer semantic content. The $\textup{CIDEr}(c_{i}, S_{i})$ score represents the overall score based on $n$-grams of various lengths.

\begin{equation}
    \textup{R}_{\textup{lcs}}=\frac{\textup{LCS}(X,Y)}{m};
    \textup{P}_{\textup{lcs}}=\frac{\textup{LCS}(X,Y)}{n};
    \textup{F}_{\textup{lcs}}=\frac{(1+\beta^{2})\textup{R}_{\textup{lcs}}\textup{P}_{\textup{lcs}}}{\textup{R}_{\textup{lcs}}+\beta^{2}\textup{P}_{\textup{lcs}}},
    \label{eq:f1-rouge}
\end{equation}
The longest common subsequence (LCS) based $\textup{F}$-measure, denoted as $\textup{F}_{\textup{lcs}}$ or ROUGE-L, is employed to evaluate the similarity between the ground truth text $X = \{x_{1}, \ldots, x_{m}\}$ of length $m$ and the predicted text $Y = \{y_{1}, \ldots, y_{n}\}$ of length $n$. The parameter $\beta$ is used to balance the relative importance of $\textup{P}_{\textup{lcs}}$ and $\textup{R}_{\textup{lcs}}$.

\begin{equation}
    \textup{METEOR}_{\textup{score}}=\frac{10PR}{R+9P}\left ( 1-0.5(\frac{\#\textup{chunks}}{\#\textup{unigrams}\_\textup{matched}})^{3} \right ),
    \label{eq:f1-meteor}
\end{equation}
where $P$ represents unigram precision, $R$ denotes unigram recall, $\#\textup{chunks}$ refers to the number of chunks, and $\#\textup{unigrams}\_\textup{matched}$ indicates the number of matched unigrams. For further details, see \cite{banerjee2005meteor}.

\begin{figure}[t!]
\includegraphics[width=1.0\linewidth]{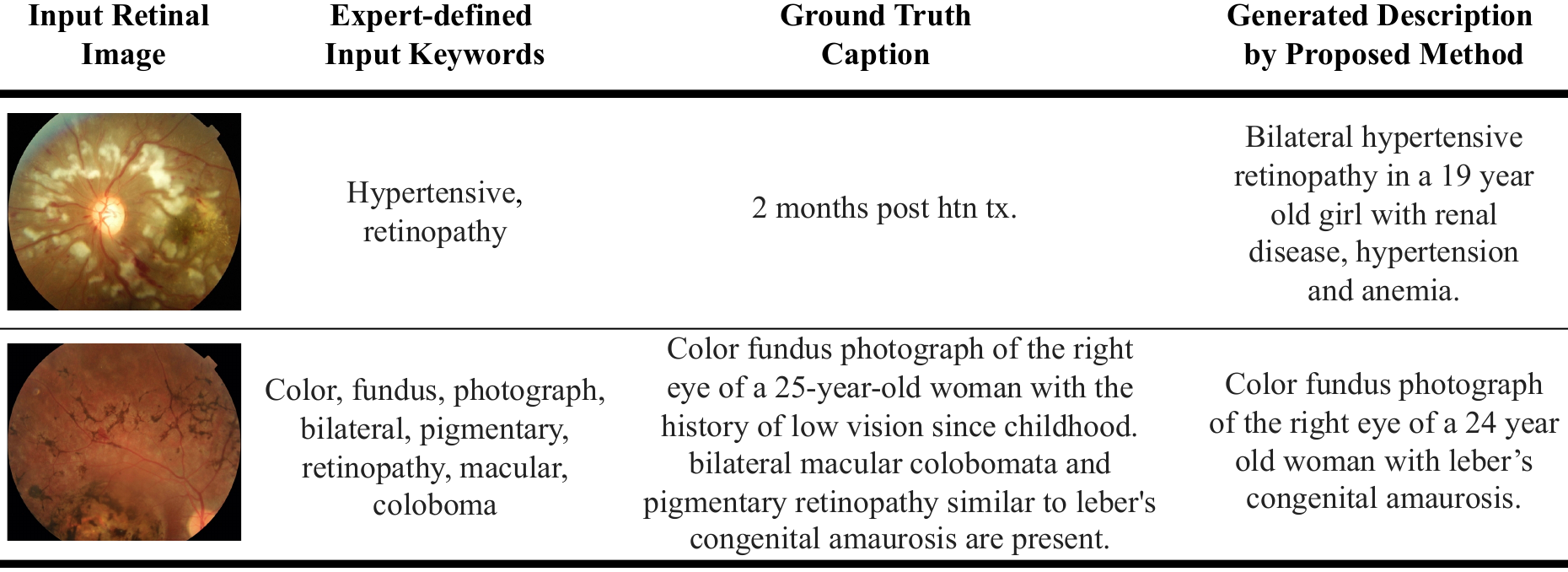}
\caption{
Qualitative results from the proposed multi-modal medical image captioning model. These results illustrate how the model effectively leverages contextualized keyword representations to generate meaningful medical descriptions. It is important to note that attributes such as ``date,'' ``skin color,'' ``gender,'' and ``age'' are included in the dataset, and the system should incorporate these into the medical descriptions through slot-filling or post-processing \cite{huang2021deepopht}. In the first ground truth caption, ``htn'' is used as an abbreviation for ``hypertensive,'' as noted by the retinal specialist.
}
\label{fig:figure_4_3}
\end{figure}

\subsection{Experimental Setup}
In this study, we utilize VGG16 and VGG19 pre-trained on ImageNet to extract image features, similar to the approach in \cite{huang2021deepopht}. For processing keywords and descriptions, we remove non-alphabetic characters, convert all remaining characters to lowercase, and replace words that appear only once with a special token, $<UNK>$. Without keywords, the vocabulary size is $4,007$, while including keywords increases the vocabulary size to $4,292$. All sentences are truncated or padded to a maximum length of $50$. Each word is encoded with a token/word embedding size of $300$. The proposed contextualized keyword encoder is based on the GPT-2 architecture \cite{radford2019language}, and we leverage the pre-trained weights of GPT-2 for initialization, which has a vocabulary size of $50,257$ and is trained on a large corpus. For the medical description generator, we use a hidden layer size of $256$ for the LSTM unit. The dataset from \cite{huang2021deepopht} is used with a default split of $60\%$ for training, $20\%$ for validation, and $20\%$ for testing. The implementation is carried out using Keras, with models trained for $2$ epochs, a batch size of $64$, a learning rate of $1e-3$, and the Adam optimizer \cite{kingma2014adam}. For the Adam optimizer, the coefficients for computing the moving averages of the gradient and its square are set to $\beta_{1}=0.9$ and $\beta_{2}=0.999$, respectively. Additionally, $\epsilon=1e-8$ is added to the denominator to enhance numerical stability.

\subsection{Effectiveness Analysis}
Tables \ref{table:table_4_1} and \ref{table:table_4_2} indicate that the proposed model, which utilizes contextualized keyword representations, significantly outperforms the baseline model using BoW \cite{huang2021deepopht}. Additionally, the model with contextualized keyword representations surpasses the one using static word embeddings \cite{pennington2014glove}. This improvement can be attributed to the ability of contextualized keyword representations to effectively capture keyword information, leading to enhancements of +$53.2$\% in BLEU-avg and +$18.6$\% in CIDEr, as well as an overall improvement in the quality of the generated medical descriptions. It is important to note that while employing the beam search algorithm with three beams enhances model performance during testing, it incurs a computational cost of approximately $12$ times greater than using a single beam (with times of $14851$ seconds compared to $1265$ seconds). Qualitative results are illustrated in Figure \ref{fig:figure_4_3}.

\section{Conclusion}
In conclusion, we introduce an innovative end-to-end deep learning model for multi-modal medical image captioning. Our approach harnesses contextualized keyword representations, a textual feature reinforcement module, and masked self-attention mechanisms to boost performance. Extensive experiments validate the model's effectiveness, showing that it substantially surpasses the baseline model, with marked improvements in BLEU and CIDEr scores.

\chapter{Non-Local Attention Enhances Retinal Image Captioning}\label{ch:ch5}

\textit{
Automatically generating medical reports from retinal images presents a significant challenge, as algorithms must produce semantically coherent descriptions for each image. Traditional methods primarily rely on the input image for generating these descriptions; however, many abstract medical concepts cannot be derived from image information alone. In Chapter \ref{ch:ch5}, we enhance the report generation process by integrating additional information. We note that ophthalmologists often record a small set of keywords early in the diagnostic process, highlighting critical information that later informs the creation of medical reports. Recognizing the importance and prevalence of these keywords, we incorporate them into the automatic report generation framework. Given that we are working with two types of inputs—expert-defined unordered keywords and images—effectively merging features from these diverse modalities poses a challenge. To address this, we propose TransFuser, a novel keyword-driven medical report generation method that utilizes a non-local attention-based multi-modal feature fusion approach. Our experiments demonstrate that TransFuser effectively captures the mutual information between keywords and image content. Furthermore, we show that our keyword-driven generation model, reinforced by TransFuser, outperforms baseline models across popular text evaluation metrics, including BLEU, CIDEr, and ROUGE.
}

\colorlet{shadecolor}{lightgray!50!} 
\begin{shaded}
\begin{changemargin}{0cm}{0.25cm} 
\small
This Chapter is based on:
\begin{itemize}
  \item \textit{``Non-Local Attention Improves Description Generation for Retinal Images''}, published in \textit{IEEE/CVF Winter Conference on Applications of Computer Vision}, 2022~\cite{huang2022non}, by \textbf{Jia-Hong Huang}, Ting-Wei Wu, Chao-Han Huck Yang, Zenglin Shi, I-Hung Lin, Jesper Tegner, and Marcel Worring.
  \item \textit{``Deep Context-Encoding Network For Retinal Image Captioning''}, published in \textit{IEEE International Conference on Image Processing}, 2021~\cite{huang2021deep,huang2021longer}, by \textbf{Jia-Hong Huang}, Ting-Wei Wu, Chao-Han Huck Yang, and Marcel Worring.
\end{itemize}

\end{changemargin}
\end{shaded}

\section{Introduction}

\begin{figure*}[t!]
\begin{center}
\includegraphics[width=1.0\linewidth]{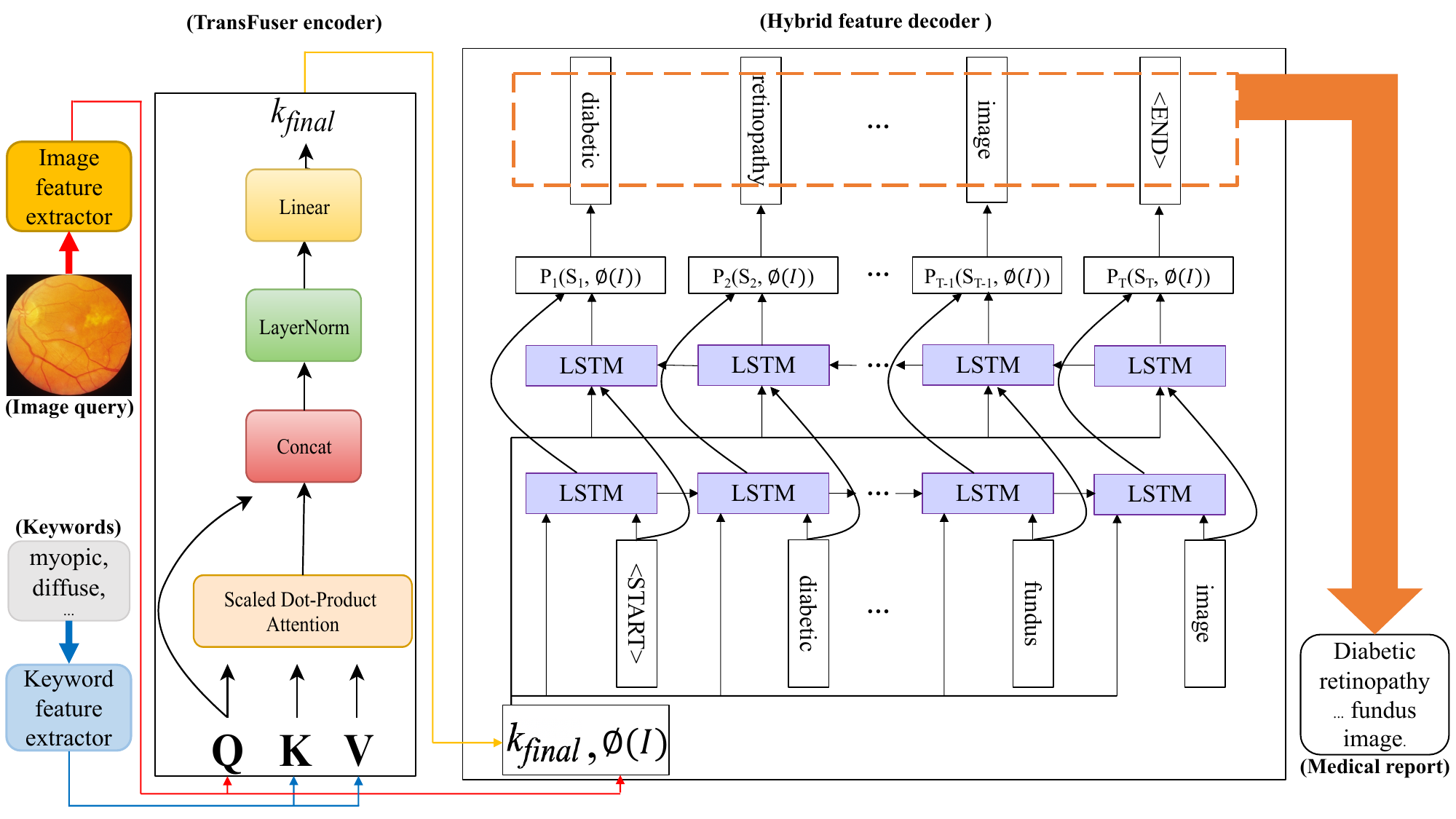}
\end{center}
   \caption{
   Flowchart of the proposed keyword-driven medical report generation model, enhanced by our proposed TransFuser architecture. The model takes two inputs: a retinal image and a set of keywords. The keywords serve to reinforce the model, enabling it to generate more accurate and meaningful descriptions for the retinal images. Within the TransFuser, the term ``Concat'' refers to concatenation, ``LayerNor'' signifies layer normalization, and ``Linear'' denotes a fully connected layer. The variable $k_{\textup{final}}$ represents an attention-based embedding vector. Here, $Q$ is the transformed image query, $K$ is composed of key vectors, $V$ consists of value vectors, $\phi(I)$ indicates the image feature vector, and $P_{i}(S_{i}, \phi(I))$ represents a probability distribution where $i=1,2,...,T$.
   }
\label{fig:figure_5_1}
\end{figure*}

Automatically generating medical reports for retinal images presents a complex challenge within the broader field of image captioning \cite{xu2015show}. This task involves algorithmically creating long, semantically coherent medical descriptions for a given image \cite{huang2021deepopht,huang2021deep,huang2021longer,huang2021contextualized}. Several technical aspects of retinal image report generation \cite{huang2021deepopht} complicate this task compared to the more extensively researched domain of natural image captioning \cite{cornia2019show,kim2019dense}. For instance, retinal images differ significantly from natural images in terms of object sizes and intricate details \cite{tierney2020comparative,huang2021deepopht}. Consequently, existing methods, such as those proposed in \cite{xu2015show} and \cite{karpathy2015deep}, which perform well on natural image datasets, often fail to generalize effectively to retinal images.

Recently, several methods \cite{jing2018automatic,li2018hybrid} have been developed for generating medical reports, primarily relying on traditional natural image captioning models that focus solely on image content. However, many abstract medical concepts or descriptions \cite{laserson2018textray,jing2018automatic} cannot be effectively derived from image information alone. To address this limitation, the authors of \cite{huang2021deepopht} introduced an average-based method that leverages expert-defined contextual information in the form of keyword sequences alongside image content to enhance description generation. While this keyword-driven average-based approach improves the medical report generation model, a significant challenge remains: effectively fusing the multi-modal information—specifically, the expert-defined keywords and the image. The average-based method may inadequately capture the mutual and interactive information between keywords and images \cite{huang2021deepopht}. This loss of interactive information can negatively impact the quality of the generated descriptions \cite{jing2018automatic,agrawal2017vqa,huang2020query,huang2019assessing,huang2019novel}. Therefore, to create more accurate and meaningful descriptions for retinal images, a specialized method capable of effectively integrating expert-defined keywords and image information is essential.

In Chapter \ref{ch:ch5}, we introduce a novel keyword-driven medical report generation method featuring a non-local attention-based keyword-image encoder, termed TransFuser, as depicted in Figures \ref{fig:figure_5_1}, \ref{fig:figure_5_2}, and \ref{fig:figure_5_3}. The TransFuser encoder integrates feature vectors from different modalities to facilitate the automatic generation of medical reports. Specifically, it encodes unordered keyword sequences alongside image content, applying varying attention weights to each keyword. This attention mechanism enables the inputs to guide the different modalities, resulting in more accurate outputs. Leveraging the non-local attention mechanism \cite{vaswani2017attention}, our proposed method effectively captures the mutual information between the image and keywords.

The authors of \cite{huang2021deepopht} introduced a state-of-the-art model along with a large-scale dataset featuring expert-defined keywords for medical report generation from retinal images. In Chapter \ref{ch:ch5}, we present the experimental results of our proposed model, which is evaluated using their dataset. Our keyword-driven generation model, enhanced by the TransFuser, demonstrates the ability to produce more accurate and meaningful descriptions for retinal images compared to baseline models. This improved performance is evidenced by significant gains in several text evaluation metrics: BLEU-avg (+32\%), CIDEr (+2.5\%), and ROUGE (+25.4\%).

\vspace{+3pt}
\noindent\textbf{Contributions.}
\begin{itemize}
    \item We introduce a new keyword-driven medical report generation method, employing a non-local attention-based keyword-image encoder, termed TransFuser. This encoder is designed to effectively integrate feature vectors from multiple modalities, specifically unordered keyword sequences and image content, to generate more accurate and semantically meaningful medical reports for retinal images.
    
    \item Our proposed method addresses the limitations of existing average-based approaches by capturing the mutual and interactive information between expert-defined keywords and image content. This improvement overcomes the challenges posed by traditional methods, which often fail to adequately fuse multi-modal information, thereby enhancing the quality of the generated descriptions.
    
    \item We demonstrate that our TransFuser-enhanced keyword-driven model achieves substantial improvements in text evaluation metrics when generating medical reports from retinal images. Specifically, the model shows significant gains in BLEU-avg, CIDEr, and ROUGE scores compared to baseline models, indicating its superior ability to produce accurate and meaningful descriptions.
    
\end{itemize}

\section{Related Work}
This section provides an overview of related image captioning methods for both natural and medical images, as well as a discussion of existing retinal image datasets.

\subsection{Caption Generation for Natural Images}

The encoder-decoder network architecture \cite{vinyals2015show,xiao2019deep,jun2020t,huang2017robustness,huang2017vqabq,huang2017robustness} is a widely used approach for image captioning. In this framework, CNNs act as the encoder, extracting global image features, while RNNs serve as the decoder, generating sequences of words. In \cite{mao2016generation}, the authors introduce a method for generating text descriptions of specific objects or regions, known as referring expressions \cite{kazemzadeh2014referitgame}. Similarly, \cite{wang2016image} presents a bidirectional LSTM-based approach that captures both past and future context to model long-term visual-language interactions.

Attention-based models have also shown strong performance in image captioning. For example, \cite{pedersoli2017areas} proposes an area-based attention model that predicts the next word along with its corresponding image regions at each RNN time step. However, most existing image captioning methods primarily rely on a single image for description generation, which can limit their ability to convey abstract concepts \cite{laserson2018textray,jing2018automatic}.

In Chapter \ref{ch:ch5}, we address this limitation by integrating expert-defined keyword sequences \cite{huang2021deepopht} into the model, enhancing its ability to generate more accurate and contextually rich descriptions.

\begin{table*}[t!]
    \caption{
    Summary of retinal datasets. The proposed DEN dataset stands out as both unique and significantly larger than other retinal datasets. As noted in \cite{huang2021deepopht}, most existing retinal datasets primarily consist of image data and are relatively small in size. `` Text*'' refers to clinical descriptions and keywords, while ``Text'' indicates clinical descriptions only.
    }
\begin{center}
\rotatebox{270}{
\scalebox{0.7}{
    \begin{tabular}{c|c|c|c|c}
    \toprule
    \textbf{Name of Dataset} & \textbf{Field of View} & \textbf{Resolution} & \textbf{Data Type} & \textbf{Number of Images}\\ \midrule
    STARE \cite{hoover2003locating} & $\approx 30^{\circ} \sim 45^{\circ}$ & $700 \times 605$ & Image + Text & 397\\ \midrule
    DIARETDB1 \cite{kauppi2007diaretdb1} & $50^{\circ}$ & $1500 \times 1152$ & Image + Text & 89\\ \midrule
    MESSIDOR \cite{decenciere2014feedback} & $45^{\circ}$ & $1440 \times 960 \sim 2304 \times 1536$ & Image + Text & 1,200\\ \midrule
    \textbf{DEN \cite{huang2021deepopht}} & $\approx \textbf{30}^{\circ} \sim \textbf{60}^{\circ}$ & \textbf{various} & \textbf{Image + Text*} & \textbf{15,709}\\ \bottomrule 
    
    \end{tabular}}
    }

    \label{table:table_5_1}
\end{center}
\end{table*}

\subsection{Caption Generation for Medical Images}

In \cite{li2018hybrid}, the authors introduce a Hybrid Retrieval-Generation Reinforced Agent for medical image captioning, which blends human prior knowledge with learning-based generation techniques. This agent employs a retrieval policy module to decide whether to generate sentences using a generation module or to retrieve specific sentences from a template database built on human expertise. The agent follows a hierarchical decision-making process to sequentially generate multiple sentences. Similarly, \cite{jing2018automatic} presents a multi-task learning framework that simultaneously predicts tags and generates captions, using an attention-based mechanism to localize regions with abnormalities and produce detailed descriptions through a hierarchical LSTM model. These approaches primarily focus on generating medical reports for chest radiology images.

However, chest radiology images and retinal images differ significantly in characteristics such as object size and detail \cite{laserson2018textray,tierney2020comparative,huang2021deepopht}. For instance, chest radiology images are generally grayscale \cite{laserson2018textray}, while retinal images are often colorful \cite{huang2021deepopht}. Most existing methods rely heavily on image inputs to generate captions.

In contrast, our proposed method is based on a CNN-RNN framework that incorporates a novel component called TransFuser to effectively integrate features from different modalities, enhancing our medical report generation model. Unlike standard sentences, keyword sequences have an unordered nature, presenting a unique challenge. Thus, combining image input with keywords while minimizing information loss remains an open question \cite{huang2021deepopht}.

\subsection{Retinal Datasets for Medical Report Generation}

Research on retinal diseases has a long history, leading to the development of various retinal datasets for computer vision tasks, such as those referenced in \cite{hoover2003locating,kauppi2007diaretdb1,decenciere2014feedback,huang2021deepopht,yang2018novel,yang2018auto,liu2019synthesizing}. The STARE dataset, introduced by the authors in \cite{hoover2003locating}, contains 397 retinal images and is primarily used for developing automatic systems to diagnose human eye diseases. In \cite{kauppi2007diaretdb1}, the authors present DIARETDB1, a dataset consisting of 89 color fundus images, of which 84 display at least mild non-proliferative signs of DR, while 5 images are classified as normal, showing no signs of DR. Additionally, the MESSIDOR dataset, proposed in \cite{decenciere2014feedback}, includes 1,200 fundus images, each accompanied by a corresponding text-based clinical description. However, it lacks manual annotations for lesion contours or locations.

Although the STARE, DIARETDB1, and MESSIDOR datasets contain retinal images and clinical descriptions, they do not feature expert-defined keywords, making them suitable for general medical report generation tasks but not specifically for keyword-driven report generation. Furthermore, other existing retinal datasets only include retinal images and are utilized solely for traditional computer vision tasks \cite{huang2021deepopht}. To validate the keyword-driven approach and train a deep medical report generation model, the authors of \cite{huang2021deepopht} have introduced the DEN dataset, which is significantly larger than the previously mentioned datasets, comprising 15,709 retinal images. Each image in the DEN dataset is paired with a corresponding expert-defined keyword sequence and text-based clinical description. Thus, DEN serves as an appropriate dataset for validating the effectiveness of the proposed model. A summary of these retinal datasets is presented in Table \ref{table:table_5_1}.

\section{Methodology}


In this section, we introduce our keyword-driven medical report generation model, enhanced by the TransFuser, as illustrated in Figure \ref{fig:figure_5_1}. We also detail the methods used to train the model with supervised keyword knowledge. First, an image and a set of keywords are input into modality-specific extractors to generate embedded vectors for both the image and the keywords. Following this information extraction, these vectors are fed into an encoder to obtain a final attention-based embedding vector, $k_{\textup{final}}$, which fuses information from both images and keywords, as described in Subsections \hyperref[kd:keyword_decoder]{5.3.1} and \hyperref[tf:transfuser]{5.3.2}. Next, we employ a bidirectional LSTM-based model as a decoder, referenced in Subsection \hyperref[hd:hybrid]{5.3.3}, to generate output words that form medical descriptions. This LSTM-based decoder utilizes the image vector extracted from the image feature extractor along with $k_{\textup{final}}$ and a decoder output token from the previous time step as inputs for generating the final sentence, as depicted in Figure \ref{fig:figure_5_1}.

\subsection{Keyword Encoder}
\label{kd:keyword_decoder}

In this subsection, we delve deeper into the role of keywords and their mechanisms in our proposed model for automatic medical report generation. Keywords are designed to represent significant image content while subtly conveying its semantic relationships. By treating an arbitrary number of keywords as a sequence, we enhance the model performance through a keyword encoder, denoted as $f(k_n, I)$, which takes $N$ keywords $k_n$ and an image $I$ as inputs, as outlined in Equation (\ref{eq:keyword}). We refer to this non-linear feature mapping process as \textit{``TransFuser,''} which functions as a hybrid approach for integrating image and keyword features. Further details on this mechanism will be provided in the following subsection.

\begin{figure}[t!]
\includegraphics[width=1.0\linewidth]{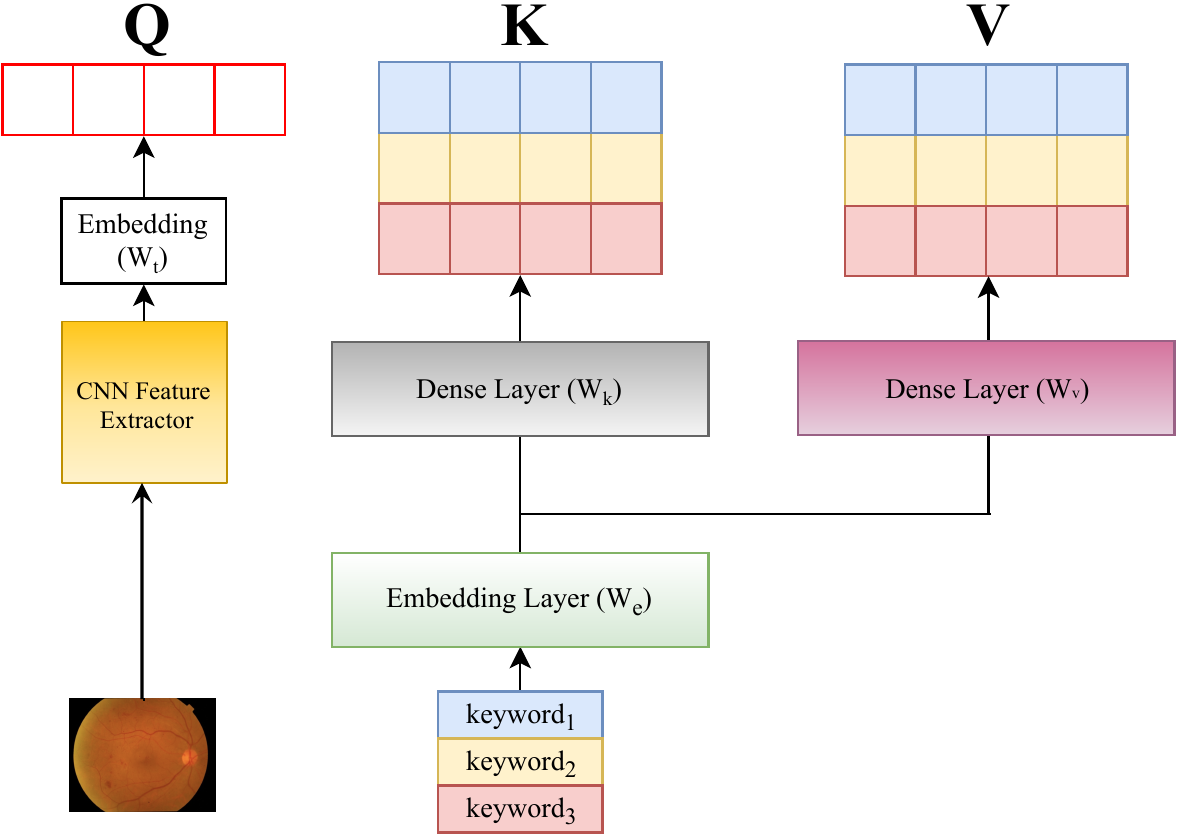}
\caption{Preprocessing of Query ($Q$), Key ($K$), and Value ($V$). Image contents are treated as the query, while the embedded keyword vectors are transformed into key and value vectors.}
\label{fig:figure_5_2}
\end{figure}

\begin{figure}[t!]
\includegraphics[width=1.0\linewidth]{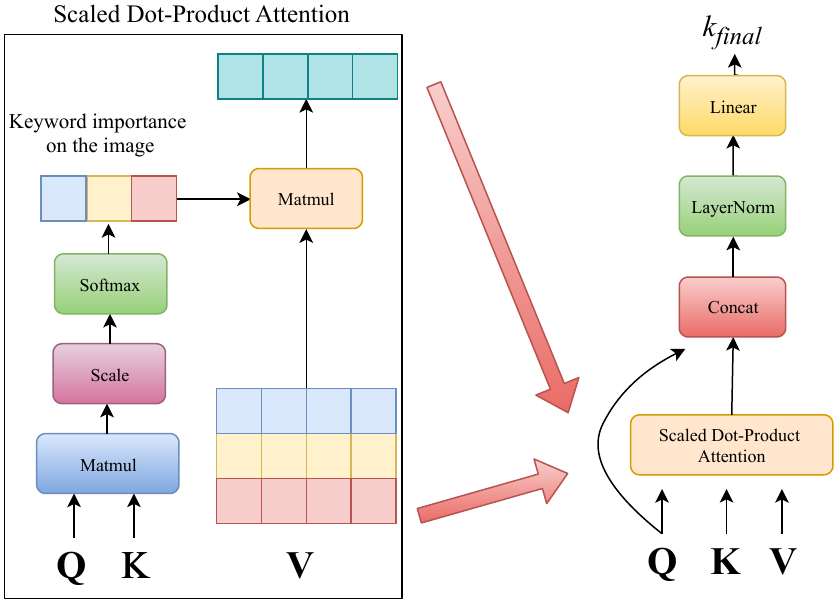}
\caption{Proposed TransFuser Workflow. Scaled dot-product attention is used to generate a final weighted embedded vector, highlighting the importance of keywords to the image vector. Ultimately, the final keyword vector is produced through a fully connected layer followed by layer normalization.}
\label{fig:figure_5_3}
\end{figure}

\begin{equation}
    k_{\textup{final}} = f(k_n, I), n \in \{0,...,N\}
	\label{eq:keyword}
\end{equation}


\subsection{TransFuser Encoder for Multi-modal Feature Fusion}
\label{tf:transfuser}

The Transformer structure \cite{vaswani2017attention,yang2021voice2series,huang2021gpt2mvs,disipio2021dawn,hu2019silco} has emerged as a leading approach for sequence modeling and transduction tasks. Its attention mechanism enables the modeling of global dependencies between input and output, effectively overcoming the memory limitations of traditional recurrent models. Inspired by this architecture and its capacity for parallelization in attention-weighted positions, we leverage its principles to integrate keyword sequences with image content, applying varying attention weights to each keyword. The scaled dot-product attention mechanism is employed to assess the importance of keywords relative to the embedded image vector. Rather than using the last decoder output as the query in an encoder-decoder attention cell, we utilize the image vector directly. Thus, the detailed formulation for $f(k_{n}, I)$, as illustrated in Equation (\ref{eq:keyword}), can be interpreted as mapping an image query $Q$, derived from image $I$, and a set of keyword key-value pairs $(K, V)$, derived from keywords $k_{n}$, to produce the output $k_{\textup{final}}$. We outline the complete procedure as follows.

\begin{equation}
    Q = W_t\times \phi(I)
	\label{eq:image_embed2}
\end{equation}
\begin{equation}
\begin{split}
    x_n = W_ek_n, n \in \{0,...,N\} \\
    K = W_k*x_n+b_k \\
    V = W_v*x_n+b_v 
	\label{eq:keyword_embed3}
\end{split}
\end{equation}
\begin{equation}
    Z = \textup{Attention}(Q,K,V) = \textup{softmax}(\frac{QK^T}{\sqrt{d_k}})V
	\label{eq:attention_6}
\end{equation}


First, we employ a CNN image embedder $\phi$ \cite{vinyals2015show,jing2018automatic,laserson2018textray,li2018hybrid,wang2018tienet,joshi2020robust,xu2015show,cornia2019show,karpathy2015deep} to extract features from the input image. We then map the image feature vector $\phi(I)$ using the embedding matrix $W_t \in \mathbb{R}^{T_H \times F}$, as indicated in Equation (\ref{eq:image_embed2}). Here, $F$ represents the size of the image feature, while $T_H$ denotes the hidden feature size within the TransFuser. The output $Q$ will serve as an image query for interaction with the keyword vectors later.

Next, we map the unordered sequence of keywords—regardless of their quantity—using the embedding matrix $W_e \in \mathbb{R}^{E \times V_k}$. In this context, $E$ refers to the word embedding size, and $V_k$ signifies the size of the vocabulary, which includes the keywords. We then utilize two linear layers $W_k, W_v \in \mathbb{R}^{T_H \times E}$ to generate the keyword key and value vectors, denoted as $K$ and $V$, respectively, as shown in Equation (\ref{eq:keyword_embed3}).

The output $Z$ is computed as a weighted sum of the value vectors $V$, with the weights determined by the importance of each keyword, calculated through dot-product attention between the image query $Q$ and the key $K$, as described in Equation (\ref{eq:attention_6}). We employ the dot-product mechanism to facilitate a faster and more space-efficient exploration of the relationship between keywords and images. Notably, we omit the positional encoding trick \cite{vaswani2017attention} to avoid introducing redundant sequential information, given the unordered nature of the keywords \cite{huang2021deepopht}.

\begin{equation}
    Z_{\textup{Norm}} = \textup{LayerNorm}(Q+Z)
	\label{eq:layernorm_6}
\end{equation}
\begin{equation}
    k_{\textup{final}} = \textup{max}(0,W_1Z_{\textup{Norm}}+b_1)W_2+b_2
	\label{eq:ffinal}
\end{equation}


Finally, we incorporate a residual shortcut with $Q$ to enhance the attention output $Z$. The resulting output denoted as $Z_{\textup{Norm}}$, is obtained after applying layer normalization, as described in Equation (\ref{eq:layernorm_6}). This normalized output is then fed into position-wise feed-forward networks, which are similarly connected after the attention sub-layer. We can now consistently utilize the final mixed vector, as shown in Equation (\ref{eq:ffinal}), and feed it back into our RNN model.

To gain a deeper understanding of the \textit{TransFuser} mechanism behind this embedding technique, we refer to Figures \ref{fig:figure_5_2} and \ref{fig:figure_5_3} for a detailed explanation. In the matrix multiplication $QK^T$, the image query vector in $Q$ interacts with each keyword-embedded vector in $K$. This interaction yields weights for each keyword to the image feature vector. After applying the scaled dot-product and softmax operations, we obtain probability-like weights that reflect each keyword's attention or relevance to the current image. Finally, these weights are multiplied by the corresponding value vector in $V$, representing their combined importance in providing attention-weighted image-keyword information.

\subsection{Hybrid Feature Decoder}
\label{hd:hybrid}

Once we have obtained the image-keyword hybrid vector $k_{\textup{final}}$, we can construct our complete model for generating image descriptions or reports. In each time step of the subsequent bidirectional LSTM decoder, we input $k_{\textup{final}}$ along with the image embedding vector $e_t$ and the preceding tokens, which are defined by $p(S_t|I, S_0,..., S_{t-1})$, where $S=(S_0,..., S_T)$ represents the true sentence describing the image. It is important to note that at each time step $t \in \{0,..., T\}$, we use the same image embedding vector $e_t$ and the hybrid vector $k_{\textup{final}}$ as inputs. We unroll the description generator as follows:

\begin{equation}
    e_t = W_d\times \phi(I), t \in \{0,...,T\}
	\label{eq:image_embed}
\end{equation}
\begin{equation}
    x_t = W_eS_t, t \in \{0,...,T\}
	\label{eq:caption_embed}
\end{equation}
\begin{equation}
    P_{t} = \textup{BiLSTM}([e_t, k_{\textup{final}}, x_t]), t \in \{0,...,T\}
    \label{eq:lstm}
\end{equation}
\begin{equation}
    L(P|I,S) = \mathbb{E}_{S\sim P_I}{[\textup{log}{P(S,I)]}}
    \label{eq:loss_6}
\end{equation}

In Equations (\ref{eq:image_embed}) and (\ref{eq:caption_embed}), we represent each word using a bag-of-words ID $S_t$. The words $S$ and the image vector $I$ are mapped to the same space: the image is processed through an image encoder $\phi$, which is a deep CNN followed by a fully connected layer $W_d \in \mathbb{R}^{E \times F}$, while the words are embedded using a word embedding matrix $W_e \in \mathbb{R}^{E \times V}$. Here, $E$ denotes the word embedding size, $F$ represents the number of image features, and $V$ indicates the vocabulary size. In Equation (\ref{eq:lstm}), at each time step, we feed the network with the image content $e_t$, the image-keyword hybrid vector $k_{\textup{final}}$, and the ground truth word vector $x_t$ to enhance its memory of the images. Additionally, we employ dropout to mitigate the effects of noise and overfitting. If we denote $P_I$ as the true medical descriptions for $I$ provided in the training set and $P(S, I)$ as the final probability distribution obtained from a fully connected layer followed by a softmax function, we can express the overall likelihood function $L(P|I, S)$ based on our medical descriptions and the given image, as shown in Equation (\ref{eq:loss_6}). Finally, we minimize the total loss, which is calculated as the sum of the negative log-likelihood at each time step.

For inference, we employ the beam search method \cite{huang2021deepopht} to generate sentences from an image. Unlike a greedy approach that selects the most likely next step at each stage of sequence construction, beam search explores all possible next steps and retains the $k$ most probable sentences. Here, $k$ is a user-defined parameter that controls the number of beams or parallel searches through the probability sequences. At each time step $t$, we consider the set of $k$ sentences generated up to that point as candidates for generating $P_{t+1}$. We then maintain the best $k$ sentences based on their cumulative probabilities. By examining multiple candidate sequences, beam search enhances the likelihood of finding a better match for the target sequence. However, this improved performance comes at the cost of reduced decoding speed.

\begin{table*}
\caption{
Evaluation results comparing keyword-driven and non-keyword-driven models for medical report generation. The highest scores in each column are highlighted, with ``w/o'' denoting non-keyword-driven baseline models and ``w/'' representing our proposed keyword-driven models. The ``BLEU-avg'' score reflects the average of BLEU-1, BLEU-2, BLEU-3, and BLEU-4 metrics. Among the non-keyword-driven models, \cite{xu2015show} achieves the top performance, while the keyword-driven model by \cite{cornia2019show} emerges as the best overall. Notably, all TransFuser-enhanced keyword-driven models consistently outperform their non-keyword-driven counterparts.
}
\centering
\rotatebox{270}{
\scalebox{0.7}{
\begin{tabular}{c|c|c|c|c|c|c|c|c}
\toprule
\multicolumn{2}{c|}{Model}                  & BLEU-1 & BLEU-2 & BLEU-3 & BLEU-4 & BLEU-avg & CIDEr & ROUGE \\ \midrule
\multirow{2}{*}{Vinyals et al. 2015 \cite{vinyals2015show}}              & w/o & 0.054  & 0.018  & 0.002  & 0.001  & 0.019    & 0.056 & 0.083 \\ \cline{2-9} 
                                    & w/ & \textbf{0.208}  & \textbf{0.124}  & \textbf{0.070}  & \textbf{0.032}  & \textbf{0.109}    & \textbf{0.319} & \textbf{0.254} \\ \midrule
\multirow{2}{*}{Jing et al. 2018 \cite{jing2018automatic}}              & w/o & 0.130  & 0.083  & 0.044  & 0.012  & 0.067    & 0.167 & 0.149 \\ \cline{2-9} 
                                    & w/ & \textbf{0.178}  & \textbf{0.107}  & \textbf{0.058}  & \textbf{0.023}  & \textbf{0.092}    & \textbf{0.330} & \textbf{0.215} \\ \midrule
\multirow{2}{*}{Laserson et al. 2018 \cite{laserson2018textray}}        & w/o & 0.105  & 0.049  & 0.009  & 0.002  & 0.041    & 0.064 & 0.127 \\ \cline{2-9} 
                                    & w/ & \textbf{0.148}  & \textbf{0.088}  & \textbf{0.050}  & \textbf{0.023}  & \textbf{0.077}    & \textbf{0.282} & \textbf{0.198} \\ \midrule
\multirow{2}{*}{Li et al. 2018 \cite{li2018hybrid}}     & w/o & 0.066  & 0.026  & 0.007  & 0.001  & 0.025    & 0.076 & 0.091  \\ \cline{2-9} 
                                    & w/ & \textbf{0.176}  & \textbf{0.106}  & \textbf{0.060}  & \textbf{0.029}  & \textbf{0.093}    & \textbf{0.285} & \textbf{0.229} \\ \midrule
\multirow{2}{*}{Wang et al. 2018 \cite{wang2018tienet}}           & w/o & 0.081  & 0.031  & 0.009  & 0.004  & 0.031    & 0.117 & 0.134  \\ \cline{2-9} 
                                    & w/ & \textbf{0.233}  & \textbf{0.152}   & \textbf{0.095}  & \textbf{0.052}  & \textbf{0.133}  & \textbf{0.369}   & \textbf{0.282}  \\ \midrule
\multirow{2}{*}{Joshi et al. 2020 \cite{joshi2020robust}}        & w/o & 0.111  & 0.060  & 0.026  & 0.006  & 0.051    & 0.066 & 0.129 \\ \cline{2-9} 
                                    & w/ & \textbf{0.166}  & \textbf{0.097}  & \textbf{0.049}  & \textbf{0.023}  & \textbf{0.084}    & \textbf{0.304} & \textbf{0.199} \\ \midrule
\multirow{2}{*}{Xu et al. 2015 \cite{xu2015show}} & w/o & 0.153 & 0.098 & 0.058 & 0.027 & 0.084 & 0.211 & 0.184 \\ \cline{2-9} 
                                    & w/ & \textbf{0.194}  & \textbf{0.122}  & \textbf{0.071}  & \textbf{0.033}  & \textbf{0.105}    & \textbf{0.340} & \textbf{0.238} \\ \midrule
\multirow{2}{*}{Cornia et al. 2019 \cite{cornia2019show}}          & w/o & 0.138  & 0.080  & 0.035  & 0.010  & 0.066    & 0.149 & 0.157  \\ \cline{2-9} 
                                    & w/ & \textbf{0.230}  & \textbf{0.150}   & \textbf{0.094}  & \textbf{0.053}  & \textbf{0.132}  & \textbf{0.370}  & \textbf{0.291}\\ \midrule
\multirow{2}{*}{Karpathy et al. 2015 \cite{karpathy2015deep}}        & w/o & 0.067  & 0.029  & 0.005  & 0.002  & 0.026    & 0.031 & 0.085 \\ \cline{2-9} 
                                    & w/ & \textbf{0.200}  & \textbf{0.126}  & \textbf{0.079}  & \textbf{0.041}  & \textbf{0.112}    & \textbf{0.296} & \textbf{0.244} \\ \bottomrule
\end{tabular}}
}
\label{table:table_5_2}
\end{table*}

\section{Experiments and Analysis}
In this section, we will evaluate our proposed keyword-driven medical report generation method using commonly employed metrics to determine its ability to produce more accurate and meaningful descriptions for retinal images. Additionally, we will assess the effectiveness of the proposed keyword-image encoder, known as TransFuser, based on the premise put forth by \cite{huang2021deepopht,jing2018automatic} that a robust deep model is beneficial in practice.

\subsection{Dataset and Performance Evaluation Metrics}

To advance research in retinal diseases, the authors \cite{huang2021deepopht} introduced the DEN dataset, a large-scale retinal image dataset featuring unique keyword annotations provided by experienced retina specialists. These keyword labels contain critical information regarding potential diseases and patient conditions derived from retinal image analysis and patient conversations, making them invaluable for ophthalmologists in writing medical reports. The DEN dataset comprises two types of images: grayscale FA and colorful CFP, totaling 15,709 images—1,811 FA and 13,898 CFP. We adhere to the dataset's established split of 60\% for training, 20\% for validation, and 20\% for testing. Each retinal image in the DEN dataset is paired with two labels: keywords and clinical descriptions, with most keyword sequences ranging from 5 to 10 words. For our experiments, we use the image and keyword labels as inputs, while the clinical descriptions serve as the ground truth for predictions. To evaluate the generated descriptions for retinal images, we employ commonly used text evaluation metrics from the medical report generation field, as referenced in \cite{papineni2002bleu,lin2004rouge,vedantam2015cider,huang2021deepopht,li2019knowledge}.

\begin{figure*}[ht]
\begin{center}
\includegraphics[width=1.0\linewidth]{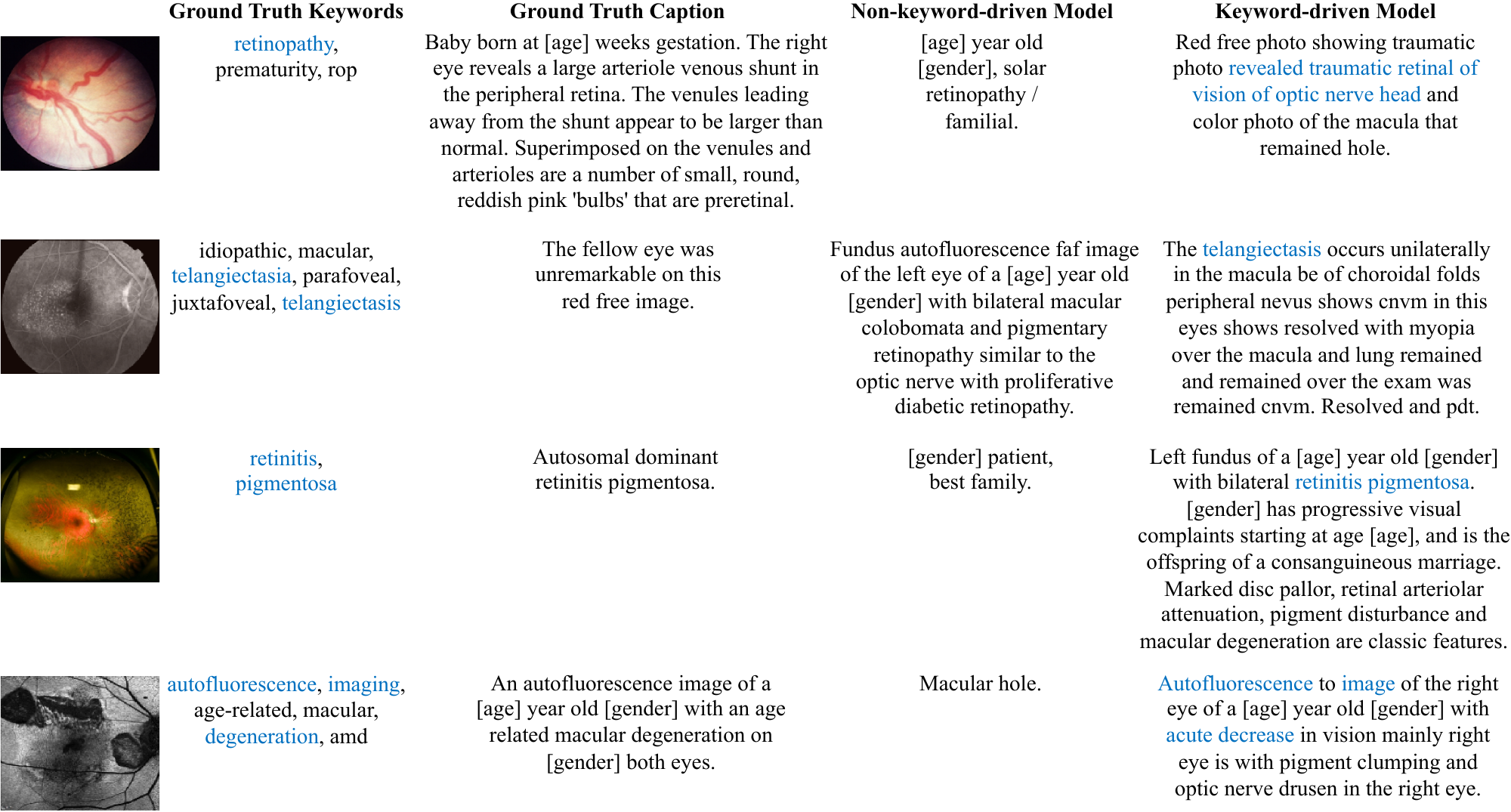}
\end{center}
   \caption{
   Generated medical reports from both keyword-driven and non-keyword-driven models. The results indicate that the keyword-driven models produce more accurate descriptions of key characteristics in retinal images. The keywords highlighted in blue illustrate the model's understanding of these crucial elements. For further details and explanations regarding the terms [age] and [gender], please refer to Subsection \hyperref[qual:quality.5.4.4]{5.4.4} and Section \hyperref[diss:discussion.5.5]{5.5}.
   }
\label{fig:figure_5_5}
\end{figure*}

\subsection{Experimental Settings}

We utilize image feature extractors $\phi$, pre-trained on ImageNet, to extract features from our proposed retinal image dataset. Most retinal images, including those in our dataset, are colorful, similar to the images in ImageNet. Thus, pre-training on ImageNet can enhance model performance from a low-level feature perspective, such as color. Initially, we resize each image to the appropriate dimensions for input into the model. We then use the layer just before the final fully connected layer to extract the embedding features for input into the main LSTM model.

To process the annotations and keywords in the dataset, we remove non-alphabetic characters, convert all remaining characters to lowercase, and replace words that appear only once with a special token, $<\textup{UNK}>$. As a result, our vocabulary size is $V=4,007$, and the vocabulary size including keywords is $V_k=4,292$. All sentences are truncated or padded to a maximum length of 50. For the word embedding layer, we use an embedding size of $E=300$ to encode the words, with a hidden layer size of $H_{\textup{LSTM}}=256$.

During training, we first input the image feature sets extracted from $\phi$ alongside the word-embedded vectors into the LSTM. Subsequently, we incorporate the keywords fused from our embedding model. In our TransFuser model, we set the hidden size for representation learning to $T_H=64$. For all models, we configure a mini-batch size of 64 and a learning rate of $1e-3$, training the model for two epochs.

\begin{table*}[t!]
\caption{
The results indicate that our proposed TransFuser outperforms the baseline models in the ``Image + Keywords'' scenario. Here, ``mul'' refers to element-wise multiplication, while ``sum'' denotes summation. The results are derived from the top-performing keyword-driven model \cite{cornia2019show} presented in Table \ref{table:table_5_2}.
}
\centering
\rotatebox{270}{
\scalebox{0.7}{
\begin{tabular}{c|c|c|c|c|c|c|c|c}
\toprule
\multicolumn{2}{c|}{Fusing method}     & BLEU-1 & BLEU-2 & BLEU-3 & BLEU-4 & BLEU-avg & CIDEr & ROUGE \\ \midrule
\multicolumn{2}{c|}{Baseline-1 (sum)}  & 0.014  & 0.002  & 0.001  & 0.000  & 0.004    & 0.019 & 0.023 \\ \midrule
\multicolumn{2}{c|}{Baseline-2 (mul)}  & 0.077  & 0.031  & 0.004  & 0.001  & 0.028    & 0.042 & 0.102  \\ \midrule
\multicolumn{2}{c|}{DeepOpht \cite{huang2021deepopht}}  & 0.184  & 0.114  & 0.068  & 0.032  & 0.100    & 0.361 & 0.232  \\ \midrule
\multicolumn{2}{c|}{Ours (TransFuser)} & \textbf{0.230}  & \textbf{0.150}   & \textbf{0.094}  & \textbf{0.053}  & \textbf{0.132}  & \textbf{0.370}  & \textbf{0.291} \\ \bottomrule
\end{tabular}}
}
\label{table:table_5_3}
\end{table*}

\begin{table*}[t!]
\caption{
The results show that the proposed TransFuser effectively captures not only the original information from both keywords and images but also the interactive information between them. These results are derived from the best-performing keyword-driven model \cite{cornia2019show}, as detailed in Table \ref{table:table_5_2}.
}
\centering
\rotatebox{270}{
\scalebox{0.7}{
\begin{tabular}{c|c|c|c|c|c|c|c|c}
\toprule
\multicolumn{2}{c|}{Input}            & BLEU-1 & BLEU-2 & BLEU-3 & BLEU-4 & BLEU-avg & CIDEr & ROUGE \\ \midrule
\multicolumn{2}{c|}{Keywords}    & 0.057  & 0.029  & 0.017  & 0.005  & 0.027    & 0.168 & 0.091  \\ \midrule
\multicolumn{2}{c|}{Image}       & 0.153  & 0.098  & 0.058  & 0.027  & 0.084    & 0.211 & 0.184 \\ \midrule
\multicolumn{2}{c|}{Image + Keywords} & \textbf{0.230}  & \textbf{0.150}   & \textbf{0.094}  & \textbf{0.053}  & \textbf{0.132}  & \textbf{0.370}  & \textbf{0.291} \\ \bottomrule
\end{tabular}}
}
\label{table:table_5_4}
\end{table*}

\subsection{Effectiveness Analysis}

\noindent\textbf{Keywords.}
Given the distinct characteristics of medical images compared to general images, along with the varying capabilities of different CNN models to capture these characteristics, we explore various CNN architectures with and without keywords to assess the effectiveness of our keyword-driven approach. In experiments, we utilize two types of models: keyword-driven and non-keyword-driven. According to Table \ref{table:table_5_2}, the model by \cite{xu2015show} exhibits the best performance among non-keyword-driven models, while the keyword-driven model by \cite{cornia2019show} ranks as the top performer overall. Notably, all keyword-driven models outperform their non-keyword-driven counterparts. 

Furthermore, our findings reveal that different CNN architectures have varying abilities to capture the characteristics of retinal images. Specifically, the best-performing keyword-driven model shows improvements of approximately 58\% in BLEU-avg, 75\% in CIDEr, and 58\% in ROUGE compared to the best non-keyword-driven model. This enhanced performance can be attributed to the fact that keywords represent critical content within the image while also providing semantic context. Therefore, in this context, keywords serve as additional information for the models. Our experimental results demonstrate that the proposed keyword-driven method surpasses the non-keyword-driven approach across commonly used evaluation metrics, as detailed in Table \ref{table:table_5_2}.

\noindent\textbf{TransFuser.}
Given the unordered nature of our keywords, traditional methods for fusing keywords and image features, such as element-wise multiplication and summation, may not be optimal. The authors of \cite{huang2021deepopht} introduced an averaging method called DeepOpht for this purpose. As shown in Table \ref{table:table_5_3}, our proposed TransFuser outperforms both the summation and element-wise multiplication baselines, as well as DeepOpht.

We observe that both TransFuser and DeepOpht significantly outperform the baselines. This improvement can be attributed to the fact that summation and element-wise multiplication may inadvertently capture the input order of the keywords, leading to potential biases due to the unordered nature of the keyword sequence. In contrast, TransFuser and DeepOpht are less susceptible to this bias. Furthermore, TransFuser surpasses DeepOpht in performance, showing increases of approximately 32\% in BLEU-avg, 2.5\% in CIDEr, and 25.4\% in ROUGE, as noted in Table \ref{table:table_5_3}. This enhancement results from TransFuser's ability to better capture the mutual and interactive information between keywords and images. More details can be found in the following subsection.

\noindent\textbf{Interaction between keywords and image.}
Table \ref{table:table_5_4} shows that the performance of the ``Image-only'' and ``Keywords-only'' baselines is inferior to that of the ``Image + Keywords'' method. This indicates that the interaction between keywords and images is essential for effective medical report generation. Our proposed TransFuser effectively captures this interaction, highlighting the relationship between keywords and images. It is worth noting that these results are derived from the best model, \cite{cornia2019show}, as presented in Table \ref{table:table_5_2}.

\subsection{Qualitative Results and Analysis}\label{qual:quality.5.4.4}

Figure \ref{fig:figure_5_5} showcases some qualitative results generated by our medical report generation model. While our models do not produce accurate ``age'' or ``gender'' information—since these attributes are not included in the content—they effectively generate accurate descriptions of the key characteristics of retinal images. Ideally, ``age'' and ``gender'' should be included in the dataset, and a comprehensive system could incorporate these details into the descriptions through post-processing or slot-filling techniques \cite{huang2021deepopht}.

\subsection{Evaluation with Retinal Specialists}

We conducted a 5-level evaluation of report and description quality, with scores ranging from 1 to 5, where a higher score indicates better quality. Due to limited resources, we randomly selected 100 samples of reports generated by our model and their corresponding ground-truth reports, which were created by ophthalmologists. Five different retinal specialists, unaware of whether the reports were model-generated or expert-generated, assessed the quality of both sets of reports. Our model-generated reports received an average score of 3.6 out of 5.0, while the ground-truth reports achieved an average score of 4.5 out of 5.0. Since the ground-truth reports are defined by ophthalmologists, these results demonstrate that our proposed model performs competitively compared to the human expert baseline.

\section{Discussion}\label{diss:discussion.5.5}
\subsection{Reasoning Ability}
Figure \ref{fig:figure_5_5} illustrates that the non-keyword-driven model often struggles to generate comprehensive and accurate conceptual descriptions for retinal images. In contrast, the proposed keyword-driven model demonstrates superior reasoning capabilities, producing longer and more conceptually accurate medical reports for retinal images.

\subsection{Does our model fully understand input keywords?}
The answer is likely no. However, as shown in Figure \ref{fig:figure_5_5}, our proposed keyword-driven model demonstrates some ability to understand input keywords. For instance, in the fourth example, the model generates the phrase ``acute decrease'' based on its comprehension of the keyword ``degeneration.'' Similarly, the first example also reflects the model's keyword understanding capability. This suggests that our approach is moving us closer to achieving fully automatic medical report generation for retinal images.

\subsection{Do features from deeper models perform well in our task?}
Our experimental results in Table \ref{table:table_5_2} and findings from \cite{huang2019a,huang2021deepopht} indicate that image features extracted from deeper networks do not necessarily lead to better performance in tasks involving multi-modal inputs. While deeper networks excel in many pure computer vision tasks, such as object detection and activity recognition, they may not be as effective for description generation within an LSTM unit. We hypothesize that additional transformations of the image features are required, and relying solely on these advanced features may hinder the performance of medical report generation.

\section{Conclusion}
In summary, we present a novel keyword-driven method for automatic medical report generation specifically for retinal images. Our approach incorporates a non-local attention-based mechanism called TransFuser, which effectively fuses features across different modalities. Experimental results demonstrate that our model generates more accurate and meaningful descriptions for retinal images, with performance improvements of approximately 32\% in BLEU-avg, 2.5\% in CIDEr, and 25.4\% in ROUGE. Additionally, our findings indicate that the TransFuser-enhanced keyword-driven method outperforms non-keyword-driven alternatives.

\chapter{Expert-defined Keywords Improve Captioning Explainability}\label{ch:ch6}

\textit{
Automatic ML systems for generating medical reports from retinal images often face challenges related to interpretability, which has impeded their broader adoption. Interpretability is closely tied to trust, as users are naturally skeptical and require clear reasoning to trust such systems. However, a universally accepted technical definition of interpretability remains elusive, making it difficult to design ML-based medical report generation systems that are fully understandable to human users.
Post-hoc explanation techniques like heat maps and saliency maps are commonly employed to improve the interpretability of ML-based medical systems, but they have significant limitations. From an ML model’s perspective, highlighted regions in an image are considered important for predictions. However, from a physician’s standpoint, even the most prominent areas on a heat map may contain both relevant and irrelevant information. Simply marking a region does not clarify what specific aspects within that area influenced the model’s decision, leading to explanations that often rely on subjective human interpretation, which can be biased.
To overcome these challenges, in Chapter \ref{ch:ch6}, we introduce a novel approach that uses expert-defined keywords as interpretability enhancers. These keywords effectively encapsulate expert domain knowledge and are easily understood by humans. Our method integrates these keywords with a specialized attention-based strategy, resulting in a more interpretable medical report generation system for retinal images. This approach not only significantly improves interpretability but also achieves state-of-the-art performance on widely used text evaluation metrics.
}

\colorlet{shadecolor}{lightgray!50!} 
\begin{shaded}
\begin{changemargin}{0cm}{0.25cm} 
\small
This Chapter is based on:
\begin{itemize}
  \item \textit{``Expert-Defined Keywords Improve Interpretability of Retinal Image Captioning''}, published in \textit{IEEE/CVF Winter Conference on Applications of Computer Vision}, 2023~\cite{wu2023expert}, by Ting-Wei Wu$^*$, \textbf{Jia-Hong Huang}$^*$, Joseph Lin, and Marcel Worring.
\end{itemize}
$^*$ indicates equal contribution.

\end{changemargin}
\end{shaded}

\begin{figure}[t!]
\begin{center}
\includegraphics[width=1.0\linewidth]{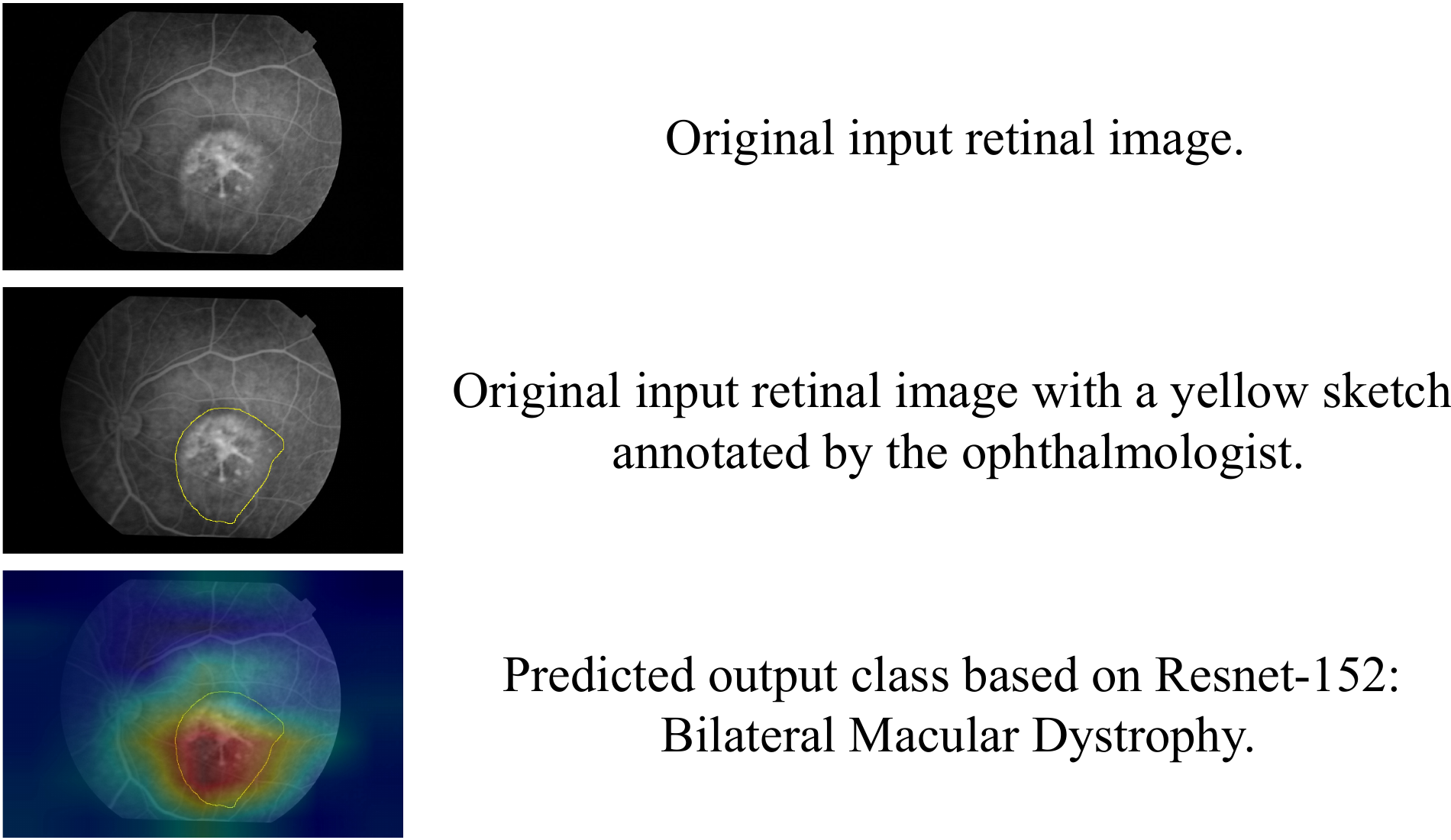}
\end{center}
  \caption{
  A heat map generated using CAM \cite{zhou2016learning}. This serves as a post-hoc explanation method. In the heat map, brighter colors (red) represent areas identified by the DNN as having higher importance, while darker colors (blue) indicate regions of lower significance. The model predicts the retinal disease as ``Bilateral Macular Dystrophy.''
}
\label{fig:figure_6_1}
\end{figure}

\section{Introduction}

Automatic ML-based medical systems, such as those for generating medical reports for retinal images, have not yet gained widespread acceptance \cite{ghassemi2021false,burkart2021survey}. A primary barrier to their adoption is the relative lack of explainability and interpretability in these systems, making it difficult for users to understand or receive explanations for machine-generated decisions.
As noted in \cite{ghassemi2021false}, several high-level definitions of interpretability have been proposed by researchers \cite{doshi2017towards,lipton2018mythos}. For instance, the authors of \cite{burkart2021survey} define interpretability, particularly in terms of attribute importance, as conveying a sense of causality to the target audience of the system. This concept of causality can only be understood when the system elucidates the underlying input-output relationships. However, there remains a lack of consensus on precise technical definitions of interpretability, making it challenging to develop a human-comprehensible ML-based medical report generation system.

To enhance interpretability, methods based on heat maps and saliency maps \cite{zhou2016learning,selvaraju2017grad,lundberg2017unified} are frequently employed to indicate how much each region of a medical image contributes to the decisions made by ML-based medical systems. However, these methods have been widely criticized in the interpretability literature \cite{adebayo2018sanity,ghassemi2021false}. For instance, consider the results derived from CAM \cite{zhou2016learning} shown in Figure \ref{fig:figure_6_1}. From the perspective of an ML model, the highlighted areas in the retinal image are deemed most critical for diagnosing or classifying retinal diseases. Conversely, ophthalmologists recognize that even the most prominent regions (marked in red) may contain both relevant and irrelevant information. As such, merely identifying these areas does not clarify what specific elements within them the model considered significant.
This raises questions about whether the ML model correctly identified the presence of the macular pattern as essential for its decision, whether it relied on the vessels, or whether it was influenced by non-clinical features, such as specific textures or pixel values that could relate more to the image acquisition process than to the underlying retinal condition. This explainability gap in widely used interpretability methods \cite{zhou2016learning,selvaraju2017grad} places the burden on humans to interpret the significance of heat maps, yet human biases can lead to overly optimistic interpretations \cite{bornstein2016artificial,ghassemi2021false}. The same issue arises in ML-based medical report generation models.

Textual data, such as sequences of expert-defined keywords, are easily understood by humans and serve as effective carriers of specialized domain knowledge. As noted in \cite{huang2021deepopht}, ophthalmologists typically compile a small set of keywords that capture essential information during the early stages of diagnosis. These keywords can be collected with relative ease \cite{huang2021deepopht,huang2021deep,huang2021contextualized,huang2022non,huck2019auto,liu2019synthesizing,yang2018novel}.
In Chapter \ref{ch:ch6}, we propose leveraging these interpretability boosters—specifically, expert-defined keywords—along with a specialized attention-based strategy to create a more comprehensible medical report generation system for retinal images. By utilizing these human-readable keywords, which embody the domain knowledge of ophthalmologists, we train our ML-based model to produce more explainable outputs. Our proposed attention-based strategy focuses on identifying the salient combinations of local features that align with keywords in a specific modality; further details can be found in Section \hyperref[methodology:method]{6.3}. Together, the expert-defined keywords and our attention-based approach significantly enhance interpretability.

\vspace{+3pt}
\noindent\textbf{Contributions.}
\begin{itemize}
	\item We propose a more explainable retinal image captioning model that utilizes interpretability boosters, specifically expert-defined keywords.
	
	\item We introduce a novel attention-based strategy within the transformer decoder that aligns human-comprehensible keywords with local image patches. This strategy significantly enhances the interpretability of our proposed method.

	\item Extensive experiments conducted on the widely used dataset reveal that integrating the context-aware transformer decoder results in performance improvements across all commonly used metrics compared to the baselines. This indicates that the semantic-grounded image representations are effective and capable of generalizing across a wide range of models.
 
\end{itemize}

\section{Related Work}
This section presents an overview of related work, focusing on methods for enhancing interpretability in image captioning, along with approaches specific to both natural and retinal image captioning.
\subsection{Current Methods for Improving Interpretability}

Efforts to provide human-comprehensible explanations for the decisions made by ML models typically fall into two categories: inherent interpretability and post-hoc interpretability \cite{ghassemi2021false}. Inherent interpretability is often found in simpler ML models that directly relate input data to output decisions; for instance, a linear regression model uses coefficients to indicate the strength and direction of relationships. However, modern AI applications often employ complex models that cannot be easily explained through straightforward input-output relationships.

In these cases, researchers focus on dissecting the decision-making processes of ML models, which falls under post-hoc interpretability \cite{zhou2016learning,selvaraju2017grad,ribeiro2016should,slack2020fooling,huang2022causal}. For example, \cite{zhou2016learning} introduces CAM, which utilizes the global average pooling layer \cite{lin2013network} to create a localizable representation that reveals the implicit attention of CNNs to specific image areas. Similarly, \cite{selvaraju2017grad} develops Grad-CAM to leverage gradients flowing into the final convolutional layer, generating a coarse localization map that highlights important regions in an image.

Heat maps have become popular tools in medical imaging, providing a straightforward way to illustrate some limitations of post-hoc interpretability techniques \cite{zhou2016learning,selvaraju2017grad,ghassemi2021false}. However, they are also known to present challenges within the broader interpretability literature \cite{adebayo2018sanity}. This issue extends to other well-known post-hoc explanation methods, such as locally interpretable model-agnostic explanations (LIME) \cite{ribeiro2016should} and Shapley values (SHAP) \cite{slack2020fooling}.

In Chapter \ref{ch:ch6}, we utilize expert-defined keywords to bridge the interpretability gap associated with heat map-based explanations. Keywords not only encapsulate essential image content but also highlight its semantic relationships, serving as effective carriers of expert domain knowledge.

\subsection{Natural Image Captioning}

The encoder-decoder paradigm is a widely used architecture for image captioning \cite{vinyals2015show,karpathy2015deep}, yielding promising results. This approach begins with a CNN to encode the image, followed by an RNN that generates the output word sequence. For instance, \cite{wang2016image} proposes a bidirectional LSTM-based method that leverages both past and future information to capture long-term interactions between visual and linguistic elements.

In another approach, \cite{pedersoli2017areas} introduces an area-based attention model that predicts the next word along with the corresponding regions of the image at each RNN time step, facilitating more accurate image captions. The work by \cite{yao2018exploring} utilizes graph convolutional networks (GCNs) \cite{scarselli2008graph} in combination with LSTM networks \cite{hochreiter1997long} to construct an encoder-decoder architecture for image captioning, where graphs are formed based on the spatial and semantic relationships among detected objects in an image.

In addition, \cite{huang2019attention} presents an attention-on-attention module that assesses the relevance between attention outputs and queries, integrating conventional attention mechanisms within both the encoder and decoder of the captioning model. Similarly, \cite{pan2020x} introduces a unified attention block that employs bilinear pooling to selectively utilize visual information, enhancing the interaction of multi-modal features through attention blocks integrated into the image encoder and sentence decoder.

While these methods primarily focus on natural images, they often generate basic image descriptions. However, retinal images differ significantly from natural images in terms of object sizes and details \cite{huang2021deepopht}. Consequently, applying these natural image-based approaches directly to retinal image captioning still requires improvement in the quality of generated medical descriptions.

\subsection{Retinal Image Captioning}

Generating medical descriptions for retinal images, known as retinal image captioning, poses a significant challenge in computer vision. This task requires the algorithmic generation of long and semantically coherent medical descriptions for each retinal image \cite{vellakani2020enhanced,mishra2020automatic,huang2021deepopht,huang2021deep,huang2021contextualized}.

In \cite{vellakani2020enhanced}, the authors present a ML-based clinical decision support system aimed at assisting ophthalmologists more effectively, primarily utilizing an LSTM-based image captioning model. Similarly, \cite{mishra2020automatic} introduces an automatic medical description generation model that combines CNNs with self-trained bidirectional LSTMs. The work in \cite{huang2021deepopht} proposes an AI-driven approach to enhance traditional retinal disease treatment processes, integrating a RDI, a CDG, and a CAM-based deep network visual explanation module. Additionally, the authors introduce a large-scale retinal image captioning dataset, DEN, to train and validate their method.

Furthermore, \cite{huang2021deep} proposes a context-driven encoding network designed to produce more accurate and meaningful medical reports for retinal images, incorporating a multi-modal input encoder and a fused feature decoder. In \cite{huang2021contextualized}, an end-to-end transformer-based model is developed for retinal image description generation, utilizing non-local attention mechanisms, feature reinforcement modules, and masked self-attention.

Despite the advancements in ML-based models for retinal image captioning, none provide clear interpretability. To create a more human-comprehensible image captioning system, we adopt an encoder-decoder framework and enhance interpretability through the integration of expert-defined keywords and a specialized attention-based strategy, as detailed in Section \hyperref[methodology:method]{6.3}.

\begin{figure*}[t!]
\begin{center}
\scalebox{1.0}{
\includegraphics[width=1.0 \linewidth]{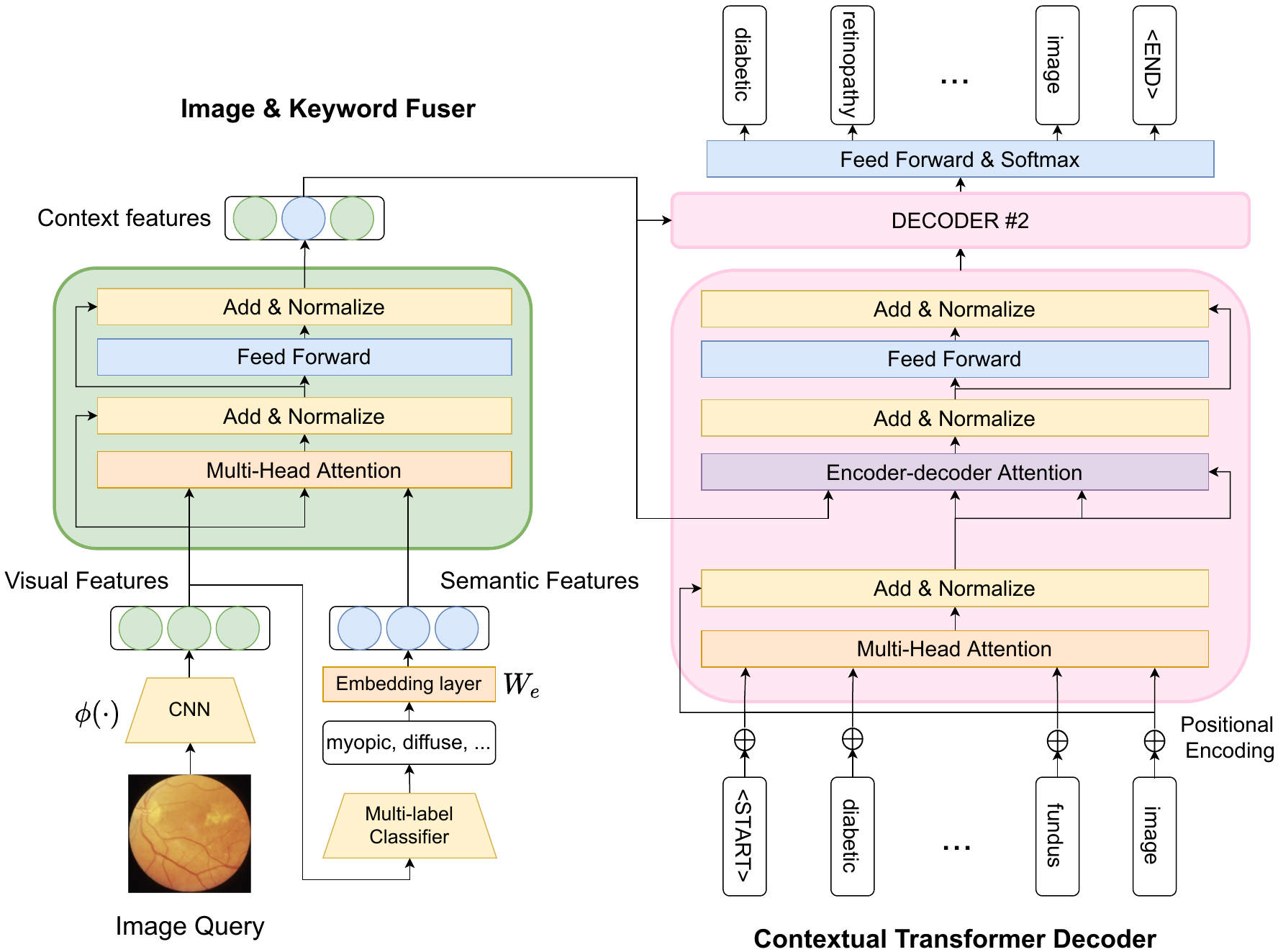}}
\end{center}
  \caption{
  Flowchart of the proposed method. Our method consists of an image/keyword fuser and a contextual transformer decoder for generating medical descriptions. Visual features are extracted from a CNN, while semantic features are obtained from a multi-label classifier. These two sets of features are then fused within a transformer block to assess the importance of each image patch. The resulting contextual features serve as the encoder output for the transformer decoder, enabling the generation of appropriate medical descriptions. Note that ``Image \& Keyword'' refers to ``Image \& Interpretability Booster.''
}
\label{fig:figure_6_2}
\end{figure*}


\section{Methodology}
\label{methodology:method}
In this section, we present the proposed explainable retinal image captioning model, as illustrated in Figure \ref{fig:figure_6_2}. The model is driven by interpretability boosters, specifically expert-defined keywords. It generates a long and semantically coherent medical description based on a given retinal image and its corresponding keywords. In Section \hyperref[methodology:method-3-1]{6.3.1}, we also consider a more general scenario in which expert-defined keywords are not provided as input. In this case, a CNN extracts visual features from the image patches, which are then processed by a multi-label classifier to predict relevant keywords. These predicted keywords are treated as ``pseudo'' expert-defined keywords, and their embedding vectors serve as the semantic features for the retinal images. Following this information extraction, both visual and semantic features are fed into a contextual transformer decoder, which sequentially generates output words for the medical descriptions. The contextual transformer decoder is similar to the standard transformer decoder \cite{vaswani2017attention}, but it incorporates an image-keyword attention-based encoder to fuse information from both images and keywords.

\subsection{Interpretability Booster Prediction}
\label{methodology:method-3-1}

According to \cite{huang2021deepopht,huang2021deep,huang2021contextualized}, ophthalmologists typically record a small set of keywords that denote important information early in the diagnosis process. As a result, expert-defined keywords are often available in this context. However, such keywords may not be as readily available in other fields, such as biology, chemistry, or physics. To address this, we not only utilize ground truth expert-defined keywords for report generation but also introduce a multi-label classifier that predicts these keywords based on the given image. It is important to note that the accuracy of keyword prediction significantly impacts model performance, as discussed in Section \hyperref[quantitative-analysis:quantitative-analysis]{6.5.1}. Given an image $I$, we extract its features $\mathbf{v} \in \mathbb{R}^{N \times H_I}$ using a CNN extractor $\phi(\cdot)$ \cite{he2016deep}, which are then input into a multi-layer perceptron (MLP) classifier to predict one or more keywords from a vocabulary of size $L$.

\begin{equation}
    p(\mathbf{l}_i = 1|\mathbf{v}) \propto e^{(W^{MLP}_{i} (\mathbf{v}))},
\end{equation}
where $\mathbf{l} \in \mathbb{R}^L$ represents the keyword vector, with $\mathbf{l}_i$ indicating the presence or absence of the $i^{\textup{th}}$ keyword. $W^{\textup{MLP}}_{i}$ denotes the weight of the MLP classifier corresponding to the $i^{\textup{th}}$ output. We select keywords for the decoding process based on the condition $p(\mathbf{l}_i = 1|\mathbf{v}) > \tau$, where $\tau$ is the confidence threshold.

\subsection{Image and Interpretability Booster Fusion}

After generating the corresponding keywords, we embed the keyword sequences with image content to capture the interactions between keywords and the image, applying different attention weights to each keyword using a self-attention mechanism. Specifically, for a given set of keywords $\{k_i\}_{i=1}^K$, where $K$ is the total number of keywords, we first preprocess them by inserting a special token ``[\textup{SEP}]'' between each keyword to form a complete sequence. We then use a GloVe embedding layer $W_e$ to obtain the keyword embedded vector $\mathbf{k} \in \mathbb{R}^{K \times H_e}$, with $H_e$ representing the embedding size.

Next, we introduce an attention feature mapping function $f(\mathbf{v}, \mathbf{k})$, as outlined in Equation (\ref{eq:keyword_7}). This function can be understood as mapping an image query $Q$ from image $I$ and a set of keyword key-value pairs $K, V$ from the keywords $\mathbf{k}$ to an output $Z$. To enhance efficiency, we leverage a dot-product mechanism, which provides a faster and more space-efficient way to explore the relationships between keywords and the image. Additionally, we omit the positional encoding trick, as we want to avoid introducing redundant sequential information given the unordered nature of the keywords.

\begin{align}
    Q &= W_q \phi(I)+b_q \\
    K &= W_k \mathbf{k}+b_k \\
    V &= W_v \mathbf{k}+b_v \\
    \notag 
    Z &= \textup{Attention}(Q,K,V) \\ 
      &= \textup{softmax}(\frac{QK^T}{\sqrt{d_k}})V
    \label{eq:keyword_7}
\end{align}
We also incorporate a residual connection, followed by layer normalization and position-wise feed-forward layers, to further enhance the model's performance.

\begin{equation}
    Z_{\textup{Norm}} = \textup{LayerNorm}(Q+Z)
	\label{eq:layernorm_7}
\end{equation}
\begin{equation}
    k_{\textup{final}} = \textup{max}(0,W_1Z_{\textup{Norm}}+b_1)W_2+b_2
	\label{eq:ffinal_7}
\end{equation}
In the matrix multiplication $QK^T$, the image query $Q$ is multiplied with each keyword embedded vector, represented as keys $K$. This process yields weights for each keyword concerning the image vector. After applying a scaling factor and the softmax function, we obtain probability-like weights for each keyword, reflecting their attention or relevance to the current image. Finally, these weights are multiplied by their corresponding values $V$ to denote the hybrid importance of the keywords in providing attention-weighted image-keyword information.

\subsection{Contextual Transformer Decoder}

The transformer is a state-of-the-art architecture for sequence modeling and transduction tasks \cite{vaswani2017attention}. Its attention mechanism enables the modeling of global dependencies between inputs and outputs, effectively overcoming the memory limitations of traditional recurrent models. Inspired by the architecture of the transformer and its ability to parallelize attention across different positions, we employ it as the primary output decoder in our system.

As illustrated in Figure \ref{fig:figure_6_2}, a contextual transformer decoder cell consists of a masked self-attention unit, an encoder-decoder attention unit, and a final feed-forward layer, resembling its conventional counterpart. We utilize the encoder-decoder structure \cite{vaswani2017attention}, with the encoder directly integrating the attention function $f(\mathbf{v}, \mathbf{k})$. This allows us to describe the decoding process as follows:

\begin{align}
    \label{eq:word_embed}
    \mathbf{x} &= W_e S \\
    \label{eq:pe}
    C_1 &= \mathbf{x} + \textup{PE}(\mathbf{x}) \\
    C'_{l-1} &= \textup{MultiHeadAtt}([C_{l-1}, C_{l-1}, C_{l-1}]) \\
    C_l &= \textup{FCN}(\textup{MultiHeadAtt}([k_{\textup{final}}, k_{\textup{final}}, C'_{l-1}]))
    \label{eq:decoder}
\end{align}

In Equation (\ref{eq:word_embed}), we represent a true sentence describing the image as $S=(S_0,..., S_T)$ and map the bag-of-word IDs into word vectors $\mathbf{x} \in \mathbb{R}^{T \times H_e}$ using the same GloVe embedding layer $W_e$. We then apply positional embedding, as described in Equation (\ref{eq:pe}), to $\mathbf{x}$ to incorporate sequential information. The semantic vector will then pass through multiple attention layer blocks. For each layer of the decoder, we input the data into a self-attention layer followed by an encoder-decoder layer to focus on the image-keyword fusion contexts. We also employ dropout to mitigate noise and reduce the risk of overfitting. Finally, we feed the output from the last layer $C_L$ into a fully connected layer to obtain the joint distribution of the decoding words.

\begin{align}
    P_L &= W_v C_L \\
    L(P|S,I,K) &= \mathbb{E}_{S\sim P_I}{[\textup{log}{P_L(S,I,K)]}}
    \label{eq:loss_7}
\end{align}
Let $P_I$ represent the true medical descriptions for image $I$ provided in the training set, and $P_L(S, I, K)$ denote the final probability distribution obtained after the fully connected layer and softmax function. We can express the overall likelihood function $L(P|S, I, K)$ based on our medical descriptions and the given image, as shown in Equation (\ref{eq:loss_7}). Our objective is to minimize the total loss, calculated as the sum of the negative log-likelihood at each time step. During inference, we employ a ``Greedy Search'' strategy, sampling words based on the maximum likelihood for each output $P_t$ from the predicted distribution $P_{t+1}$ until $P_{t+1}$ equals the special end-of-sentence token.

\section{Experiments}

In this section, we outline the commonly used datasets for retinal image captioning, along with the evaluation metrics employed. Additionally, we provide summaries of the baseline models and the experimental setup utilized in our study.

\subsection{Dataset}

DEN \cite{huang2021deepopht} is a widely recognized benchmark dataset for retinal image captioning, consisting of 15,709 retinal images. Each image is labeled with two types of annotations: expert-defined keywords and concise clinical descriptions, typically 5 to 10 words in length. These labels are provided by experienced retinal specialists following comprehensive image analysis and patient consultations. In this study, we expand the DEN dataset by incorporating an additional 3,145 expert-annotated retinal images, collected using the same rigorous methodology as the original dataset. This expansion brings the total number of images in our dataset to 18,854. For our experiments, we partition the dataset into training, validation, and testing sets in an 80/10/10 split, resulting in 15,083 images for training, 1,885 for validation, and 1,886 for testing.

\begin{table*}[t!]
\caption{
Evaluation results of the proposed model in comparison to several competitive baselines, utilizing expert-defined keywords as ground truth. The term ``BLEU-avg'' refers to the average scores of BLEU-1, BLEU-2, BLEU-3, and BLEU-4. Notably, all keyword-driven models outperform their non-keyword-driven counterparts.
}
\centering
\rotatebox{270}{
\scalebox{0.7}{
\begin{tabular}{c|c|c|c|c|c|c|c|c}
\toprule
\textbf{Model}                                    & \textbf{BLEU-1} & \textbf{BLEU-2} & \textbf{BLEU-3} & \textbf{BLEU-4} & \textbf{BLEU-avg} & \textbf{ROUGE} & \textbf{CIDEr} & \textbf{METEOR} \\ \midrule
LSTM \cite{wang2016image}                & 0.2273	& 0.1650	& 0.1224	& 0.1017	& 0.1541 & 0.2533	& 0.1102	& 0.2437 \\ \midrule
Show and tell \cite{xu2016show}          & 0.4234	& 0.3583	& 0.3002	& 0.2757	& 0.3394 & 0.4463	& 0.3029	& 0.4335 \\ \midrule
Semantic Att \cite{you2016image}         & 0.5904	& 0.5100	& 0.4360	& 0.3969	& 0.4833 & 0.6228	& 0.4460	& 0.6056 \\
\midrule
ContexGPT \cite{huang2021contextualized}  & 0.6254  & 0.5500    & 0.4758    & 0.4344    & 0.5214 & 0.6602   & 0.4951    & 0.6390 \\
\midrule
CoAtt \cite{jing2018automatic}     & 0.6712	& 0.5950	& 0.5211	& 0.4817	& 0.5673 & 0.6988	& 0.5419	& 0.6798 \\ \midrule
H-CoAtt \cite{lu2017hierarchical}         & 0.6718	& 0.5956	& 0.5201	& 0.4829	& 0.5676 & 0.7045	& 0.5417	& 0.6864 \\ \midrule
DeepContex \cite{huang2021deep}           & 0.6749  & 0.6036    & 0.5307    & 0.4890    & 0.5745 & 0.7020   & 0.5496    & 0.6835 \\
\midrule
MIA \cite{liu2019aligning}                & 0.6877	& 0.6138	& 0.5421	& 0.5000	& 0.5859 & 0.7195	& 0.5596	& 0.7006 \\ \midrule
Ours                              & \textbf{0.6969}	& \textbf{0.6195}	& \textbf{0.5496}	& \textbf{0.5008}	& \textbf{0.5892} & \textbf{0.7252}	& \textbf{0.5650}	& \textbf{0.7044} \\ \bottomrule
\end{tabular}}
}
\label{table:table_6_1}
\end{table*}

\begin{table*}[t!]
\caption{
Evaluation results of the proposed model alongside several competitive baselines, utilizing predicted keywords, or pseudo expert-defined keywords. The term ``BLEU-avg'' represents the average scores of BLEU-1, BLEU-2, BLEU-3, and BLEU-4. All keyword-driven models demonstrate superior performance compared to their non-keyword-driven counterparts.
}
\centering
\rotatebox{270}{
\scalebox{0.7}{
\begin{tabular}{c|c|c|c|c|c|c|c|c}
\toprule
\textbf{Model}                                    & \textbf{BLEU-1} & \textbf{BLEU-2} & \textbf{BLEU-3} & \textbf{BLEU-4} & \textbf{BLEU-avg} & \textbf{ROUGE} & \textbf{CIDEr} & \textbf{METEOR} \\ \midrule
LSTM \cite{wang2016image}           & 0.2273	& 0.1650	& 0.1224	& 0.1017	& 0.1541	& 0.2533	& 0.1102	& 0.2437 \\ \midrule
Show and tell \cite{xu2016show}     & 0.4234	& 0.3583	& 0.3002	& 0.2757	& 0.3394	& 0.4463	& 0.3029	& 0.4335 \\ \midrule
H-CoAtt \cite{lu2017hierarchical}    & 0.4465	& 0.3822	& 0.3285	& 0.2969	& 0.3636	& 0.4788	& 0.3409	& 0.4564 \\
\midrule
ContexGPT \cite{huang2021contextualized} & 0.4493 & 0.3744  & 0.3109    & 0.2800	& 0.3536    & 0.4771    & 0.3171    & 0.4588 \\ 
\midrule
Semantic Att \cite{you2016image}     & 0.4541	& 0.3771	& 0.3117	& 0.2777	& 0.3552	& 0.4785	& 0.3118	& 0.4610 \\ \midrule
CoAtt \cite{jing2018automatic}& 0.4647	& 0.4038	& 0.3479	& 0.3162	& 0.3831	& 0.4906	& 0.3563	& 0.4759 \\ 
\midrule
DeepContex \cite{huang2021deep}      & 0.4683   & 0.3966    & 0.3302    & 0.2969    & 0.3730    & 0.4941    & 0.3341    & 0.4803 \\
\midrule
MIA \cite{liu2019aligning}           & 0.5077	& 0.4446	& 0.3861	& 0.3514	& 0.4224	& 0.5326	& 0.3897	& 0.5163 \\ \midrule
Ours                         & \textbf{0.5268}	& \textbf{0.4600}	& \textbf{0.3915}	& \textbf{0.3634}	& \textbf{0.4354}	& \textbf{0.5482}	& \textbf{0.4105}	& \textbf{0.5316} \\ \bottomrule
\end{tabular}}
}
\label{table:table_6_2}
\end{table*}

\begin{table*}[t!]
\caption{
The results indicate the performance decline that occurs when expert-defined keywords are unavailable, specifically in the scenario labeled ``With predicted keywords.''
}
\centering
\rotatebox{270}{
\scalebox{0.7}{
\begin{tabular}{c|c|c|c|c|c|c|c|c}
\toprule
\textbf{Model}                                    & \textbf{BLEU-1} & \textbf{BLEU-2} & \textbf{BLEU-3} & \textbf{BLEU-4} & \textbf{BLEU-avg} & \textbf{ROUGE} & \textbf{CIDEr} & \textbf{METEOR} \\ \midrule
With predicted keywords                          & 0.5268 & 0.4600 & 0.3915 & 0.3634 & 0.4354 & 0.5482 & 0.4105 & 0.5316 \\ \midrule
With expert-defined keywords                       & \textbf{0.6969}	& \textbf{0.6195}	& \textbf{0.5496}	& \textbf{0.5008}	& \textbf{0.5892} & \textbf{0.7252}	& \textbf{0.5650}	& \textbf{0.7044} \\ \bottomrule
\end{tabular}}
}
\label{table:table_6_3}
\end{table*}

\subsection{Performance Evaluation Metrics}

In our experiments, we utilize widely accepted text evaluation metrics \cite{papineni2002bleu,lin2004rouge,vedantam2015cider,banerjee2005meteor,huang2019assessing,huang2019novel,huang2017vqabq,huang2017robustness,huang2017robustness,di2022dawn} that are commonly used in the field of retinal image captioning \cite{li2019knowledge} to assess the generated medical descriptions for retinal images. While these automatic evaluation metrics are popular in both natural and retinal image captioning tasks, their inherent characteristics \cite{papineni2002bleu,lin2004rouge,vedantam2015cider,banerjee2005meteor,huang2021deepopht,huang2021longer} make them more applicable to natural image captioning than to retinal image captioning. Therefore, in this study, we also perform a human expert evaluation of the proposed method, as detailed in Section \hyperref[human:human]{6.5.3}.

\subsection{Baseline Models}

We compare the proposed method with several competitive image captioning models. 
\begin{itemize}
    
    \item \textbf{LSTM} \cite{wang2016image} utilizes a deep CNN combined with a BiLSTM architecture for image captioning.

    \item \textbf{Show and tell} \cite{xu2016show} employs an attention mechanism that selectively focuses on specific regions of the original image while generating descriptions.

    \item \textbf{Semantic Att} \cite{you2016image} forecasts a set of visual attributes that are integrated with the hidden states at both the input and output stages of an RNN caption generator.
    
    \item \textbf{CoAtt} \cite{jing2018automatic} employs a co-attention mechanism to create joint context vectors, which are utilized for generating medical descriptions based on specific topics.
    
    \item \textbf{H-CoAtt} \cite{lu2017hierarchical} introduces a co-attention model designed for VQA tasks, which hierarchically analyzes questions in relation to visual features.
    
    \item \textbf{ContexGPT} \cite{huang2021contextualized} utilizes a non-local attention mechanism, masked self-attention, and a feature reinforcement module to create a retinal image captioning network.

    \item \textbf{DeepContex} \cite{huang2021deep} introduces a context-driven encoding network specifically designed for retinal image captioning.

    \item \textbf{MIA} \cite{liu2019aligning} introduces a mutual iterative attention mechanism that simultaneously captures interactions between images and keywords for both image captioning and VQA.
\end{itemize}

\subsection{Experimental Setup}
We use ResNet50 \cite{he2016deep}, pre-trained on ImageNet, as our retinal image feature extractor, denoted as $\phi$. Initially, we resize the images to the appropriate dimensions for model input, and we extract visual features from the layer just before the final fully connected layer. For processing the annotations and keywords in the dataset, we remove non-alphabetic characters, convert all remaining characters to lowercase, and replace any words that appear only once with a special token, $\langle \textup{UNK} \rangle$. This preprocessing results in a vocabulary size of 3,524. All sentences are truncated or padded to a maximum length of 50. For keyword prediction, we set the threshold $\tau = 0.5$. The word embedding layer uses an embedding size of $H_e = 300$ to encode the words. We implement two transformer blocks, each with 8 attention heads, a hidden size of 2,048 for the fully connected layer, and a hidden size of 64. Finally, we set the mini-batch size to 64 and the learning rate to 1e-4, training all models for 10 epochs.

\begin{table*}[t!]
\caption{
Results of the ablation study on the proposed model structure. The terms ``Image only'' and ``Keywords only'' indicate scenarios where either the image features or keyword features are fed into the model independently. ``Image+Keywords (concat)'' refers to the approach where the image and keyword vectors are simply concatenated before being input into the transformer decoder. In contrast, ``Image+Keywords (coatt)'' represents the complete architecture of the proposed method.
}
\small
\centering
\rotatebox{270}{
\scalebox{0.7}{
\begin{tabular}{c|c|c|c|c|c|c|c|c|c}
\toprule
\multicolumn{2}{c|}{\textbf{Input}}               & \textbf{BLEU-1} & \textbf{BLEU-2} & \textbf{BLEU-3} & \textbf{BLEU-4} & \textbf{B-avg} & \textbf{ROUGE} & \textbf{CIDEr} & \textbf{METEOR} \\ \midrule
\multicolumn{2}{c|}{Image only}               & 0.4357	& 0.3651	& 0.3041	& 0.2773	& 0.3455	& 0.4608	& 0.3067	& 0.4454  \\ \midrule
\multicolumn{2}{c|}{Keywords only}            & 0.5568	& 0.4970	& 0.4322	& 0.3971	& 0.4708	& 0.6110	& 0.4618	& 0.5881 \\ \midrule
\multicolumn{2}{c|}{Image+Keywords (concat)}         & 0.6527	& 0.5752	& 0.4984	& 0.4626	& 0.5472	& 0.6783	& 0.5166	& 0.6643 \\ \midrule
\multicolumn{2}{c|}{Image+Keywords (coatt)}       & \textbf{0.6969}	& \textbf{0.6195}	& \textbf{0.5496}	& \textbf{0.5008}	& \textbf{0.5892} & \textbf{0.7252}	& \textbf{0.5650}	& \textbf{0.7044} 
\\ \bottomrule
\end{tabular}}
}
\label{table:table_6_4}
\end{table*}

\section{Results and Discussion}
\subsection{Quantitative Analysis}
\label{quantitative-analysis:quantitative-analysis}

\noindent\textbf{With expert-defined keywords.}
We begin by presenting the results of medical description generation, using both retinal images and their corresponding expert-defined keywords (i.e., ground truth keywords), as shown in Table \ref{table:table_6_1}. Notably, the vanilla LSTM decoder performs significantly worse than other models that incorporate attention mechanisms, which is expected given its limitations in capturing image dependencies.

The introduction of expert-defined keywords during the generation process leads to substantial improvements in performance across all metrics for keyword-driven models, starting with Semantic Att \cite{you2016image}. This underscores the advantage of using keywords to guide the model towards more accurate predictions. These human-comprehensible keywords also enhance interpretability, as discussed in Section \hyperref[human:human]{6.5.3} regarding human expert evaluation.

Further enhancements from the co-attention mechanism between images and keywords reinforce our confidence that the models effectively focus on integrated representations of both visual and semantic concepts. By leveraging mutual attention weights from the images and keywords, our model, with a transformer decoder instead of LSTM, outperforms all baselines. This architecture allows for comprehensive referencing of previous tokens and fusion concepts when generating the next token. Overall, we observe an increase of 74\% in BLEU average, 63\% in ROUGE, 87\% in CIDEr, and 63\% in METEOR compared to non-keyword-driven attention models \cite{xu2016show}.

\begin{figure*}[t!]
\begin{center}
\includegraphics[width=1.0\linewidth]{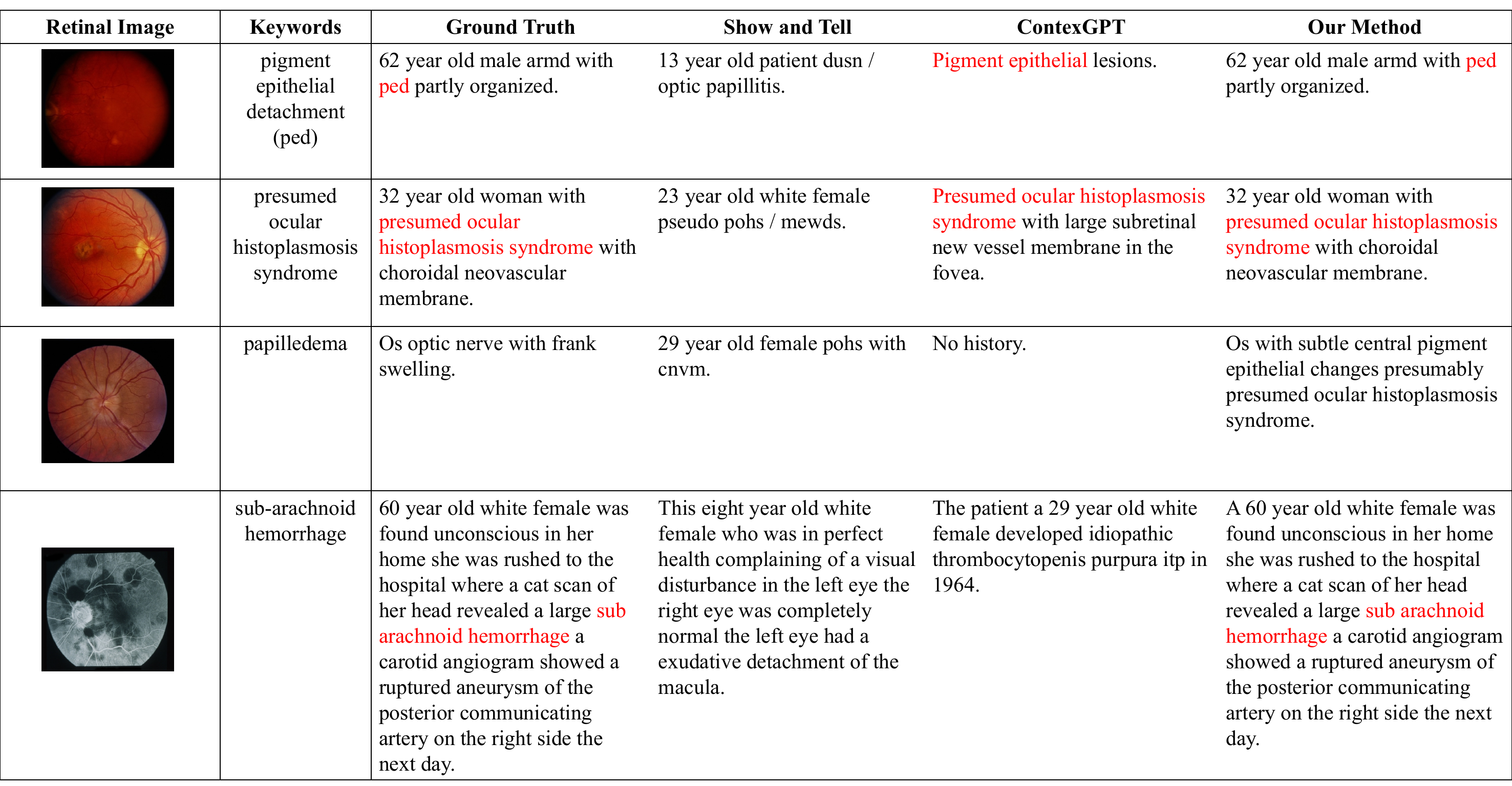}
\end{center}
  \caption{
  Illustration of the descriptions generated by the proposed model alongside two baseline models \cite{xu2016show, huang2021contextualized}.
  }
\label{fig:figure_6_3}
\end{figure*}

\begin{figure*}[t!]
\begin{center}
\includegraphics[width=1.0\linewidth]{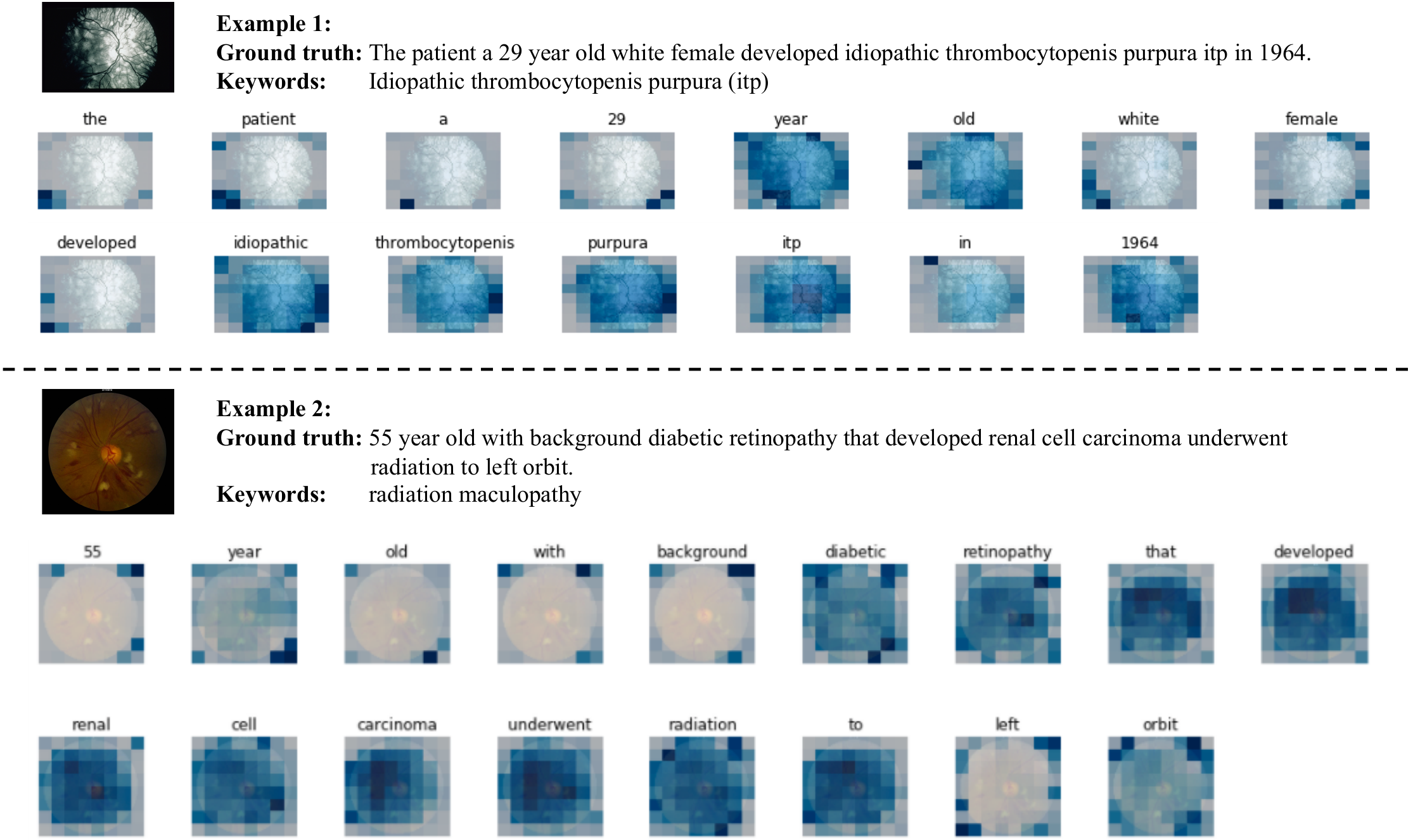}
\end{center}
  \caption{
  Visualization of image attention throughout the text generation process. Each example includes an input image, ground truth descriptions, and predicted keywords. The predicted words are displayed on top of the image attention for each time step.
  }
\label{fig:figure_6_4}
\end{figure*}

\noindent\textbf{Keyword prediction.}
To create a more general setting, we also present experimental results using predicted keywords—i.e., pseudo-expert-defined keywords—generated by our pre-trained multi-label classifier, as shown in Tables \ref{table:table_6_2} and \ref{table:table_6_3}. We observe that the benefits of keyword fusion diminish due to some inaccurately predicted keywords. This presents a significant challenge in medical description generation, as there are 3,465 keyword options in the DEN dataset, with an undetermined number of keywords applicable to each image. Nevertheless, we still see an overall improvement across all metrics when incorporating the predicted keyword contexts. Our approach notably surpasses several co-attention-based methods, which tend to overfit the training data due to their complexity. The single cross-attention mechanism embedded in our transformer decoder, which connects visual and semantic concepts, proves more resilient to keyword noise across multiple-layer transitions.

\noindent\textbf{Co-attention between image and keywords.}
Table \ref{table:table_6_4} reveals that the performance of the ``Image only'' and ``Keyword only'' baselines is significantly inferior to the ``Image+Keywords'' methods. This indicates that the interaction between keywords and images is essential for generating effective medical reports. To further illustrate the advantages of attention mechanisms, we include another baseline where image and keyword features are simply concatenated without any additional fusion. The results show a substantial decline in performance when attention mechanisms are not utilized.

\subsection{Qualitative Results and Analysis}

\noindent\textbf{Comparison with the classic attention model.}
We present qualitative results from three medical description generation models, including ours, as shown in Figure \ref{fig:figure_6_3}. The ``Show and Tell'' model \cite{xu2016show} does not utilize any keywords, while ContexGPT \cite{huang2021contextualized} serves as a keyword-oriented baseline. In the first two images, both ContexGPT and our model generate descriptions that effectively reference the predicted keywords. However, our model matches the ground truth descriptions precisely. In contrast, the Show and Tell model lacks explicit textual attributes, resulting in inaccuracies in symptom names and detailed descriptions. Similarly, ContexGPT struggles to maintain accurate semantic information, even with keyword guidance. These observations highlight the necessity of our pre-trained keyword predictor, which first provides a coarse tagging of the retinal image before elaborating on the details—an approach that is particularly crucial and intuitive in medical contexts compared to general domains.

In the third image, the keywords are not explicitly mentioned in the ground truth; instead, it focuses on symptom illustration. Both baseline models generate irrelevant descriptions for the image, while our model offers a more detailed account of the phenomena associated with the symptoms. Additionally, in the last image, we further showcase the robustness of our model, as it generates longer, more coherent descriptions through effective context fusion, in contrast to the less structured expressions produced by the other baselines.

\noindent\textbf{Does our model fully understand the fused concepts?}
To gain insights into how our model leverages the fused visual and semantic concepts for token sampling, we visualize the attention weights on the input image at each time step, as shown in Figure \ref{fig:figure_6_4}. Each image is divided into 64 patches, and each patch is assigned a weight corresponding to the current word and the keyword-fused image patch. Our model shows reduced sensitivity to words specifying numbers (e.g., 29, 55) or colors (e.g., white), indicating that it relies primarily on the input image and keywords. However, it places significant emphasis on specific image regions when predicting medical keywords. We observe a consistent trend in saliency among consecutive words, which suggests that our model identifies abnormalities in the particular image. These highlighted regions demonstrate promising interpretability, revealing how our model processes keywords and image patches to generate coherent sequences.

\subsection{Evaluations with Retinal Specialists}
\label{human:human}

We employed a 5-level report quality evaluation scale, ranging from 1 to 5, where higher scores indicate better quality. Due to limited research resources, we randomly selected 100 samples from our model-generated reports and their corresponding ground-truth reports. Five different retinal specialists evaluated the quality of both the model-generated and ground-truth reports without knowing which was which. Our findings revealed an average score of 4.0 out of 5.0 for the model-generated reports and an average score of 4.3 out of 5.0 for the ground-truth reports. Given that the ground-truth reports were authored by ophthalmologists, these results suggest that our proposed method performs competitively compared to human expert baselines.

We applied the same evaluation setup to assess interpretability. In the first scenario, we presented the 100 generated reports without their corresponding keywords to the retinal specialists, yielding an interpretability score of 4.0 out of 5.0. In the second scenario, the same reports were presented alongside their keywords, resulting in an interpretability score of 4.6 out of 5.0. This indicates that the inclusion of keywords significantly enhances the interpretability of the generated reports.

\section{Conclusion}
In summary, we propose an explainable retinal report generation method using expert keywords and a novel attention-based strategy. This approach effectively predicts essential technical keywords and integrates them for enhanced word sampling. Experimental results demonstrate that our model generates more accurate and meaningful descriptions for retinal images, achieving improvements of approximately 74\% in BLEU average, 63\% in ROUGE, 87\% in CIDEr, and 63\% in METEOR compared to non-keyword attention-based baselines. Attention visualization reveals intriguing patterns associated with potential symptoms in specific image regions.

\chapter{Summary and Conclusions}\label{ch:conclusion}
\section{Summary}

This thesis addresses the critical challenge of automating medical report generation for retinal images, a task driven by the rising incidence of retinal diseases and the increasing demand for ophthalmic care, which far exceeds the capacity of the available workforce. The primary research question guiding this work is: \textit{How can we improve automatic medical report generation to optimize the efficiency and effectiveness of traditional treatment processes for retinal diseases?}
In \textbf{Chapter~\ref{ch:ch2}}, we present an AI-based method that integrates DNNs to improve retinal disease diagnosis with enhanced report generation and visual explanations. In \textbf{Chapter~\ref{ch:ch4}}, we introduce a multi-modal model incorporating contextualized keyword representations and masked self-attention to boost model performance. In \textbf{Chapter~\ref{ch:ch5}}, we propose TransFuser, an attention-based feature fusion technique that effectively merges keywords and image data for improved report generation. Finally, in \textbf{Chapter~\ref{ch:ch6}}, we enhance interpretability by using expert-defined keywords with a specialized attention-based strategy. A brief overview of each chapter is provided below:

\textbf{Chapter~\ref{ch:ch2}: Report Generation with Deep Models and Visual Explanations.} 
This chapter presents an AI-based method to enhance retinal disease diagnosis by improving both efficiency and accuracy. Our approach features a DNN module that includes a disease identifier, CDG, and visual explanation component. We validate its performance using a large-scale retinal disease image dataset and a manually annotated dataset by ophthalmologists. Results show that our method effectively generates meaningful image descriptions and relevant visual explanations.

\textbf{Chapter~\ref{ch:ch4}: Contextualized Features for Multi-modal Retinal Image Captioning.} 
In this chapter, we tackle the limitations of traditional medical image captioning models that struggle with abstract concepts due to their reliance on single-image inputs. We propose a multi-modal model that combines contextualized keyword representations, textual feature reinforcement, and masked self-attention. This approach significantly enhances performance, outperforming state-of-the-art methods on widely used metrics such as BLEU-avg and CIDEr.

\textbf{Chapter~\ref{ch:ch5}: Non-Local Attention Enhances Retinal Image Captioning.} 
This chapter enhances retinal image report generation by incorporating expert-defined keywords, commonly recorded early in diagnosis by ophthalmologists. To effectively merge these keywords with image data, we introduce TransFuser, an attention-based multi-modal feature fusion method. Our experiments demonstrate that TransFuser captures essential mutual information between keywords and images, resulting in a keyword-driven generation model that surpasses baseline methods across metrics like BLEU, CIDEr, and ROUGE.

\textbf{Chapter~\ref{ch:ch6}: Expert-defined Keywords Improve Captioning Interpretability.} 
Interpretability is crucial for trust in ML systems for medical report generation, yet traditional techniques like heat maps often fall short. In this chapter, we propose a novel approach that uses expert-defined keywords combined with a specialized attention-based strategy to enhance interpretability. These keywords, coupled with a specialized attention-based strategy, encapsulate domain knowledge in a way that users easily understand. Our method not only makes the system more interpretable but also achieves state-of-the-art results on key text evaluation metrics.

\section{Conclusions}
In conclusion, this thesis addresses the pressing challenge posed by the increasing prevalence of retinal diseases and the growing demand for ophthalmic care, which far exceeds the available workforce. By investigating the potential of AI in automating medical report generation from retinal images, we propose a novel multi-modal deep learning framework that significantly enhances the efficiency and accuracy of the diagnostic process. Traditional methods of generating medical reports are often labor-intensive and error-prone, straining the resources of ophthalmologists. In contrast, our AI-based approach not only automates this process but also identifies subtle patterns within large volumes of image data, allowing for more precise diagnoses.

The proposed methods tackle key challenges in automated report generation by capturing the intricate interactions between textual keywords and retinal images, leading to the creation of more comprehensive medical reports. Our enhanced keyword representation techniques improve the precision of medical terminology, while the transformer-based architecture overcomes the limitations of RNN-based models in capturing long-range dependencies in medical descriptions. Moreover, we prioritize interpretability in AI systems, ensuring that our model builds trust and fosters acceptance in clinical practice.

Rigorous evaluations demonstrate that our proposed model achieves state-of-the-art performance across multiple metrics, underscoring AI's potential to revolutionize retinal disease diagnosis. By automating medical report generation, we aim to alleviate the workload of ophthalmologists, enabling them to concentrate on more complex cases and ultimately enhancing clinical efficiency and patient care. As we move forward, our research lays a solid foundation for further exploration of AI-driven solutions in healthcare, emphasizing the need for ongoing advancements in interpretability and accuracy to ensure effective integration into clinical workflows.



\newpage

\bibliographystyle{abbrv}
\bibliography{thesis}


\appendix
\chapter{Complete List of Publications}
\label{ch:pub}

\begin{itemize}
  \item \textbf{Jia-Hong Huang}, Yixian Shen, Evangelos Kanoulas. ``GradNormLoRP: Gradient Weight-normalized Low-rank Projection 
        for Efficient LLM Training'', in submission to \textbf{Neural Information Processing Systems (NeurIPS)}, 2024.
  \item \textbf{Jia-Hong Huang}, Weijuan Xi, Eunyoung Kim, Wonsik Kim, Amelio Vázquez-Reina. ``CSR-ASD: Cross-modal Signal 
        Reprogramming Enhancing Audio-Visual Active Speaker Detection'', in submission to \textbf{European Conference on Computer Vision (ECCV)}, 2024.
  \item Hongyi Zhu*, \textbf{Jia-Hong Huang*}, Yixian Shen, Shuai Wang, Stevan Rudinac, Evangelos Kanoulas. ``Enhancing Text-to-Video Retrieval Through Multi-Modal Multi-Turn Conversation'', in submission to \textbf{ACM Multimedia (ACMMM)}, 2024.
  \item Shuai Wang, David W Zhang, \textbf{Jia-Hong Huang}, Stevan Rudinac, Monika Kackovic, Nachoem Wijnberg, Marcel Worring. 
        ``Ada-HGNN: Adaptive Sampling for Scalable Hypergraph Neural Networks'', in submission to \textbf{ACM International Conference on Information and Knowledge Management (CIKM)}, 2024. \cite{wang2024ada}
  \item \textbf{Jia-Hong Huang}, Chao-Han Huck Yang, Pin-Yu Chen, Min-Hung Chen, Marcel Worring. ``Conditional Modeling 
        Based Automatic Video Summarization'', in submission to \textbf{ACM Transactions on Multimedia Computing, Communications, and Applications (TOMM)}, 2024. \cite{huang2023conditional}
  \item \textbf{Jia-Hong Huang}, Modar Alfadly, Bernard Ghanem, Marcel Worring. ``Improving Visual Question Answering Models 
        through Robustness Analysis and In-Context Learning with a Chain of Basic Questions'', in submission to \textbf{IEEE 
        Transactions on Multimedia (TMM)}, 2024. \cite{huang2019assessing,huang2023improving}
  \item \textbf{Jia-Hong Huang}. ``Personalized Video Summarization using Text-Based Queries and Conditional
Modeling'', 
        \textbf{University of Amsterdam, Doctoral Thesis}, 2024. \cite{huang2024personalized}
  \item \textbf{Jia-Hong Huang*}, Hongyi Zhu*, Yixian Shen, Stevan Rudinac, Alessio M. Pacces, Evangelos Kanoulas. ``A Novel 
        Evaluation Framework for Image2Text Generation'', \textbf{International ACM SIGIR Conference on Research and Development in Information Retrieval (SIGIR)} LLM4Eval Workshop, 2024. \cite{huang2024novel}
  \item \textbf{Jia-Hong Huang*}, Chao-Chun Yang*, Yixian Shen, Alessio M. Pacces, Evangelos Kanoulas. ``Optimizing Numerical 
        Estimation and Operational Efficiency in the Legal Domain through Large Language Models'', \textbf{ACM International Conference on Information and Knowledge Management (CIKM)}, 2024. \cite{huang2024optimizing}
  \item \textbf{Jia-Hong Huang}. ``Multi-modal Video Summarization'', \textbf{ACM International Conference on Multimedia 
        Retrieval (ICMR)}, 2024. \cite{huang2024multi}
  \item Hongyi Zhu*, \textbf{Jia-Hong Huang*}, Stevan Rudinac, Evangelos Kanoulas. ``Enhancing Interactive Image Retrieval With 
        Query Rewriting Using Large Language Models and Vision Language Models'', \textbf{ACM International Conference on Multimedia Retrieval (ICMR)}, 2024. \cite{zhu2024enhancing}
  \item Weijia Zhang*, \textbf{Jia-Hong Huang*}, Svitlana Vakulenko, Yumo Xu, Thilina Rajapakse, Evangelos Kanoulas. ``Beyond 
        Relevant Documents: A Knowledge-Intensive Approach for Query-Focused Summarization using Large Language Models'', \textbf{International Conference on Pattern Recognition (ICPR)}, 2024. \cite{zhang2024beyond}
  \item Weijia Zhang, Mohammad Aliannejadi, Jiahuan Pei, Yifei Yuan, \textbf{Jia-Hong Huang}, Evangelos Kanoulas. ``A Comparative 
        Analysis of Faithfulness Metrics and Humans in Citation Evaluation'', \textbf{International ACM SIGIR Conference on Research and Development in Information Retrieval (SIGIR)} LLM4Eval Workshop, 2024. \cite{zhang2024comparative}
  \item Weijia Zhang, Mohammad Aliannejadi, Yifei Yuan, Jiahuan Pei, \textbf{Jia-Hong Huang}, Evangelos Kanoulas. ``Towards Fine- 
        Grained Citation Evaluation in Generated Text: A Comparative Analysis of Faithfulness Metrics'', \textbf{International Natural Language Generation Conference (INLG)}, 2024. [\textbf{Oral}] \cite{zhang2024towards}
  \item Weijia Zhang, Vaishali Pal, \textbf{Jia-Hong Huang}, Evangelos Kanoulas, Maarten de Rijke. ``Beyond Relevant Documents: A 
        Knowledge-Intensive Approach for Query-Focused Summarization using Large Language Models'', \textbf{European Conference on Artificial Intelligence (ECAI)}, 2024. \cite{zhang2024qfmts}
  \item \textbf{Jia-Hong Huang}, Luka Murn, Marta Mrak, Marcel Worring. ``Query-based Video Summarization with Pseudo Label 
        Supervision'', \textbf{IEEE International Conference on Image Processing (ICIP)}, 2023. \cite{huang2023query}
  \item \textbf{Jia-Hong Huang}, Chao-Han Huck Yang, Pin-Yu Chen, Min-Hung Chen, Marcel Worring. ``Causalainer: Causal    
        Explainer for Automatic Video Summarization'', \textbf{IEEE/CVF Conference on Computer Vision and Pattern Recognition (CVPR)} New Frontiers in Visual Language Reasoning Workshop, 2023. \cite{huang2023causalainer}
  \item Ting-Wei Wu*, \textbf{Jia-Hong Huang*}, Joseph Lin, Marcel Worring. ``Expert-defined Keywords Improve 
        Interpretability of Retinal Image Captioning'', \textbf{IEEE/CVF Winter Conference on Applications of Computer Vision (WACV)}, 2023. [\textbf{Oral}]. \cite{wu2023expert}
  \item \textbf{Jia-Hong Huang}, Chao-Han Huck Yang, Pin-Yu Chen, Andrew Brown, Marcel Worring. ``Causal Video Summarizer 
        for Video Exploration'', \textbf{IEEE International Conference on Multimedia and Expo (ICME)}, 2022. \cite{huang2022causal}
  \item Riccardo Di Sipio, \textbf{Jia-Hong Huang}, Samuel Yen-Chi Chen, Stefano Mangini, Marcel Worring. ``The Dawn of 
        Quantum Natural Language Processing'', \textbf{IEEE International Conference on Acoustics, Speech and Signal Processing (ICASSP)}, 2022. \cite{di2022dawn}
  \item \textbf{Jia-Hong Huang}, Ting-Wei Wu, Chao-Han Huck Yang, Zenglin Shi, I-Hung Lin, Jesper Tegner, Marcel Worring. 
        ``Non-local Attention Improves Description Generation for Retinal Images'', \textbf{IEEE/CVF Winter Conference on Applications of Computer Vision (WACV)}, 2022. [\textbf{Oral}]. \cite{huang2022non}
  \item \textbf{Jia-Hong Huang}, Marta Mrak, Luka Murn, Marcel Worring. ``GPT2MVS: Generative Pre-trained Transformer-2 for 
        Multi-modal Video Summarization'', \textbf{ACM International Conference on Multimedia Retrieval (ICMR)}, 2021. [\textbf{Oral}]. \cite{huang2021gpt2mvs}
  \item \textbf{Jia-Hong Huang}, Ting-Wei Wu, Marcel Worring. ``Contextualized Keyword Representations for Multi-modal 
        Retinal Image Captioning'', \textbf{ACM International Conference on Multimedia Retrieval (ICMR)}, 2021. \cite{huang2021contextualized}
  \item \textbf{Jia-Hong Huang}, Ting-Wei Wu, Chao-Han Huck Yang, Marcel Worring. ``Deep Context-Encoding Network for 
        Retinal Image Captioning'', \textbf{IEEE International Conference on Image Processing (ICIP)}, 2021. \cite{huang2021deep,huang2021longer}
  \item \textbf{Jia-Hong Huang}, Chao-Han Huck Yang, Fangyu Liu, Meng Tian, Yi-Chieh Liu, Ting-Wei Wu, I-Hung Lin, Kang 
        Wang, Hiromasa Morikawa, Herng-Hua Chang, Jesper Tegner, Marcel Worring. ``DeepOpht: Medical Report Generation for Retinal Images via Deep Models and Visual Explanation'', \textbf{IEEE/CVF Winter Conference on Applications of Computer Vision (WACV)}, 2021. \cite{huang2021deepopht}
  \item \textbf{Jia-Hong Huang}, Marcel Worring. ``Query-controllable Video Summarization'', \textbf{ACM International 
        Conference on Multimedia Retrieval (ICMR)}, 2020. \cite{huang2020query}
  \item \textbf{Jia-Hong Huang}, Cuong Duc Dao, Modar Alfadly, Bernard Ghanem. ``A Novel Framework for Robustness Analysis 
        of Visual QA Models'', \textbf{AAAI Conference on Artificial Intelligence (AAAI)}, 2019. [\textbf{Oral}] \cite{huang2019novel}
  \item Tao Hu, Pascal Mettes, \textbf{Jia-Hong Huang}, Cees G. M. Snoek. ``SILCO: Show a Few Images, Localize the Common 
        Object'', \textbf{IEEE/CVF International Conference on Computer Vision (ICCV)}, 2019. \cite{hu2019silco}
  \item \textbf{Jia-Hong Huang}, Cuong Duc Dao, Modar Alfadly, Chao-Han Huck Yang, Bernard Ghanem. ``Robustness Analysis of  
        isual QA Models by Basic Questions'', \textbf{IEEE/CVF Conference on Computer Vision and Pattern Recognition (CVPR)} Visual Dialog Workshop, 2018. \cite{huang2018robustness}
  \item Chao-Han Huck Yang*, \textbf{Jia-Hong Huang*}, Fangyu Liu, Fang-Yi Chiu, Mengya Gao, Weifeng Lyu, I-Hung Lin, Jesper 
        Tegner. ``A Novel Hybrid Machine Learning Model for Auto-Classification of Retinal Diseases'', \textbf{International Conference on Machine Learning (ICML)} Computational Biology Workshop, 2018. \cite{yang2018novel}
  \item Yi-Chieh Liu, Hao-Hsiang Yang, Chao-Han Huck Yang, \textbf{Jia-Hong Huang}, Meng Tian, Hiromasa Morikawa, Yi-Chang 
        James Tsai, Jesper Tegner. ``Synthesizing New Retinal Symptom Images by Multiple Generative Models'', \textbf{Asian Conference on Computer Vision (ACCV)} Artificial Intelligence Applied to Retinal Image Analysis Workshop, 2018. [\textbf{Oral}] \cite{liu2019synthesizing}
  \item Chao-Han Huck Yang*, Fangyu Liu*, \textbf{Jia-Hong Huang*}, Meng Tian, I-Hung Lin, Yi-Chieh Liu, Hiromasa Morikawa, 
        Hao-Hsiang Yang, Jesper Tegner. ``Auto-Classification of Retinal Diseases in the Limit of Sparse Data Using a Two-Streams Machine Learning Model'', \textbf{Asian Conference on Computer Vision (ACCV)} Artificial Intelligence Applied to Retinal Image Analysis Workshop, 2018. \cite{huck2019auto}
  \item \textbf{Jia-Hong Huang}, Modar Alfadly, Bernard Ghanem. ``VQABQ: Visual Question Answering by Basic Questions'', 
        \textbf{IEEE/CVF Conference on Computer Vision and Pattern Recognition (CVPR)} Visual Question Answering Workshop, 2017. \cite{huang2017vqabq,huang2017robustness}
\end{itemize}

\chapter{Samenvatting}
Deze thesis richt zich op de kritieke uitdaging van het automatiseren van de generatie van medische rapporten voor retinale beelden, een taak die wordt gedreven door de toenemende incidentie van retinale ziekten en de groeiende vraag naar oogheelkundige zorg, die ver boven de capaciteit van de beschikbare arbeidskrachten uitstijgt. De centrale onderzoeksvraag die dit werk begeleidt is: \textit{Hoe kunnen we de automatische generatie van medische rapporten verbeteren om de efficiëntie en effectiviteit van traditionele behandelingsprocessen voor retinale ziekten te optimaliseren?} 
In \textbf{Hoofdstuk~\ref{ch:ch2}} presenteren we een AI-gebaseerde methode die diepe neurale netwerken integreert om de diagnose van retinale ziekten te verbeteren met verbeterde rapportgeneratie en visuele uitleg. In \textbf{Hoofdstuk~\ref{ch:ch4}} introduceren we een multimodaal model dat contextuele keywordrepresentaties en gemaskeerde zelfaandacht combineert om de prestaties van het model te verbeteren. In \textbf{Hoofdstuk~\ref{ch:ch5}} stellen we TransFuser voor, een aandacht-gebaseerde functie-fusietechniek die effectief keywords en afbeeldingsdata samenvoegt voor verbeterde rapportgeneratie. Ten slotte verbeteren we in \textbf{Hoofdstuk~\ref{ch:ch6}} de interpreteerbaarheid door gebruik te maken van door experts gedefinieerde keywords met een gespecialiseerde aandacht-gebaseerde strategie. Hieronder volgt een beknopt overzicht van elk hoofdstuk:

\textbf{Hoofdstuk~\ref{ch:ch2}: Rapportgeneratie met Diepe Modellen en Visuele Uitleg.} 
Dit hoofdstuk presenteert een AI-gebaseerde methode om de diagnose van retinale ziekten te verbeteren door zowel de efficiëntie als de nauwkeurigheid te verhogen. Onze aanpak omvat een diep neurale netwerk (DNN)-module die een ziekte-identificator, generator voor klinische beschrijvingen en een visuele uitlegcomponent bevat. We valideren de prestaties met behulp van een grootschalige dataset van retinale ziekten en een handmatig geannoteerde dataset door oogartsen. De resultaten tonen aan dat onze methode effectief betekenisvolle afbeeldingsbeschrijvingen en relevante visuele uitleg genereert.

\textbf{Hoofdstuk~\ref{ch:ch4}: Contextuele Kenmerken voor Multimodale Captioning van Retinale Beelden.} 
In dit hoofdstuk pakken we de beperkingen aan van traditionele modellen voor het captionen van medische beelden die moeite hebben met abstracte concepten door hun afhankelijkheid van single-image inputs. We stellen een multimodaal model voor dat contextuele keywordrepresentaties, tekstuele versterking van functies en gemaskeerde zelfaandacht combineert. Deze aanpak verbetert de prestaties aanzienlijk en overtreft state-of-the-art methoden op veelgebruikte metrics zoals BLEU-avg en CIDEr.

\textbf{Hoofdstuk~\ref{ch:ch5}: Non-Local Aandacht Verbetert Captioning van Retinale Beelden.} 
Dit hoofdstuk verbetert de generatie van rapporten voor retinale beelden door gebruik te maken van door experts gedefinieerde keywords, die vaak vroeg in de diagnose door oogartsen worden genoteerd. Om deze keywords effectief te combineren met afbeeldingsdata, introduceren we TransFuser, een aandacht-gebaseerde multimodale functie-fusiemethode. Onze experimenten tonen aan dat TransFuser essentiële wederzijdse informatie tussen keywords en afbeeldingen vastlegt, wat resulteert in een keyword-gedreven generatie model dat baseline-methoden overtreft op metrics zoals BLEU, CIDEr en ROUGE.

\textbf{Hoofdstuk~\ref{ch:ch6}: Door Experts Gedefinieerde Keywords Verbeteren de Interpreteerbaarheid van Captioning.} 
Interpreteerbaarheid is cruciaal voor vertrouwen in ML-systemen voor medische rapportgeneratie, maar traditionele technieken zoals heatmaps schieten vaak tekort. In dit hoofdstuk stellen we een nieuwe aanpak voor die gebruikmaakt van door experts gedefinieerde keywords in combinatie met een gespecialiseerde aandacht-gebaseerde strategie om de interpreteerbaarheid te verbeteren. Deze keywords, gecombineerd met een gespecialiseerde aandacht-gebaseerde strategie, vangen domeinkennis op een manier die voor gebruikers gemakkelijk te begrijpen is. Onze methode maakt het systeem niet alleen beter interpreteerbaar, maar behaalt ook state-of-the-art resultaten op belangrijke tekst-evaluatiemetrics.

\end{document}